\PassOptionsToPackage{pdfpagelabels=false}{hyperref} 
\documentclass[useAMS,usenatbib, usedcolumn]{mnras}
\pdfoutput=1 
\usepackage{amsmath, amssymb, amstext}
\usepackage{aas_macros}
\usepackage{natbib}
\usepackage{rotating}
\usepackage{xspace}
\usepackage{booktabs}
\usepackage{multirow}
\usepackage{color}
\usepackage{verbatim}
\usepackage[caption=false]{subfig}
\input{hyperlink-year-only-natbib-patch}

\newcommand{\hMsun}{{h^{-1}M_{\odot}}\xspace}

\newcommand{\Rvir}{\ensuremath{R_\mathrm{vir}}\xspace}
\newcommand{\Ob}{\ensuremath{\Omega_{\rm b}}}
\newcommand{\Ol}{\ensuremath{\Omega_\Lambda}}

\newcommand{\Om}{\ensuremath{\Omega_{\mathrm m}}}

\newcommand{\hod}{{\small{HOD}}\xspace}
\newcommand{\ngal}{\ensuremath{n_\mathrm{gal}}\xspace}
\newcommand{\wprp}{\ensuremath{w_p(r_p)}\xspace}
\newcommand{\gmf}{\ensuremath{n(N)}\xspace}
\newcommand{\Nsatavg}{\ensuremath{\langle N_\mathrm{sat} \rangle}}
\newcommand{\Ncenavg}{\ensuremath{\langle N_\mathrm{cen} \rangle}}

\newcommand{\Nsat}{\ensuremath{N_\mathrm{sat}}}
\newcommand{\Mmin}{\ensuremath{M_\mathrm{min}}}
\newcommand{\erf}{{\ensuremath{\mathrm{erf}}\xspace}}
\newcommand{\logM}{ \ensuremath{\log M}\xspace}
\newcommand{\logMmin}{ \ensuremath{\log M_\mathrm{min}}\xspace}
\newcommand{\logMone}{\ensuremath{\log M_{1}}\xspace}
\newcommand{\siglogM}{\ensuremath{\sigma_{\log M}}\xspace}
\newcommand{\siglogL}{\ensuremath{\sigma_{\log \mathrm{L}}}\xspace}
\newcommand{\hMpc}{\ensuremath{{h^{-1}\mathrm{Mpc}}\xspace}}
\newcommand{\hkpc}{\ensuremath{{h^{-1}\mathrm{kpc}}\xspace}}
\newcommand*{\bigchi}{{\Large{\ensuremath{\chi}}}}
\newcommand{\pvalue}{\ensuremath{p-}value\xspace}
\newcommand{\pvalues}{\ensuremath{p-}values\xspace}

\newcommand{\emcee}{\texttt emcee}

\newcommand{\LCDM}{\Lambda\mathrm{CDM}}
\newcommand*{\mailto}[1]{\href{mailto:#1}{#1}}

\hypersetup{
    pdfauthor={Manodeep Sinha},
    pdftitle={Towards Accurate Modelling of Galaxy Clustering on Small Scales: Testing the Standard $\LCDM$ + Halo Model},
    pdfkeywords={cosmology: theory --- cosmology: dark matter --- cosmology: large-scale structure of Universe 
--- galaxies: evolution --- galaxies: halos --- galaxies: groups: general --- methods: numerical},
    colorlinks=true,
    citecolor=blue,
    linkcolor=magenta,
    urlcolor=cyan}

\title[Towards Accurate Modelling of Small-Scale Galaxy Clustering]{Towards Accurate Modelling of Galaxy Clustering on Small Scales: Testing the Standard $\LCDM$ + Halo Model}
\author[Sinha et al.]{%
Manodeep Sinha$^{1, 2, 6}$\thanks{E-mail: \mailto{msinha@swin.edu.au}},
Andreas A. Berlind$^{1}$,
Cameron K.\ McBride$^{3}$,
\newauthor 
Roman Scoccimarro$^{4}$,
Jennifer A.\ Piscionere$^{1, 2}$, \& 
Benjamin D.\ Wibking$^{1,5}$\vspace*{0.2em} \\ 
$^{1}$Department of Physics and Astronomy, Vanderbilt University, Nashville, TN 37235, USA\\
$^{2}$Centre for Astrophysics \& Supercomputing, Swinburne University of Technology, 1 Alfred St., Hawthorn, VIC 3122, Australia\\
$^{3}$Harvard-Smithsonian Center for Astrophysics, 60 Garden St., Cambridge, MA 02138\\
$^{4}$Center for Cosmology and Particle Physics, New York University, New York, NY 10003\\
$^{5}$Department of Astronomy, The Ohio State University, Columbus, OH 43210\\
$^{6}$ARC Centre of Excellence for All Sky Astrophysics in 3 Dimensions (ASTRO 3D)
}

\date{Accepted XXX. Received YYY; in original form ZZZ}

\pubyear{2017}

\begin{document}
\label{firstpage}
\pagerange{\pageref{firstpage}--\pageref{lastpage}}
\maketitle

\begin{abstract}
Interpreting the small-scale clustering of galaxies with halo models can elucidate the connection between 
galaxies and dark matter halos. Unfortunately, the modelling is typically not sufficiently accurate for ruling 
out models statistically. It is thus difficult to use the information encoded in small scales to test 
cosmological models or probe subtle features of the galaxy-halo connection. In this paper, we attempt to 
push halo modelling into the ``accurate'' regime with a fully numerical mock-based methodology and careful 
treatment of statistical and systematic errors. With our forward-modelling
approach, we can incorporate clustering statistics beyond the traditional
two-point statistics. We use this modelling methodology to test the 
standard $\Lambda$CDM + halo model against the clustering of SDSS DR7 galaxies. Specifically, we use 
the projected correlation function, group multiplicity function and galaxy number density as constraints. We 
find that while the model fits each statistic separately, it struggles to fit them simultaneously. 
Adding group statistics leads to a more stringent test of the model and significantly tighter constraints on 
model parameters. We explore the impact of varying the adopted halo definition and cosmological model 
and find that changing the cosmology makes a significant difference. The most successful model we tried 
(Planck cosmology with Mvir halos) matches the clustering of low luminosity galaxies, but exhibits a 
$2.3\sigma$ tension with the clustering of luminous galaxies, thus providing evidence that the ``standard'' 
halo model needs to be extended. This work opens the door to adding interesting freedom to the halo model and including 
additional clustering statistics as constraints.
\end{abstract}


\section{INTRODUCTION}

\setcounter{footnote}{0}

The observed spatial distribution of galaxies contains a richness of information about the initial conditions 
and subsequent evolution of the matter density field in the universe. Moreover, the dependence of the 
spatial distribution of galaxies on their observed properties contains information about the physics of galaxy 
formation and evolution. The study of galaxy clustering has thus proved to be a fruitful avenue for constraining 
cosmology and galaxy formation theory and it has provided the primary motivation for the construction of 
large astronomical galaxy surveys.

The specific information content of galaxy clustering depends on the physical scales considered. On large 
scales, galaxy clustering provides fairly clean constraints on cosmology because galaxies are simple tracers 
of the matter density field \citep[e.g.,][]{scherrer_weinberg_98,narayanan_etal_00}. Prime examples of such
constraints are measurements of the large-scale galaxy power spectrum \citep[e.g.,][]{tegmark_etal_04a} and 
the Baryon Acoustic Oscillation (BAO) signature \citep{eisenstein_etal_05}. On small scales ($\lesssim 10\hMpc$),
galaxy clustering depends both on cosmology and on the detailed relationship between the galaxy and dark 
matter density fields, i.e., the {\it bias}, which is nontrivial and is set by the physics of galaxy formation.
Although in principle galaxy clustering measurements on small scales have the potential to constrain both 
cosmological and galaxy formation theories, in practice this is extremely challenging because the process of
galaxy formation is very complex and we currently lack a complete predictive theory for it. Moreover, the most 
accurate theories that do exist require computationally expensive hydrodynamic simulations to make predictions, 
and are thus not suited for exploring and constraining large parameter spaces.

Halo based models that begin with the assumption that galaxies form and live inside dark mater halos provided
the breakthrough that made it possible to quantitatively model galaxy clustering on small scales. These models 
rely on the fact that the statistical properties of dark matter halos are easy to predict with collision-less N-body 
simulations where the only important physical process is gravity. Halo models then adopt a parameterisation 
to connect galaxies to halos, thus bypassing the need to understand galaxy formation physics. In the halo model
framework, there is a convenient conceptual and operational division between the roles of cosmology and galaxy
formation: cosmology dictates the dark matter halo distribution while galaxy formation determines how exactly
galaxies occupy halos. This division is not perfect since gas physics can affect the properties of halos 
\citep[e.g.,][]{cui_etal_12}; however, this is a second order effect.

The connection between galaxy halo occupation and galaxy clustering was first made by semi-analytic models of
galaxy formation that modelled the formation and evolution of galaxies inside halos. Since these halos resided
within a larger density field in cosmological N-body simulations, it was straightforward to predict the clustering
of the semi-analytic galaxies \citep{kauffmann_etal_97,kauffmann_etal_99,baugh_etal_99}. \citet{jing_etal_98}
and \citet{benson_etal_00} took this a step further by realising that galaxy
clustering did not necessarily depend on all the details of galaxy formation, but rather only cared about halo occupation statistics as a function of halo 
mass. A series of papers then built upon the earlier work of \citet{neyman_scott_52} and \citet{scherrer_bertchinger_91} 
to develop a full analytic machinery for combining parameterised halo properties with occupation statistics to 
calculate the correlation function and power spectrum of galaxies on all scales \citep[e.g.,][]{peacock_smith_00,
seljak_00,scoccimarro_etal_01,cooray_sheth_02}. These analytic models became generally known as the 
``halo model". \citet{berlind_weinberg_02} focused on the ``halo occupation distribution" (\hod), which is the 
complete parameterisation connecting galaxies of a given class to halos, and they investigated how the \hod 
affects several galaxy clustering statistics and laid out a road-map for empirically constraining the \hod with 
measurements of galaxy clustering using data from large surveys.

The halo model has since been used by a large number of studies to model galaxy clustering data in several
galaxy redshift surveys: the 2dF Galaxy Redshift Survey \citep[2dFGRS;][]{colless_etal_01}, the Sloan Digital 
Sky Survey \citep[SDSS;][]{york_etal_00}, the 6dF Galaxy Redshift Survey \citep[6dFGRS;][]{jones_etal_04}, 
and the SDSS III Baryon Oscillation Spectroscopic Survey \citep[BOSS;][]{dawson_etal_13}. Some studies 
investigated the two-point correlation function of low redshift galaxies \citep{magliocchetti_porciani_03,
zehavi_etal_04,collister_lahav_05,zehavi_etal_05,tinker_etal_05,zehavi_etal_11,watson_etal_12,beutler_etal_13,
piscionere_etal_14}, others investigated the same for red galaxies \citep{blake_etal_08,brown_etal_08,zheng_etal_09,
watson_etal_10,white_etal_11,parejko_etal_13,nikoloudakis_etal_13,guo_etal_14,reid_etal_14,guo_etal_15a},
or other special classes of galaxies such as AGN or radio galaxies \citep{wake_etal_08,mandelbaum_etal_09,
richardson_etal_13}. Many studies modelled the two-point correlation function of Lyman Break and other types 
of high redshift galaxies \citep{bullock_etal_02,moustakas_somerville_02,hamana_etal_04,zheng_04,lee_etal_06,
tinker_etal_10,jose_etal_13,kim_etal_14}, while others combined modelling of galaxies at different redshifts in 
order to learn about the co-evolution of galaxies and halos \citep{yan_etal_03,cooray_06,zheng_etal_07,
tinker_wetzel_10,abbas_etal_10,wake_etal_11,tinker_etal_13}. Though the vast majority of studies have focused 
on the two-point correlation function, a few have applied the halo model to other clustering statistics, like the 
three-point correlation function \citep{marin_etal_11}, or the galaxy-mass cross-correlation function as measured 
by galaxy-galaxy lensing \citep{guzik_seljak_02,cacciato_etal_13}. Closely related to the \hod approach is the 
conditional luminosity function (CLF) approach that includes galaxy luminosity in the halo occupation parameterisation 
\citep[e.g.,][]{yang_etal_03,vandenbosch_etal_03,vale_ostriker_04,leauthaud_etal_11,leauthaud_etal_12}.

All of these studies successfully used the halo model to translate clustering statistics into constraints on the 
relation between galaxy properties and the dark matter halos they inhabit. For the most part, the constraints are focused
on the relation between the luminosity or stellar mass of galaxies and the mass of their halos 
\citep[e.g.,][]{zehavi_etal_11}. However, some studies focused on other aspects of the \hod, such as the radial 
distribution of galaxies within halos \citep[e.g.,][]{watson_etal_10,watson_etal_12,piscionere_etal_14}, the 
velocity distribution of galaxies within halos \citep{guo_etal_15a,guo_etal_15b,guo_etal_15c}, or galaxy assembly bias 
\citep{zentner_etal_16}. These results have been quite illuminating and have helped to explain many observed features of 
the galaxy population, such as the morphology-density relation and the different clustering strengths of different galaxy 
types. Halo modelling across different redshifts has even constrained the competing roles of merging and star formation 
in the stellar mass buildup of galaxies \citep[e.g.,][]{tinker_etal_13}.

In most halo model analyses, the statistical methodology employed is sophisticated. Clustering uncertainties
and their correlations are quantified in estimated covariance matrices, parameter searches are executed using 
Monte Carlo Markov Chain (MCMC) methods, model parameter values are reported with their full probability 
distributions, and the goodness of fit for the halo model is typically reported using the $\chi^2$ statistic. However,
the systematic errors in these analyses have been largely unquantified. Systematic errors exist in the estimation
of covariance matrices that typically use the Jackknife method \citep{norberg_etal_09}. Systematic errors also
exist in the halo model itself, since its implementation is almost always analytic and it contains several 
approximations. Without a robust characterisation of systematic errors, it is impossible to interpret the published 
results in a statistical sense. For example, many of the published works contain model fits with reported values
of $\chi^2$ that are high enough to warrant ruling the models out (e.g., some of the high luminosity samples
in \citealt{zehavi_etal_11}), but without a proper accounting of systematic errors in the modelling, it is not possible 
to determine whether the models have actually been ruled out. This problem is exacerbated by the ever shrinking
statistical errors in clustering measurements due to the growing sample sizes provided by current galaxy surveys. 
In order to trust the goodness of fit reported by a given study, systematic errors in the modelling must be kept 
smaller than the statistical errors in the data measurements.

Since systematic errors in modelling are for the most part not quantified by published studies, many of the results in 
the literature should be interpreted with caution. In general, qualitative trends found are likely correct, but precise 
parameter values, error bars, or goodness of fit estimates are not necessarily reliable. This does not pose a problem 
for most published works because their goal was to uncover general trends rather than to test specific models. 
However, if we wish to use the halo model to probe more subtle features of the \hod, like the presence of assembly 
bias \citep[e.g., see][]{zentner_etal_14}, or if we wish to use galaxy clustering on small scales to constrain cosmological 
models, we will need to make our modelling methodology more accurate. A few studies have demonstrated the 
power of small scales to constrain cosmology \citep[e.g.,][]{abazajian_etal_05,vandenbosch_etal_07,cacciato_etal_13}, 
but they will only be able to compete with more established probes of cosmology if the modelling is sufficiently accurate.

In this paper, we attempt to push the modelling of galaxy clustering on small scales into the ``accurate regime",
where results can be trusted enough to confirm or rule out physical models. To be precise, by ``accurate''
we mean that {\em given a set of measured statistics from a galaxy survey, and a galaxy-halo connection model being tested},
we can produce a reliable goodness of fit {\em and} reliable posterior probabilities for the model parameters. To achieve this, we need
to: i) minimize the theoretical errors in the predicted distribution of halos for the
assumed cosmological model, ii) forward-model the measured clustering statistics to incorporate all observational systematic errors,
and iii) robustly estimate the statistical errors and covariances in the measured clustering statistics
{\em from the model}. In other words, our goal is to 
accomplish for small scales what is already routinely done in the study of large scale clustering. This is an ambitious 
goal and we can only tackle it by adopting a fully numerical modelling framework that is based on large numbers 
of realistic mock galaxy catalogues. This data-intensive approach requires substantial computational effort. However, 
we are motivated to do this by the immense amount of information present in galaxy surveys on small scales, 
which is currently not being harnessed. In this first paper, we primarily assume that the cosmological
model is known and we adopt the most widely used formulation of the \hod to model the clustering of SDSS 
galaxies. Our objective is thus to test the standard $\LCDM$ + halo model against the small scale clustering of 
SDSS galaxies. Specifically, we use measurements of the projected correlation function and the group multiplicity 
function for two luminosity threshold samples from the SDSS DR7 \citep{abazajian_etal_09} dataset. However, this 
work is also intended to open the door for future studies where we will use additional clustering statistics and larger 
galaxy samples to test extensions of the standard \hod as well as variations in cosmology.

The layout of this paper is as follows. In \S~\ref{SDSS} we describe the SDSS data samples that we use. In 
\S~\ref{Mocks} we describe the N-body simulations and mock galaxy catalogues that make up the workhorse of
our modelling methodology. In \S~\ref{Clustering} we present our galaxy clustering measurements and in \S~\ref{Errors} 
we present their associated errors and correlation matrices. In \S~\ref{Fitting} we describe our model fitting methodology 
and present all of our modelling results. In \S~\ref{Improvements} we discuss future improvements to our methodology. 
Finally, in \S~\ref{Summary} we summarise our results.

\section{SDSS Galaxy Samples} \label{SDSS}

The Sloan Digital Sky Survey \citep{york_etal_00} completed its original imaging and spectroscopic goals in 2008
with its seventh data release \citep[DR7;][]{abazajian_etal_09}. The DR7 spectroscopic main galaxy sample 
\citep{strauss_etal_02} is complete down to an apparent $r$-band Petrosian magnitude limit of 17.77 and contains 
over 900,000 galaxies. In this work, we use the large-scale structure samples from the NYU Value Added Galaxy 
Catalog \citep[NYU-VAGC;][]{blanton_etal_05}. Specifically, we use a parent sample of just over 530,000 galaxies,
which only covers the northern part of the SDSS footprint and is cut back to $r<17.6$ so that it is complete down to
that magnitude limit across the sky. The reason for restricting the data to the northern footprint is that we can 
construct more independent mock catalogues using the north-only survey volume. Galaxy absolute magnitudes have 
been k-corrected to rest-frame magnitudes at redshift $z=0.1$ \citep{blanton_etal_03b} and corrected for passive 
luminosity evolution using the simple model described by \citet{blanton_etal_06}.

The SDSS spectroscopic sample has an incompleteness due to the mechanical restriction that spectroscopic fibres 
cannot be placed closer to each other than their own thickness. This fibre collision constraint makes it impossible to 
obtain redshifts for both galaxies in pairs that are closer than 55$\arcsec$ on the sky. This restriction results in 
$\sim$7\% of targeted galaxies not having a measured redshift. We assign fibre collided galaxies the redshift of the 
galaxy they collided with (i.e., the ``nearest neighbour correction"; \citealt{zehavi_etal_02}). This correction recovers 
the true correlation function well on scales larger than the physical scale corresponding to to 55$\arcsec$, which
for the outer redshift limit of our samples corresponds to 0.1$\hMpc$. There is some additional incompleteness due 
to bright foreground stars blocking background galaxies, but this is at the 1\% level. 

In this study we use two volume-limited subsamples of the full SDSS redshift sample that are each complete in a 
specified redshift range down to a limiting $r$-band absolute magnitude threshold. We construct each sample by 
choosing redshift limits $z_\mathrm{min}$ and $z_\mathrm{max}$ and only keeping galaxies whose evolved, 
redshifted spectra would still make the redshift survey's apparent magnitude and surface brightness cuts at these 
limiting redshifts. Our low-luminosity sample is complete down to an $r$-band absolute magnitude of $-19$ in the 
redshift range $0.02-0.067$, while our high-luminosity sample is complete down to an $r$-band absolute magnitude 
of $-21$ in the redshift range $0.02-0.165$. The redshift limits, median redshift, effective volume, and number density 
of these two samples are listed in Table~\ref{table:samples}. The volumes and number densities of the samples are
corrected for survey incompleteness.

Throughout this paper, co-moving distances and absolute magnitudes for SDSS galaxies are calculated adopting a flat 
$\LCDM$ cosmological model with $\Om=0.25$ and $h=1$. Our distances thus have units of $\hMpc$ and our absolute 
magnitudes are actually $M_r + 5 \mathrm{log}h$. We keep these data samples fixed regardless of what cosmological 
model we adopt when fitting to clustering statistics. Ideally, a change of cosmology in the model should also be reflected
in the data measurements. However, we have checked that clustering statistics only change by $\sim1\%$ when 
switching cosmological models, which is negligible compared to the errors in our measurements.

\begin{table}
\renewcommand{\arraystretch}{1.20}
\caption{SDSS Volume-limited Sample Parameters. The first column lists the absolute magnitude threshold of each sample at $z=0.1$.
The second, third and fourth columns list the minimum, maximum and the median redshifts, respectively.
The fifth column lists the effective volume of each sample, and the last column lists the galaxy
number density.}
\begin{center}
\begin{tabular}{cccccc}
\toprule
\multirow{2}{*}{$M_r^\mathrm{lim}$}            &
\multirow{2}{*}{$z_\mathrm{min}$}              &
\multirow{2}{*}{$z_\mathrm{max}$}              &
\multirow{2}{*}{$z_\mathrm{median}$}           &
\multicolumn{1}{c}{$V_\mathrm{eff}$}           &
\multicolumn{1}{c}{$n_g$}                      \\
\multicolumn{4}{c}{}                              &
\multicolumn{1}{c}{$10^6h^{-3}\mathrm{Mpc}^3$}    &
\multicolumn{1}{c}{$h^{3}\mathrm{Mpc}^{-3}$}     \\
\midrule
$-19$ & 0.02 & 0.067 & 0.054 &  5.555 & 0.0149 \\
$-21$ & 0.02 & 0.165 & 0.132 & 78.374 & 0.0012 \\
\bottomrule
\end{tabular}
\end{center}
\label{table:samples}
\end{table}

\section{Mock Galaxy Catalogues} \label{Mocks}

The key to accurate forward halo modelling of galaxy clustering on small scales is to use mock catalogues both for 
predicting observed statistics and for estimating errors and
covariances. Broadly, we can divide the challenge in creating an accurate
model into three distinct pieces -- 1) creating a reliable prediction of the
halo population for a cosmological model, 2)  comparing identical galaxy
clustering statistics between the predicted model and the observed data
3) correctly estimating covariances between all clustering statistics from the model.  We take the
following approach to ensure that systematic errors in the modelling are sub-dominant compared to
observational errors, i.e., an accurate model. To obtain a reliable halo
population, we use cosmological N-body simulations of sufficient volume and
resolution. To compare indentical clustering statistics,  
we create realistic mock catalogues (from the N-body simulations) that include systematic
effects like survey geometry and redshift-space distortions. In addition, we use the exact
same codes, when necessary, to measure the clustering statistics on both mock
galaxies and observed data. 
To correctly estimate covariances, we use a large, independent ensemble of these realistic
mocks instead of estimating errors from the observed dataset. 
Skipping any one of these three steps compromises the ``accurate'' part
of the model. For instance, assuming an 
universal halo mass function, or an analytic non-linear bias fitting formula,
or a Navarro-Frenk-White density profile for the halo has an
impact on the predicted \wprp.
Not including survey
geometry in the modelling changes both the amplitude and the slope of the group multiplicity function~\citep{berlind_etal_06}.
Using jack-knife resampling to build covariance matrices introduces
scale-dependent, systematic biases~\citep{norberg_etal_09}. With our adopted
methodology, we avoid all these systematic effects and push the small-scale
modelling towards the accurate regime. 

In this section we describe the simulations and mock catalogue pipeline in detail.
 
\subsection{N-body Simulations and Halo Catalogues} \label{Nbody}

\subsubsection{For building covariance matrices} \label{Nbody_covar}

The bulk of simulations that we use are from the Large Suite of Dark Matter Simulations project \citep[LasDamas;][]
{McBride_etal_09}. The LasDamas project focused on running many independent N-body realisations with the same 
cosmology but different initial phases. The cosmological parameters were roughly motivated by the WMAP3 constraints
\citep{spergel_etal_07} and are $\Om=0.25$, $\Ol=0.75$, $\Ob=0.04$, $h=0.7$, $\sigma_8=0.8$, and $n_s=1.0$.
The LasDamas simulations were designed to model SDSS galaxies and consist of four different volume and resolution
configurations that correspond to different luminosity samples. In this work, we use the Consuelo and Carmen 
configurations, which were designed to model galaxy samples with $r$-band absolute magnitude thresholds of
$-19$ and $-21$, respectively. 

All the simulations were seeded with second order Lagrangian perturbation theory initial conditions using the code 
\texttt{2LPTIC} \citep{scoccimarro_98, crocce_etal_06} and were evolved using the N-body code \texttt{GADGET-2} 
\citep{springel_05}. Each Consuelo simulation evolved $1400^3$ dark matter particles in a cubic volume of 
$420\hMpc$ on a side, from a starting redshift of $z_\mathrm{init}=99$ to $z=0$, with a gravitational force softening
length of $8\hkpc$. Each Carmen simulation evolved $1120^3$ dark matter particles in a cubic volume of 
$1000\hMpc$ on a side, from a starting redshift of $z_\mathrm{init}=49$ to $z=0$, with a gravitational force softening
length of $25\hkpc$. The resulting particle masses of the Consuelo and Carmen simulations are $1.87\times10^9\hMsun$
and $4.938\times10^{10}\hMsun$, respectively. For the purpose of estimating covariance matrices, we use 50 
realisations of each of these two boxes, which yield 200 mock catalogues per luminosity sample. 

We identify halos in the dark matter distributions using the simulation outputs corresponding to the median redshifts
of the $-19$ and $-21$ samples, which are $z=0.054$ and $z=0.132$, respectively. For the fifty Consuelo and fifty 
Carmen simulations, halos were identified with the \texttt{ntropy-fofsv} code \citep{gardner_etal_07}, which employs a 
friends-of-friends \citep[FoF;][]{davis_etal_85} algorithm. The FoF linking length was chosen to be 0.2 times the mean 
inter-particle separation. Finally, we apply the \citet{warren_etal_06} correction to the FoF halo masses. For the purpose 
of placing central galaxies in halos, we define the FoF halo centre to be at the location of the deepest part of the halo's 
gravitational potential well. The mock catalogues that we use to construct covariance matrices are based on these FoF halo 
catalogues.

The mass resolutions of these simulations result in 150 particles in halos of mass $\logM=11.45$ (for Consuelo) and 122 
particles in halos of mass $\logM=12.78$ (for Carmen) which, according to \citet{zehavi_etal_11}, are the typical 
minimum masses we expect to host galaxies in our two luminosity samples. If we also consider the scatter in these 
minimum masses (the \siglogM parameter in \citealt{zehavi_etal_11}), then halos with masses as low as $\logM=11.26$ 
(for Consuelo) and $\logM=12.10$ (for Carmen) will sometimes host a galaxy, which contain 97 and 25 particles, 
respectively. This is acceptable resolution for our purposes because these halos only rarely host a single central galaxy 
and so we only need to roughly resolve their bulk properties (position, velocity, mass). Moreover, we expect that errors 
due to resolution will not impact the scatter among clustering measurements from different simulation realisations as 
much as it will impact the clustering measurements themselves. 

\begin{table*}
\renewcommand{\arraystretch}{1.20}
\caption{Simulation parameters. The table lists the properties of the simulations used to estimate covariance matrices and to model clustering statistics in our MCMC chains for our two luminosity samples. Columns 3-9 list the cosmological model, simulation name, box size, number of particles, particle mass, force resolution, and number of boxes used.}
\begin{center}
\begin{tabular}{ccccccccc}
\toprule
\multirow{2}{*}{Use Type}            &
\multirow{2}{*}{Sample}              &
\multirow{2}{*}{Cosmology}           &
\multirow{2}{*}{Simulation}          &
\multicolumn{1}{c}{$L_\mathrm{box}$} &
\multirow{2}{*}{$N_\mathrm{part}$} &
\multicolumn{1}{c}{$m_\mathrm{part}$} &
\multicolumn{1}{c}{$\epsilon$}        &
\multirow{2}{*}{$Number$}         \\
\multicolumn{4}{c}{}          &
\multicolumn{1}{c}{$\hMpc$}   &
\multicolumn{1}{c}{}          &
\multicolumn{1}{c}{$\hMsun$}  &
\multicolumn{1}{c}{$\hkpc$}   &
\multicolumn{1}{c}{}          \\
\midrule
\multirow{2}{*} {Correlation matrix} & $-19$ & LasDamas & Consuelo     &  420 & $1400^3$ & $1.87\times10^9$    &  8 & 50 \\
                                     & $-21$ & LasDamas & Carmen       & 1000 & $1120^3$ & $4.94\times10^{10}$ & 25 & 50 \\
\cmidrule(lr{5pt}){1-9}
\multirow{4}{*} {MCMC}             & $-19$ & LasDamas & Consuelo     &  420 & $1400^3$ & $1.87\times10^9$   &  8 & 2 \\
                                   & $-21$ & LasDamas & CarmenHD     & 1000 & $2240^3$ & $6.17\times10^9$   & 12 & 2 \\
                                   & $-19$ & Planck   & Consuelo-plk &  420 & $1400^3$ & $2.26\times10^9$   &  8 & 2 \\
                                   & $-21$ & Planck   & CarmenHD-plk & 1000 & $2240^3$ & $7.46\times10^9$   & 12 & 2 \\
\bottomrule
\end{tabular}
\end{center}
\label{table:simulations}
\end{table*}

\subsubsection{For MCMC parameter exploration} \label{Nbody_mcmc}

For the purpose of generating predictions of clustering statistics within our MCMC framework, we have run a few 
additional simulations. The demands on simulation resolution are more stringent in this case because, during the halo 
model parameter search, sometimes galaxies are placed in significantly lower mass halos than they are in the fiducial 
model that we use to construct covariance matrices. The simulations used in the MCMC parameter search must 
therefore resolve halos down to lower masses. If we adopt the 2-$\sigma$ low value of \logMmin and 2-$\sigma$ high 
value of \siglogM found by \citet{zehavi_etal_11}, we can obtain a conservative estimate of the lowest mass halos we 
must resolve. These masses are $\logM=10.9$ and~11.6 for the $-19$ and $-21$ samples, which result in 42 and 8 
particles in Consuelo and Carmen simulations, respectively. The Carmen simulations do not therefore have adequate 
resolution to serve as the basis for modelling the clustering of the $M_r<-21$ SDSS sample. We have thus run a higher 
resolution version of Carmen that we name CarmenHD, which contains $2240^3$ particles and resolves halos that have 
more than five times smaller mass than Carmen. In order to probe a variation in cosmology, we have also run new 
versions of Consuelo and CarmenHD that adopt the recent set of cosmological parameters given by the Planck 
experiment \citep{planck_etal_14}. Specifically, these simulations, named Consuelo-plk and CarmenHD-plk, adopt the 
following parameter values: $\Om=0.302$, $\Ol=0.698$, $\Ob=0.048$, $h=0.681$, $\sigma_8=0.828$, and $n_s=0.96$. 
We have run two realisations each of CarmenHD, Consuelo-plk, and CarmenHD-plk since we use two boxes within our 
MCMC modelling framework. The box sizes, particle masses, force resolutions and other summary information of all the 
simulations described above are listed in Table~\ref{table:simulations}.

For the two Consuelo, CarmenHD, Consuelo-plk, and CarmenHD-plk simulations that we use in our MCMC modelling, 
halos are identified with a spherical over-density \citep[SO;][]{lacey_cole_94} algorithm using the \texttt{ROCKSTAR} 
code \citep{behroozi_etal_13}. We use two sets of SO halo definitions: the M200b definition where halos are spheres of 
density 200 times the mean density of the universe and the Mvir definition where halos have a density that depends on 
cosmology and redshift, as given by \citet{bryan_norman_98}. We use these SO halo catalogues to predict galaxy 
clustering within our MCMC parameter searches. In the case of an SO halo, the halo centre is by definition at the centre 
of the halo sphere\footnote{We note that the correlation matrices are derived from FoF halos, while the modelling is
  done with SO halos. This is because we did not have the SO halo catalogs for all the simulations while the paper was
  being prepared. We have since checked that our results do not change qualitatively with a correlation matrix derived from SO halos}

When evaluating the likelihood function during a model parameter exploration, it is desirable that the statistical errors in
the model (due to cosmic variance) are much smaller than the errors in the SDSS data so that they do not add 
appreciably to the error budget. If this is not the case, it is equivalent to degrading the statistical power of the galaxy 
survey. Therefore, the clustering statistics calculated from the model should come from a larger volume of mock galaxies 
than the SDSS sample being modelled. On the other hand, generating too large a mock 
volume on the fly within a MCMC parameter search is computationally intractable. We achieve this balance by using 
either a single realisation simulation cube in our modelling, or two realisations if it is necessary to faithfully reproduce the 
SDSS sample geometry. A single Consuelo or Carmen box contains about 13 times more total volume than the SDSS 
$-19$ or $-21$ samples, respectively. Alternatively, if we need to include the full SDSS sample geometry in the mocks, 
two Consuelo or Carmen boxes can generate eight times more mock volume than their respective SDSS samples. This 
is sufficiently large to keep statistical errors in the modelling sub-dominant. However, we reduce the model errors further 
by carefully selecting the two simulation boxes we use. We do this by selecting the two boxes that exhibit a clustering 
signal closest to the mean of all 50 boxes. We do this in the following way. First, we populate all 50 Consuelo and 
Carmen boxes with the \citet{zehavi_etal_11} \hod model for $-19$ and $-20$ threshold samples, respectively. We then 
measure the projected correlation function \wprp on each of these mocks, adopting one of the axes of the cube as the 
line-of-sight direction. We measure the mean clustering $\overline{w_p}(r_p)$ from all 50 boxes as well as their standard 
deviation $\sigma_{w_p}(r_p)$. We then compute a $\chi^2$ statistic for each box
\begin{equation}
\chi_{\mathrm{box}}^2 = \sum_{r_p}  \left(\frac{w_{p,\mathrm{box}}(r_p) - \overline{w_p}(r_p)}{\sigma_{w_p}(r_p)}\right)^2, \;  \forall \; \mathrm{box} \; \in\; [1,50].
\end{equation}
Finally, we identify the two realisations of Consuelo and Carmen that have the lowest values of $\chi^2$, which are 
essentially the two boxes whose random phases result in clustering that is closest to the mean of all 50 boxes. We use
these same exact phases in all our modelling boxes: the same two sets of phases for the two Consuelo and two 
Consuelo-plk simulations and the same two sets of phases for the two CarmenHD and two CarmenHD-plk simulations.
This procedure results in a significant reduction in cosmic variance errors. The average box-to-box scatter in $\wprp$ 
across all scales is $\sim10\%$ for Consuelo and $\sim6\%$ for Carmen, while the average difference between our best
two boxes and the mean of all 50 is only $\sim5\%$ for Consuelo and $\sim3\%$ for Carmen, representing a factor of
two reduction in cosmic variance errors. This is the accuracy we would normally obtain with a much larger simulation 
volume.

\subsection{From Halos to Galaxies} \label{HOD}

We populate dark matter halos in our simulations with mock galaxies using the `Halo Occupation Distribution' (\hod)
framework. In this framework, the number, positions, and velocities of galaxies within a halo are described statistically
given a set of parameterised prescriptions. The key advantage of the \hod approach is that if the parameterisation is 
sufficiently flexible it allows us to marginalise over the full uncertainty of galaxy formation theory. 

In this work, we adopt the `vanilla' \hod model of \citet{zheng_etal_07}, which has become somewhat of an industry 
standard in \hod modelling. Motivated by theoretical results \citep{kravtsov_etal_04,zheng_etal_05}, we split the galaxies 
into centrals and satellites within their halos. The mean number of central galaxies as a function of halo mass
$M$ is given by\footnote{$\log$ refers to $\log_{10}$ everywhere in the text}
\begin{equation}
\Ncenavg =  \frac{1}{2} \left[ 1+\erf\left(\frac{\log M - \logMmin}{\siglogM} \right) \right] ,
\label{eqn:Ncen}
\end{equation}
where $\Mmin$ sets the minimum halo mass that can host a central galaxy, $\siglogM$ sets the scatter around this 
minimum halo mass, and $\erf(x)$ is the error function, $\erf(x) = \dfrac{2}{\sqrt{\pi}} \int_0^x \exp(-y^2) dy$. The 
motivation behind this particular analytic form of \Ncenavg\ comes from assuming a log-normal distribution of central 
galaxy luminosity at fixed halo mass. If the luminosity-mass relation is a power law $L\propto M^p$ with a log-normal 
scatter $\siglogL$, then the scatter in mass at fixed luminosity is $\siglogM=\sqrt{2} \siglogL/p$. The parameter 
$\siglogM$ in equation~(\ref{eqn:Ncen}) represents this scatter at the luminosity limit of the sample and thus controls 
how quickly the probability of containing a galaxy above the luminosity limit rises from zero to one as halo mass 
increases. $\Mmin$ is the mass at which this probability is one half, i.e., where half the halos in the universe contain a 
central galaxy above the sample luminosity threshold, while the other half do not. The relation between central galaxy 
luminosity and halo mass is constrained by abundances to follow a roughly double power-law form with a steep slope at 
low mass, a transition near $M=10^{12}\hMsun$, and a shallow slope at high mass \citep[e.g.,][]{vale_ostriker_04}. For 
the $-19$ and $-21$ luminosity thresholds we consider in this paper, we expect the slope $p$ to be approximately 
$\sim1$ and $\sim0.3$, respectively. Previous work on satellite kinematics has constrained $\siglogL$ to be 
approximately $\sim0.15$ dex~\citep{more_etal_09}. We thus expect values of $\siglogM$ that are $\sim 0.2$ and 
$\sim 0.7$ for our faint and bright samples, respectively.

In each halo we place a number of satellite galaxies drawn from a
Poisson\footnote{The function for generating Poisson random numbers in `Numerical 
Recipes in C' code suffers from a floating underflow bug that sets in 
when \Nsat\ exceeds 708 -- as can happen for faint samples. We used the `GSL' routine which does not suffer from 
that particular bug.} distribution with a mean given by
\begin{equation}
\Nsatavg = \Ncenavg \times \left(\frac{M - M_0}{M_1} \right)^\alpha ,
\label{eqn:Nsat}
\end{equation}
where $M_0$ is the halo mass below which there are no satellite galaxies, $\alpha$ is the slope of the power-law 
occupation function at high masses, and $M_1$ sets the mass scale where halos contain one satellite 
galaxy on average. The actual halo mass for which \Nsatavg$=1$ is slightly higher than $M_1$ and also depends on the other 
parameter values. For some \hod parameter combinations, it is possible to have halos that do not receive a central 
galaxy and yet still have a chance to get a satellite. We do not allow such cases in our modelling. Halos are only eligible
to receive satellite galaxies if they already have a central.

Once we have specified the number of centrals and satellites in a given halo, we need to give them positions and 
velocities within the halo. In this 'vanilla' \hod model we place the central galaxy at the centre of its halo. We then assign 
to it the mean velocity of the whole halo, which assumes that it is at rest relative to the halo. For the satellites, we choose 
their positions and velocities to be equal to those of randomly selected dark matter particles within the halo. This 
assumes that satellite galaxies trace the spatial and velocity distribution of dark matter within their halo.
 
Though the \hod is a statistical model that allows us to parameterise our ignorance of the detailed physics of galaxy 
formation, the various features of the \hod do contain information about galaxy formation. For example, the mass 
scale and scatter of the central galaxy occupation at low masses (parameterised by $\Mmin$ and $\siglogM$) depend
on the luminosity-mass relation in the vicinity of the luminosity threshold of the sample. The luminosity-mass relation 
depends on the efficiency with which halos are able to cool gas and form stars as a function of mass. This efficiency is
affected by several factors, such as supernova feedback, pre-heating caused by reionization, cold versus hot accretion 
of gas into the halo, merger history, environment, etc. On the other hand, the satellite occupation at higher masses 
depends primarily on the balance between the accretion and destruction of subhalos, which depends on the distributions 
of their infall times, masses and orbits, and on the physics of dynamical friction and tidal stripping 
\citep[e.g.,][]{watson_etal_11}. 

To summarise, the vanilla \hod model we use contains five free parameters that control the mean number of galaxies
as a function of halo mass: $\Mmin$, $\siglogM$, $M_0$, $M_1$, and $\alpha$. The vanilla model also makes the 
following set of simplifying assumptions:
\newline
1. All galaxies live inside dark matter halos, using whatever halo definition we have adopted.
\newline 
2. The number of galaxies in a halo only depends on the mass of the halo and not on other halo properties, such as age 
or concentration. In other words, there is no galaxy assembly bias.
\newline
3. The functional forms of \Ncenavg\ and \Nsatavg\ are given by equation~(\ref{eqn:Ncen}) and equation~(\ref{eqn:Nsat}). 
\newline
4. The probability distribution $P(\Nsat|\Nsatavg)$ of the number of satellite galaxies $\Nsat$ given the mean number 
\Nsatavg\ is a Poisson distribution. 
\newline
5. The central galaxy in each halo lives at the halo centre and moves with the mean halo velocity. In other words, there
is no central spatial or velocity bias.
\newline
6. Satellite galaxies in each halo trace the spatial and velocity distribution of dark matter within the halo. In other words,
there is no satellite spatial or velocity bias.

These assumptions are all correct to first order; however, most of them are likely incorrect in detail. For example, galaxy
assembly bias is usually considered for colour-selected samples, but it is probably also present to some degree for 
luminosity threshold samples \citep[e.g.,][]{zentner_etal_14,zentner_etal_16}. We assume that the scatter in $\Nsat$ at 
fixed mass is Poissonian. However, recent work with high-resolution N-body simulations suggests that, at least for 
subhalos, this assumption does not hold for all host-to-satellite mass ratios~\citep{mbk_etal_10,mao_etal_2015}. 
Finally, we assume that there is no spatial or velocity bias for centrals or satellites. However, in theory we expect some 
bias for satellites since they experience dynamical effects (friction and stripping) that do not affect individual dark matter 
particles. Moreover, there is some observational evidence that both centrals and satellites have some degree of velocity
bias \citep{vandenbosch_etal_05,guo_etal_15c}. These are the sorts of interesting second-order features of the \hod that
can only be probed with an accurate modelling framework like the one we present in this paper. In this first work, we 
adopt the simple vanilla \hod model to test whether it can be ruled out by SDSS galaxy clustering data. Our strategy for 
future work is to add freedom to the model as demanded by the data. For example, if the vanilla \hod model is ruled out, 
then we will be justified to add freedom to it by relaxing the assumptions described above. If the model is not ruled out, 
then we will keep adding new clustering statistics until it is. Combining multiple statistics is a clear way forward to break 
the degeneracies in the \hod model. 

\subsection{Adding Survey Realism} \label{Realism}

Once we have a simulation box populated with galaxies according to a given \hod model, we build mock catalogues that
contain the main observational systematic effects that plague the galaxy clustering statistics we wish to model, namely 
redshift space distortions and sample geometry. First, we move the galaxy distribution from Cartesian to spherical 
coordinates by placing the mock observer at the centre of the simulation cube and converting Cartesian positions of 
galaxies into RA, DEC, and co-moving distances. We then compute the line-of-sight peculiar velocities of galaxies and 
compute galaxy redshifts as $1+z = (1+z_\mathrm{cosm})(1+z_\mathrm{doppler})$, where $z_\mathrm{cosm}$ is the 
cosmological redshift and $z_\mathrm{doppler}$ is the redshift due to the radial peculiar velocity. Finally, we throw out 
mock galaxies that lie outside the redshift limits or the sky footprint of the SDSS sample. Since the volume of the SDSS 
sample is much smaller than that of the simulation used to model it, we can extract multiple mock catalogues from a single 
simulation cube. By applying a series of rotations on the box before converting it to spherical coordinates, we are able to 
extract four entirely independent mock catalogues from each box. Our 50 simulations for each volume-limited sample can 
thus produce 200 independent mock catalogues that we can use to construct covariance matrices.

Aside from survey geometry and redshift distortions, the other main source of systematic error present in the SDSS 
clustering data is incompleteness due to fibre collisions. Unfortunately, adding fibre collisions to our mock catalogues is not 
a trivial exercise. First, the mocks only represent volume-limited samples, whereas in the SDSS a galaxy can collide with 
any other galaxy in the full flux-limited sample. Second, the severity of fibre collisions depends on sky location in a 
complicated way due to the tiling of the survey footprint with spectroscopic plates \citep{blanton_etal_03a}. In regions 
where spectroscopic plates overlap, fibre collisions are substantially reduced, and these overlap regions correlate
with the surface density of spectroscopic targets. For these reasons, we do not attempt to model fibre collisions, but 
we simply rely on the nearest-neighbour correction \citep{zehavi_etal_02}. We then restrict our clustering analysis to 
regimes where the correction works well, which include scales larger than 0.1$\hMpc$ in the correlation function 
\citep{zehavi_etal_02} and groups with at least 5 members in the multiplicity function \citep{berlind_etal_06}.

\section{Clustering Measurements} \label{Clustering}

Our approach stems from a very simple and powerful idea: if we have a correct model describing the galaxy distribution, 
then we should be able to reproduce any clustering statistic that encodes information about the galaxy density field. If
the model fails to provide a good fit to clustering statistics, then one or more of its assumptions need revising, e.g., the
cosmological model, the halo definition, the features of the \hod, etc. If the model is able to provide a good fit to some set
of clustering statistics, then we can keep adding more statistics until it fails. The galaxy clustering statistics we use do not
need to have a physical basis or even be well-established. We could define a new arbitrary statistic and use that instead. 
Since we can use the same codes and methods to compute statistics on both the data and the model, any statistic that 
can be measured from the data will do. What we want is statistics that contain a high degree of information about the 
galaxy density field while still being computationally tractable. In this work we use two fairly traditional statistics, the projected
two-point correlation function \wprp and the group multiplicity function \gmf, along with the overall number density of 
galaxies. In future work, we plan to extend to additional statistics, as well as perform more detailed studies to determine 
the optimal statistics that should be used for constraining a given model.

\subsection{The projected correlation function $\wprp$} \label{wprp}

\begin{figure*}
\centering
\includegraphics[clip=true,width=0.8\linewidth]{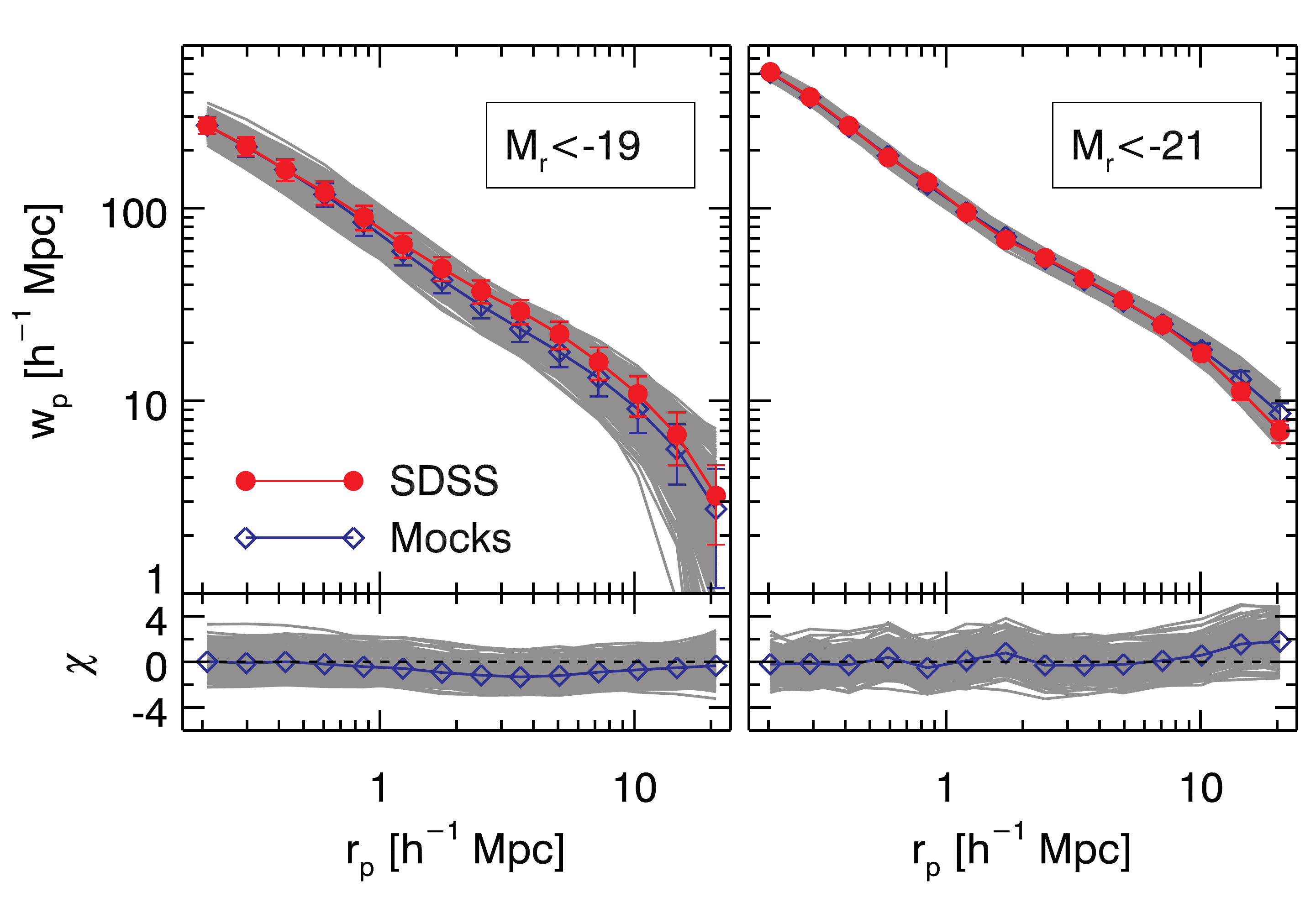}
\caption{\small Projected correlation function \wprp measurements for both SDSS and 
mock galaxies in the case of the $M_r<-19$ ({\it left panel}) and $M_r<-21$ ({\it right 
panel}) samples. Red points show measurements for SDSS galaxies with error bars 
estimated via jackknife resampling on the sky. Grey lines show individual results for 
each of 200 mock galaxy catalogues. Blue points and error bars show the mean and 
standard deviation of \wprp for the mock catalogues. The bottom section of each panel 
shows $\chi$, which is the difference between the mock and SDSS measurements 
divided by the standard deviation of the mocks.}
\label{fig:sdss_wp_fit_for_corr_matrix}
\end{figure*}  

The two-point correlation function is the most widely used galaxy clustering statistic and the one that is typically modelled 
with \hod models. The three-dimensional correlation function $\xi(r)$ is the excess number of galaxy pairs above what is 
expected for a random distribution of points, as a function of pair separation $r$. To deal with redshift distortions, galaxy 
pairs are decomposed into their line-of-sight and projected components $\pi$ and $r_p$, yielding the function 
$\xi(r_p,\pi)$. This is then integrated over $\pi$ to get the projected correlation function 
\begin{equation}
\wprp = 2 \int_{0}^{\pi_\mathrm{max}} \xi(r_p,\pi) d\pi.
\label{eqn:wprp}
\end{equation}
Since we will need to compute clustering statistics on the fly for mock catalogues within our MCMC framework, it is 
imperative that the computation be fast and efficient. We have developed and used a blazing fast new code 
\texttt{Corrfunc}\footnote{\url{https://github.com/manodeep/Corrfunc}} \citep{corrfunc_sinha_lehman} that can compute 
\wprp for either simulation cubes or mock surveys in parallel. With this code we can measure \wprp out to 20$\hMpc$ for 
$10^6$ galaxies in $\sim 6$ and ~3 seconds of wallclock time on a single CPU core, for the number densities of the 
$M_r<-19$ and $-21$ samples, respectively.

We use the \citet{zehavi_etal_02} definitions of $r_p$ and $\pi$ and we count pairs in 14 equally spaced logarithmic bins 
of $r_p$ whose centres range from $0.2-20\hMpc$. We estimate $\xi(r_p,\pi)$ with the Landy-Szalay estimator 
$(DD-2DR+RR)/RR$ \citep{landy_szalay_93}, where $DD$, $DR$, and $RR$ are the properly normalised numbers of 
data-data, data-random, and random-random pairs, respectively. We use a random catalogue that contains $\sim10$ times 
more points than its corresponding galaxy sample. Finally, we calculate \wprp by integrating out to $\pi_\mathrm{max}$ 
of 40 and 80$\hMpc$ for the $M_r<-19$ and $-21$ samples, respectively. The red points in 
Fig.~\ref{fig:sdss_wp_fit_for_corr_matrix} show our measurements for the two SDSS samples. Errors are calculated 
via jackknife resampling using 50 distinct regions on the sky (see \citealt{mcbride_etal_11} for jackknifing details). These 
jackknife errors are only used in the initial steps for constructing a mock covariance matrix, as we describe in 
\S~\ref{Errors}. Once we have a mock covariance matrix, we use that for all the main analysis in this paper. 
Fig.~\ref{fig:sdss_wp_fit_for_corr_matrix} also shows \wprp for a set of mock catalogues (grey lines and blue points), 
which we describe in \S~\ref{Errors}.

We measure \wprp as described above for the SDSS samples and for the 200 mock catalogues per sample that are used 
to construct covariance matrices. However, we follow a slightly different approach when computing \wprp within a \hod
parameter search. In this case, we do not use the Landy-Szalay estimator applied to mock catalogues that have the same 
spherical geometry as the SDSS. Rather, we use the much simpler ``natural'' estimator $DD/RR-1$ on full mock cubes 
that employ the plane parallel approximation, i.e., where one of the axes of the cube serves as the line-of-sight direction. 
Specifically, we use the single ``minimum cosmic variance'' box described in \S~\ref{Nbody_mcmc}. This methodology is 
much faster computationally because it does not require a computation of the $DR$ term in Landy-Szalay. Additionally, it 
does not contain any noise from a random catalogue because in a cubic geometry with periodic boundary conditions the 
$RR$ term can be calculated analytically. We have verified that this way of computing \wprp does not introduce 
any systematic errors in the modelling.

\subsection{The Group Multiplicity Function $\gmf$} \label{nN}

\begin{figure*}
\centering
\includegraphics[clip=true,width=0.8\linewidth]{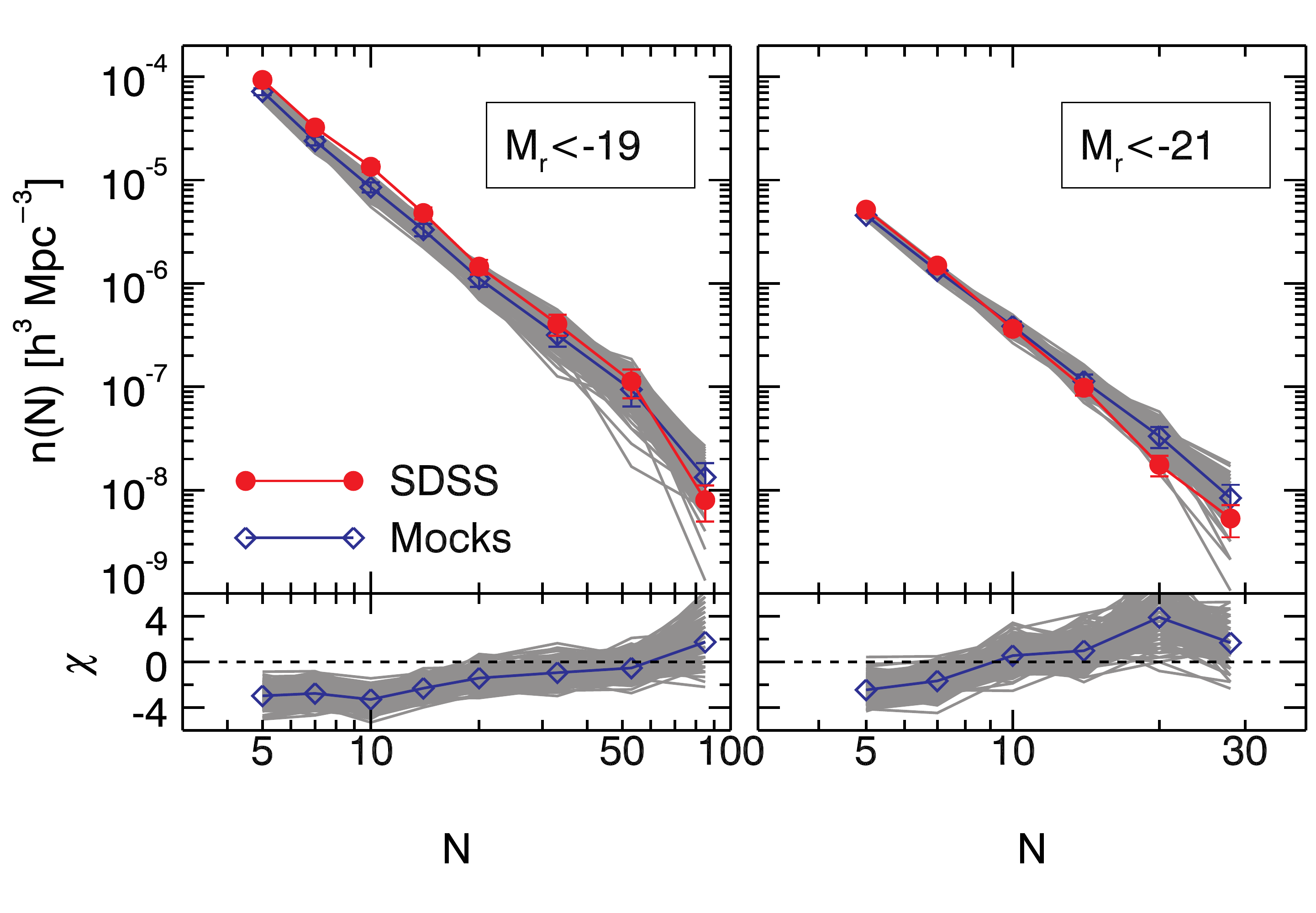}
\caption{\small Group multiplicity function \gmf measurements for both SDSS and mock 
galaxies in the case of the $M_r<-19$ ({\it left panel}) and $M_r<-21$ ({\it right panel}) 
samples. Red points show measurements for SDSS galaxies. Grey lines show individual 
results for each of 200 mock galaxy catalogues. Blue points and error bars show the mean 
and standard deviation of \gmf for the mock catalogues. Error bars on the red (SDSS) points
are estimated from scaling the mock fractional errors to the SDSS measurements. The 
bottom section of each panel shows $\chi$, which is the difference between the mock and 
SDSS measurements divided by the standard deviation of the mocks.}
\label{fig:sdss_gmf_fit_for_corr_matrix}
\end{figure*}  

Galaxy group statistics are naturally well suited to constraining halo models since groups are often systems of galaxies 
that occupy the same dark matter halo. \citet{berlind_weinberg_02} proposed using the group multiplicity function, 
defined as the abundance of galaxy groups as a function of their richness, to empirically measure the \hod. If galaxy 
groups are equivalent to halos, then the multiplicity function \gmf is directly related to the probability $P(N|M)$ that a halo 
of mass $M$ contains $N$ galaxies
\begin{equation}
\gmf = \int_0^\infty \frac{dn}{dM}P(N|M)dM ,
\label{eqn:nN}
\end{equation}
where $dn/dM$ is the halo mass function. Unfortunately, group catalogues are plagued by severe systematic effects where 
single halos are split into multiple groups and multiple halos are merged into a single group \citep[e.g.,][]{berlind_etal_06,
campbell_etal_15}. The measured group multiplicity function is thus very different from the halo multiplicity function given 
by equation~(\ref{eqn:nN}). The only way to accurately use group statistics is with a fully numerical modelling procedure 
where groups are identified directly in mock catalogues and group errors thus affect both data and model equally. 
\citet{hearin_etal_13} used the group multiplicity function as measured in mock catalogues to test subhalo abundance 
matching (SHAM) models and demonstrated that it contains complementary information to the correlation function. 
However, the multiplicity function has not been used in a full \hod modelling of galaxy survey data because it is 
computationally difficult to include mock catalogue construction and analysis within the MCMC modelling procedure. This 
work represents the first such analysis to-date.

We use the \citet{berlind_etal_06} friends-of-friends algorithm for identifying groups in both SDSS and mock data. 
According to the algorithm, two galaxies are linked together if their projected and line-of-sight separations are both less 
than a corresponding linking length. A galaxy group then consists of all the galaxies that are linked to each other in this 
way. We adopt the \citet{berlind_etal_06} linking lengths of $b_\perp=0.14$ and $b_\parallel=0.75$, which are given in 
units of the mean inter-galaxy separation $n_g^{-1/3}$, where $n_g$ is the sample number density. These linking lengths 
were specifically optimised to produce a multiplicity function that is as unbiased as possible relative to the true halo 
multiplicity function. However, this is mostly irrelevant for our study because group finding errors are equally present in 
both data and model. From our perspective, the group multiplicity function is just a different clustering statistic and any 
set of linking lengths would work. Using the sample densities listed in Table~\ref{table:samples}, the co-moving linking 
lengths for our two SDSS samples are $(r_\perp,r_\parallel) = (0.57,3.05)\hMpc$ for the $M_r<-19$ sample and 
$(r_\perp,r_\parallel) = (1.32,7.06)\hMpc$ for the $M_r<-21$ sample. The co-moving linking lengths that we use on mock 
samples adjust according to the varying number density.

\begin{figure*}
\centering
\includegraphics[clip=true,width=0.95\linewidth]{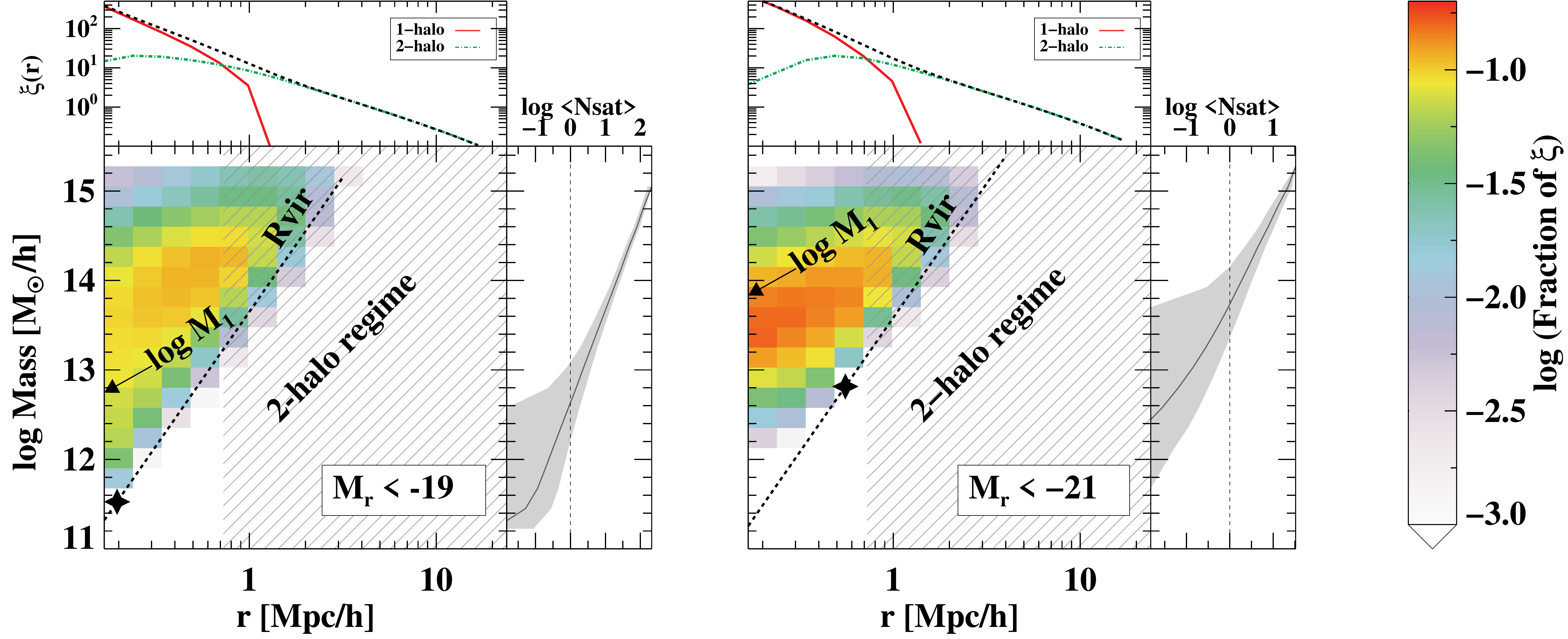}
\caption{\small Fractional contribution to 1-halo galaxy pair counts as a function of halo 
mass and scale in mock catalogues for the $M_r<-19$ ({\it left figure}) and $M_r<-21$ 
({\it right figure}) samples. The main panel for each sample shows two-dimensional 
bins of scale and halo mass, where the colour of each cell denotes the ratio of the number 
of 1-halo galaxy pairs in that bin $DD(r,M)$ divided by the number of all 1-halo galaxy 
pairs at that scale $DD(r)$. Each vertical column of cells thus shows the fractional 
distribution of halo masses that contribute 1-halo pairs for a particular scale. Also shown 
in the panel is the value of \logMmin ({\it star symbol}), \logMone ({\it arrow}), and the 
virial radius as a function of mass ({\it dotted line}). The hatched region delineates the 
scales where the 2-halo term dominates the overall clustering signal. The top panel for 
each sample shows the 1-halo and 2-halo terms of the correlation function $\xi(r)$ 
separately, while the right panel shows the mean number of satellite galaxies 
$\langle N_{\rm sat}(M) \rangle$ ({\it solid line}) and its 1-$\sigma$ scatter ({\it shaded region}).}
\label{fig:one_halo_fractions_corrmatrix}
\end{figure*}

After running the group finder and generating a group catalogue, we measure \gmf in bins of richness $N$. For the 
$M_r<-19$ sample, we adopt the following eight bins of $N$: $(5-6)$, $(7-9)$, $(10-13)$, $(14-19)$, $(20-32)$, 
$(33-52)$, $(53-84)$, $(85-220)$. For the $M_r<-21$ sample, we adopt the following six bins of $N$: $(5-6)$, $(7-9)$, 
$(10-13)$, $(14-19)$, $(20-27)$, $(28-40)$. \gmf is then simply the co-moving number density of all groups that have a 
number of members in the range given by each bin. The red points in Fig.~\ref{fig:sdss_gmf_fit_for_corr_matrix} show 
our measurements of the group multiplicity function for the two SDSS samples. 
Fig.~\ref{fig:sdss_gmf_fit_for_corr_matrix} also shows \gmf for a set of mock catalogues (grey lines and blue points), 
which we describe in \S~\ref{Errors}. Unlike in Fig.~\ref{fig:sdss_wp_fit_for_corr_matrix}, the displayed error bars 
on the SDSS measurements are not calculated via jackknife resampling, but rather from the standard deviation of mock measurements. 

\subsection{The Signal in the Correlation Function} \label{xi_signal}

Before we move on to study the covariance matrix of clustering measurements, it is useful to explore where the signal in
the galaxy correlation function comes from, in the context of the halo model. In the halo model, the correlation function 
has two terms, one on small scales that counts pairs of galaxies within the same halo (1-halo) and one on large scales 
that counts pairs of galaxies in different halos (2-halo). On large scales, the correlation function is simply a weighted 
version of the correlation function of halos, where halos of a given mass are weighted by the mean number of galaxies 
for that mass $\langle N\rangle_M$. In this discussion we focus on small scales and we investigate what mass halos 
dominate the number of galaxy pairs at each physical scale. We do not discuss the sensitivity of $\xi$ to changes in the 
\hod, which has been presented in other studies \citep[e.g.,][]{berlind_weinberg_02,watson_etal_11}.

The correlation function on small scales depends on the mean number of galaxy pairs per halo mass 
$\langle N(N-1)\rangle_M$, as well as the spatial distribution of these pairs within halos. Specifically, the 1-halo term of 
$\xi(r)$ is proportional to \citep{berlind_weinberg_02}:
\begin{equation}
r^2 \xi_{1h}(r) \propto \int\limits_0^\infty dM \frac{dn}{dM} \frac{\langle N(N-1) \rangle_M}{2} 
\frac{F^\prime (r/R_\mathrm{vir})}{R_{\mathrm{vir}}(M)},
\label{eqn:1-halo}
\end{equation}
where $dn/dM$ is the halo mass function, \Rvir is the virial radius of a halo, and $F^\prime \left(r\right)$ is the 
fractional distribution of galaxy pair separations within a halo. We can determine what mass regime dominates the 
correlation function by considering how each of the above terms scale with mass. The halo mass function scales as 
$\sim M^{-1}$ for $M\ll M^\star$ and $\sim\exp(-M/M^\star)$ for $M\gg M^\star$, where $M^\star$ is approximately equal 
to $2\times10^{14}\hMsun$.\footnote{This is not to be confused with the characteristic nonlinear collapse mass in 
Press-Schecter theory, which is much smaller.} The mean number of galaxy pairs per halo scales as 
$\langle\Nsat\rangle^2 \sim M^2$ for $M>M_1$. The virial radius scales as $\sim M^{1/3}$. Finally, the spatial distribution 
function $F^\prime (r/R_\mathrm{vir})$ is relatively insensitive to mass. Combining these terms, we find that the total 
number of pairs contributed from a given mass scales as $\sim M^{2/3}$ for $M\ll M^\star$ and $\sim\exp(-M)$ for 
$M\gg M^\star$. As a result, halos of mass $M^\star$ should dominate the signal in the 1-halo term of the correlation 
function and this result should hold for any sample where $M_1$ is less than $M^\star$.

We next investigate this question in more depth using mock catalogues for our $M_r<-19$ and $M_r<-21$ galaxy samples.
Fig.~\ref{fig:one_halo_fractions_corrmatrix} shows the fraction of galaxy 1-halo pairs as a function of mass and scale.
In each pixel showing a bin of pair separation $r$ and halo mass $M$, the colour of the pixel represents the value of 
$DD(r,M)$, the number of galaxy pairs in that bin, divided by $DD(r)$, the total number of pairs at that scale. In other 
words, each vertical column of pixels shows the normalised fractional distribution of pair counts as a function of halo 
mass that contribute to a given scale. The top panel of the figure shows the correlation function $\xi(r)$ with its 1- and 2-
halo breakdown, and the right panel shows the mean number of satellite galaxies $\langle\Nsat(M)\rangle$. Finally, the 
figure marks the halo virial radius at each mass, as well as the value of $M_1$. 
Fig.~\ref{fig:one_halo_fractions_corrmatrix} confirms that the majority of galaxy 1-halo pairs come from the cluster 
regime and that this is true both when $M_1$ is much lower than $M^\star$ (as in the $-19$ sample), and when $M_1$ 
is of order $M^\star$ (as in the $-21$ sample). In the former case each of these clusters contains many satellites and 
hence several pairs, while in the latter case each cluster only typically contains a single pair. This will help explain the 
structure of the correlation matrix that we discuss in \S~\ref{Matrix}.  The halo masses that dominate the 1-halo pairs 
naturally increase with scale since pairs must fit within the size of the halo.

\section{Covariance Matrices} \label{Errors}

In order to perform accurate modelling of galaxy clustering measurements, it is not sufficient to have an accurate model; 
we must also have accurate estimates of the errors on the measurements, as well as the correlations between these 
errors. In the case of galaxy clustering on small scales, the covariance matrix is typically estimated from the data itself 
via jackknife resampling using contiguous regions on the sky \citep[e.g.,][]{zehavi_etal_02}. However, jackknife 
resampling does not accurately represent cosmic variance since it is limited to the scale of the jackknife subsamples 
rather than the full size of the galaxy sample. More importantly, \citet{norberg_etal_09} showed that, even on small 
scales, jackknife errors are biased in a scale dependent way. A more robust way of estimating the covariance matrix is to 
use a large number of independent realisations of the full sample, which can be done with mock catalogues.

In addition to the systematic problems with using jackknife errors, there is a more fundamental reason to use mock 
catalogues for error estimation. When we run a MCMC, we compute a likelihood function that is essentially equal to 
$P(\mathrm{data} | \mathrm{model})$, the probability that a dataset like the SDSS could be observed given the model 
being tested. The correct way to estimate this probability is to generate a large number of independent realisations of the 
model, each of which has the same volume and geometry as the SDSS, and check the fraction of them whose clustering 
measurements are further from their mean than the SDSS measurements. It is thus more correct to estimate errors from 
the model being tested than from the observed data set. In other words, the distribution of clustering measurements 
obtained from realisations of the model is the correct error distribution of the data given the model being tested.
For both of
these reasons, we use mock catalogues to estimate covariance matrices.
 
\subsection{Methodology}

To estimate the covariance matrix of each SDSS sample, we use the 200 independent mock catalogues described in
\S~\ref{Mocks}. However, in order to construct the mock catalogues, we need to first choose an \hod model that produces 
a mock galaxy distribution with similar clustering properties as the SDSS data. This, in turn, requires a covariance matrix.
We adopt the following iterative procedure to solve this chicken and egg problem. First, we construct initial covariance
matrices for the 1 \ngal value and the 14 \wprp bins of the $M_r<-19$ and~$-21$ samples using jackknife resampling on 
the sky. Specifically, we use 50 distinct contiguous regions to construct the jackknife samples, as described by 
\citet{mcbride_etal_11}. We then use the \citet{nelder_mead_65} downhill simplex algorithm to find the best-fit \hod 
model to our measurements of the projected correlation function. Each time we need to evaluate \wprp for a given set of 
trial \hod parameters, we use the ``minimum cosmic variance box'' as described in \S~\ref{wprp}. We use these best-fit 
\hod models to construct 200 mock catalogues for each SDSS sample using the methods detailed in \S~\ref{Mocks}.

Fig.~\ref{fig:sdss_wp_fit_for_corr_matrix} and Fig.~\ref{fig:sdss_gmf_fit_for_corr_matrix} show the projected correlation
functions and group multiplicity functions for each of these mock catalogues (grey lines) as well as the average and 
standard deviation of the 200 mocks (blue points and error bars). The figures show that the two-point clustering of mock 
catalogues agrees well with the clustering of the SDSS samples. The group multiplicity function of the mocks roughly 
agrees with the SDSS, but the agreement is not perfect. This is not surprising given that the multiplicity function was not 
used in the fit that determined the fiducial mock \hod model. We emphasise that it is not essential that these mocks 
perfectly match the SDSS clustering; what matters is that the variance among the 200 mocks correctly captures the 
errors and covariances of our clustering statistics. For this purpose it is sufficient to get the clustering approximately right. 
Fig.~\ref{fig:sdss_wp_fit_for_corr_matrix} shows that the errors estimated from the standard deviation of 200 mocks 
are approximately in agreement with the errors estimated from jackknife resampling of the SDSS samples on the sky. 

Armed with measurements of the global number density $n_g$, \wprp and \gmf for each of 200 mocks, we can construct 
the joint covariance matrix for each of our two samples. The matrices have dimensions $23\times23$ and $21\times21$ 
for the $M_r<-19$ and $M_r<-21$ samples and they are calculated as
\begin{equation}
C_{ij} = \frac{1}{N-1}\sum_1^N(y_i-\overline{y_i})(y_j-\overline{y_j}) ,
\end{equation}
where the sum is over the $N=200$ mocks, $y_i$ and $y_j$ are two of the measurements (e.g., the number density and 
one of the multiplicity function bins), and $\overline{y_i}$ and $\overline{y_j}$ are the mean measurements over the 200 
mocks. We also compute the correlation matrices, which are simply the covariance matrices normalised by their diagonal 
elements
\begin{equation}
\mathcal{R}_{ij} = \frac{C_{ij}}{\sqrt{C_{ii}C_{jj}}} .
\label{eqn:Rij}
\end{equation}
The correlation matrix has values between $-1$ and 1, with the diagonal elements being equal to unity by definition.

\subsection{The Joint Correlation Matrix} \label{Matrix}

\begin{figure*}
\centering
\includegraphics[clip=true,width=0.9\linewidth]{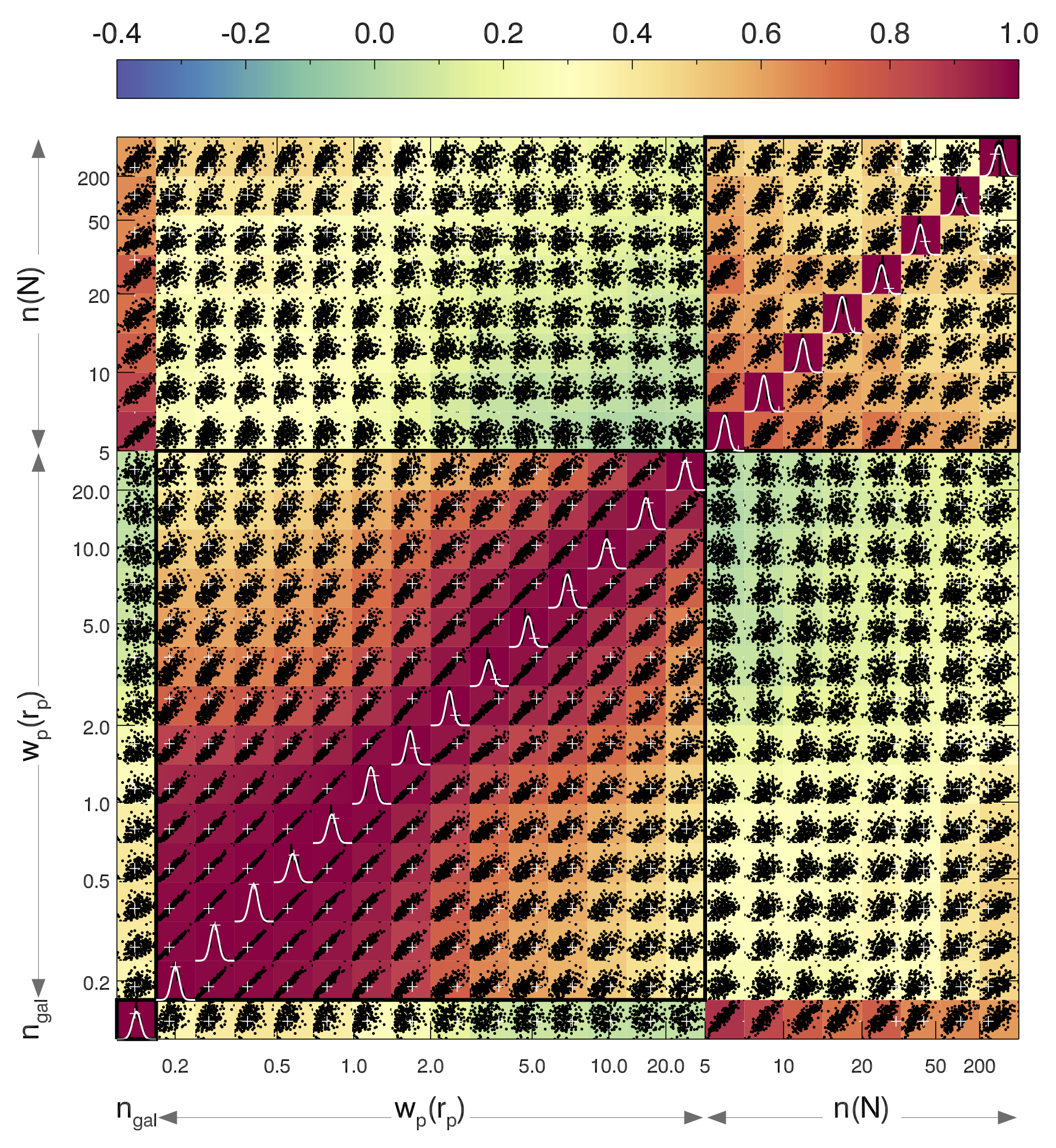}
\caption{\small The correlation matrix $\mathcal{R}$ for the number density, \ngal 
(first row/column), projected correlation function, $w_p$ (14 bins), and group multiplicity 
function, $n(N)$ (8 bins), estimated from 200 independent mock galaxy catalogues of the 
$M_r<-19$ sample. Each cell shows the correlation between two measurements (for 
example, the third bin of \wprp with the second bin of \gmf), with the colour of the cell 
denoting the correlation coefficient as given by equation~\ref{eqn:Rij}. Also shown within 
each cell is the corresponding scatter plot of 200 mock values ({\it black dots}) and 
the SDSS measurements for comparison ({\it white cross}). Each diagonal cell shows 
the distribution of mock values for that measurement ({\it black histogram}) as well as 
a Gaussian fit to the distribution ({\it white line}).}
\label{fig:corr_matrix_Mr19}
\end{figure*}

\begin{figure*}
\centering
\includegraphics[clip=true,width=0.9\linewidth]{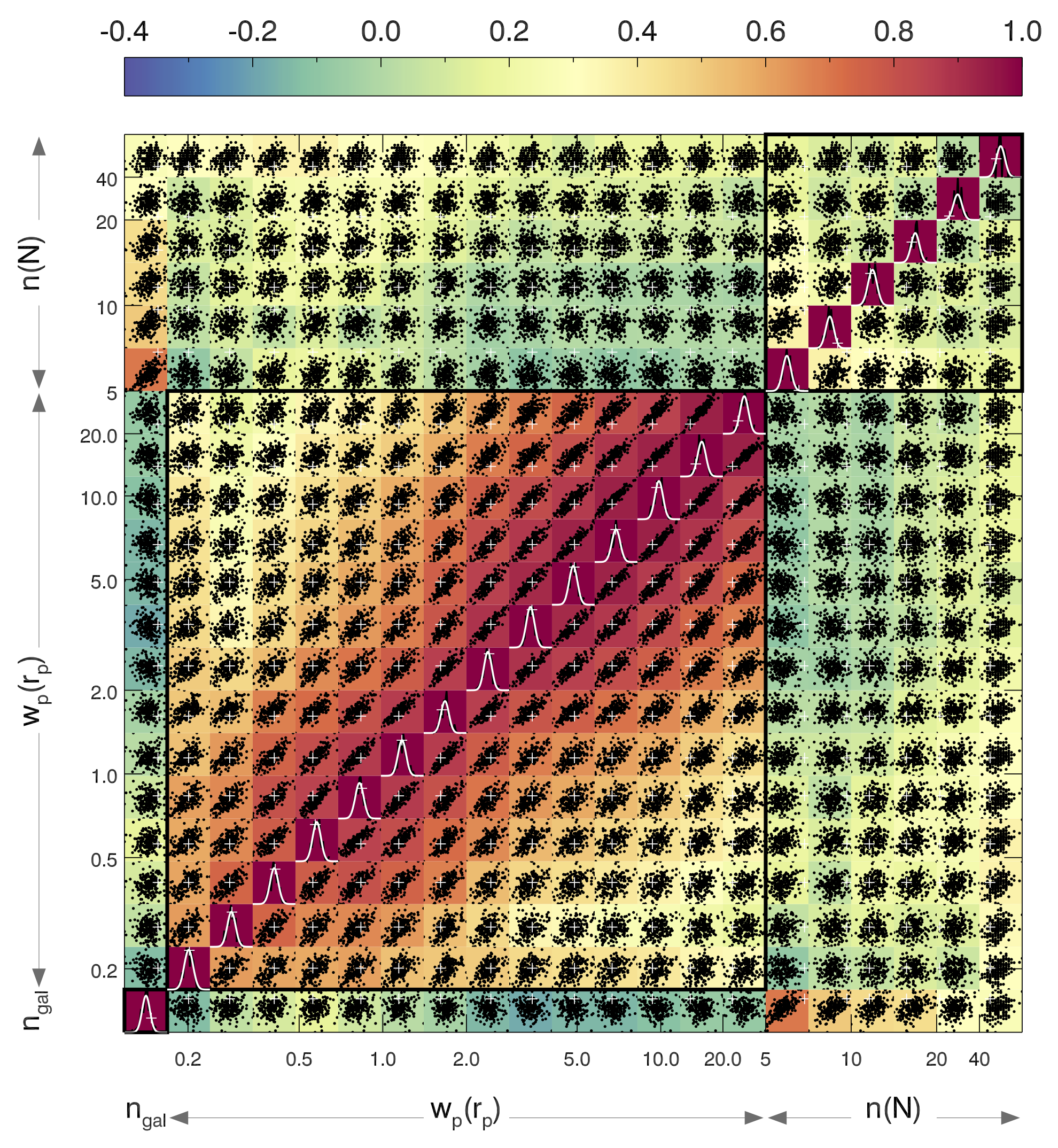}
\caption{\small The correlation matrix $\mathcal{R}$ for the number density, \ngal 
(first row/column), projected correlation function, $w_p$ (14 bins), and group 
multiplicity function, $n(N)$ (6 bins), estimated from 200 independent mock galaxy 
catalogues of the $M_r<-21$ sample. All features of the plot are the same as in 
Fig.~\ref{fig:corr_matrix_Mr19}.}
\label{fig:corr_matrix_Mr21}
\end{figure*}

Fig.~\ref{fig:corr_matrix_Mr19} shows the joint correlation matrix for the $M_r<-19$ sample. The x- and y-axes 
represent the various bins of the statistics we use -- \ngal, \wprp and \gmf. The square blocks of cells along the diagonal 
moving from the bottom left to the top right ($1\times1$, $14\times14$, and $8\times8$) show the correlation matrices of 
the individual statistics separately, while the rest of the matrix shows the correlations between different statistics. For 
example, the first column and bottom row are identical and show the correlation between \ngal and all the bins of the 
other statistics. Fig.~\ref{fig:corr_matrix_Mr21} shows the same for the $M_r<-21$ sample and differs only in the 
dimension of the \gmf portion of the matrix, which is $6\times6$. The colour of each cell in the matrix denotes the 
correlation coefficient $\mathcal{R}_{ij}$, as given by equation~(\ref{eqn:Rij}). Within all the non-diagonal cells, we show 
the actual measurements from the 200 mock catalogues as black dots. For example, in the second cell of the bottom row of 
the matrix, the black dots show the scatter plot between \ngal and the smallest scale bin of \wprp. These mock data 
provide a more in-depth way to understand the correlation matrix. Cells where
the mock values appear highly correlated (i.e., the black dots resemble a straight line)
have $\mathcal{R}_{ij}$ values close to unity, while cells where the mock values seem to be distributed randomly have 
correlation coefficients close to zero. Since the correlation coefficient $\mathcal{R}_{ij}$ only measures linear trends, it is 
possible in theory to have a scenario where the two variables are highly correlated (e.g., if they are constrained to be on 
a circle), but the associated correlation coefficient is zero. However, this is not a concern in this case since the 
cells with low $\mathcal{R}_{ij}$ shown in Fig.~\ref{fig:corr_matrix_Mr19} and Fig.~\ref{fig:corr_matrix_Mr21} do not show any obvious non-linear correlations.

Within the diagonal cells of the correlation matrices shown in Fig~\ref{fig:corr_matrix_Mr19}
and Fig.~\ref{fig:corr_matrix_Mr21} we display the probability distribution function of each clustering measurement from the 
200 mock values (black histograms), along with the best-fit Gaussian function (white lines). We find that the distribution 
of mock values is well described by a Gaussian for all the bins in \ngal, \wprp, and \gmf. Since the error distributions are 
Gaussian, we can use the $\bigchi^2$ technique to evaluate the goodness of fit. Finally, in all the cells we have also 
displayed the SDSS measurements for comparison (white crosses). Since our correlation matrices were created with 
mocks that were optimised to fit \ngal and \wprp, all of the SDSS values for these statistics are consistent with the mock 
values shown by the black dots. However, we did not optimise the mocks to fit \gmf, hence in quite a few of the \gmf 
cells, the SDSS and mock measurements do not overlap.

Let us now examine the structure in the joint correlation matrices, focusing first on \wprp and then on \gmf. The portions 
of the correlation matrices that correspond to \wprp reveal that neighbouring bins are very highly correlated. This is 
especially true for small scales in the $M_r<-19$ sample, where the first 6 bins contain almost no independent 
information (they have correlation coefficients greater than 0.9). On larger scales, correlations remain this high, but only 
for a couple neighbouring bins on each side of a given bin. In general, one must look very far to find bins that exhibit weak 
correlations. Only the smallest and largest scales of \wprp that we consider are relatively uncorrelated with each other. 
The overall degree of correlation is significantly less in the $M_r<-21$ sample, though even there neighbouring couple 
bins exhibit correlation coefficients higher than 0.8. The largest difference occurs at small scales, where the $M_r<-21$ 
sample displays much weaker correlations than the $M_r<-19$ sample. These results are in agreement with previous 
works. In particular, \citet{mcbride_etal_11} show their correlation matrices for both the redshift-space and the projected 
correlation function, revealing that much of the strong correlations on small scales are due to projection. This occurs 
because each projected scale includes pairs of galaxies from a wide range of physical scales, resulting in a high degree 
of scale mixing that causes different projected scales to contain very similar information. Even without projection, 
neighbouring scales in the correlation function are always correlated because they share the same underlying Fourier 
density modes. The $M_r<-21$ sample displays weaker correlations than the $M_r<-19$ sample for two main reasons. 
First, it is a lower density sample and thus shot noise, which is inherently uncorrelated, contributes more to the error 
budget. Second, as we discussed in \S~\ref{xi_signal}, most 1-halo pairs come from cluster-sized halos in both samples. 
In the $M_r<-21$ sample these halos typically only contribute a single pair, while in the $M_r<-19$ sample they 
contribute many pairs each. As the number of clusters fluctuates from mock to mock, in the low luminosity sample these 
fluctuations will enhance or suppress pairs at all 1-halo scales simultaneously, thus correlating the scales strongly. In the 
luminous sample this will occur at a much lesser extent.

\begin{figure*}
\centering
\includegraphics[clip=true,width=0.8\linewidth]{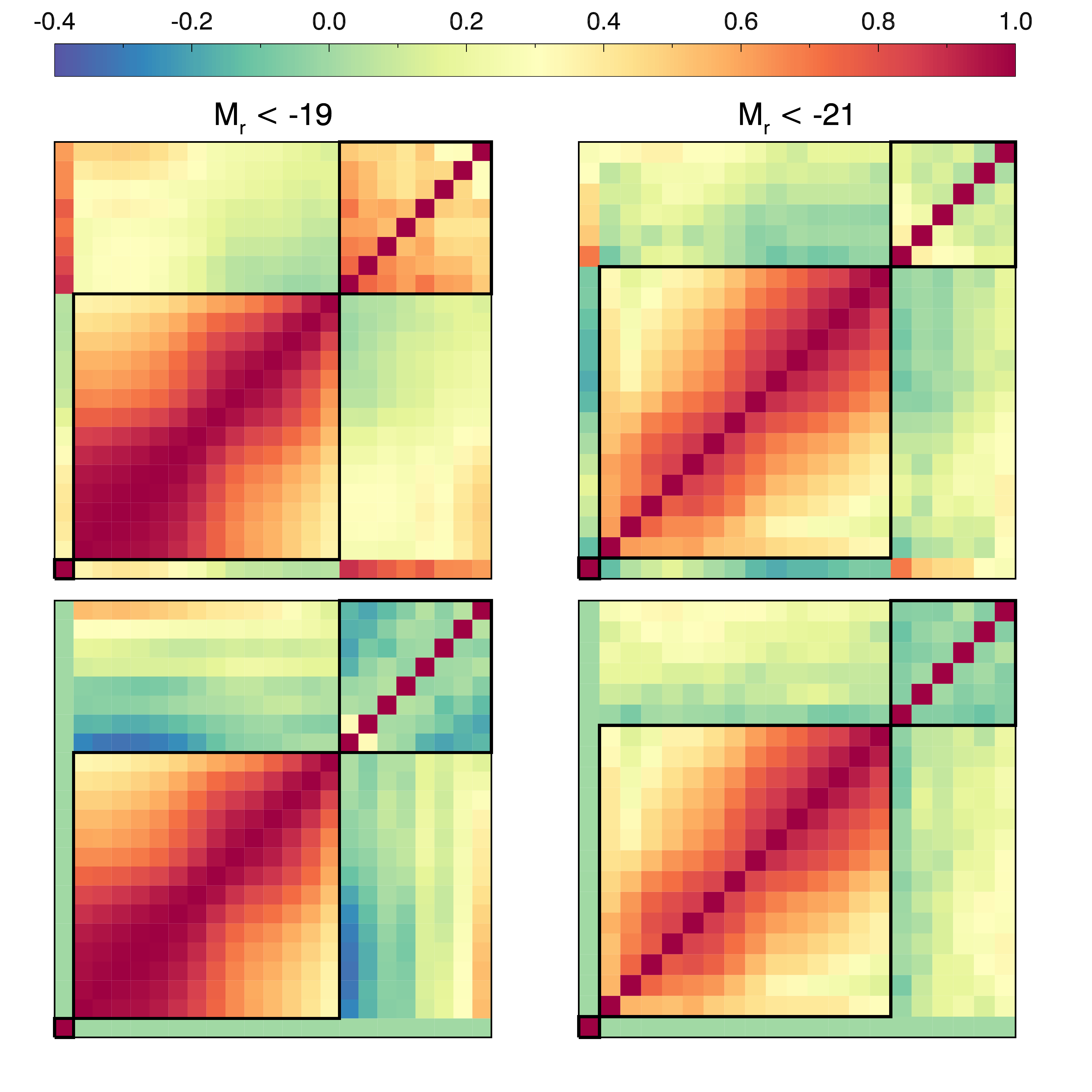}%
\caption{\small The effect of number density correlations on the correlation matrix 
for the $M_r<-19$ ({\it left panels}) and the $M_r<-21$ ({\it right panels}) samples.
Top panels show the fiducial correlation matrices as shown in Figs.~\ref{fig:corr_matrix_Mr19} 
and~\ref{fig:corr_matrix_Mr21}. Bottom panels show correlation matrices created from 
the same initial 200 mocks, but after randomly downsampling the mocks to the same 
galaxy number density, \ngal (the lowest amongst the 200 mocks). By matching the \ngal 
in all the mocks, we remove the effect of two measurements showing a correlation due to 
their individual correlations with \ngal. Since the correlation function does not depend on 
density, randomly downsampling each of the mocks has almost no effect on the correlation 
coefficients in the \wprp part of the matrix. However, the correlations in the \gmf bins 
change drastically. Most of the positive correlations seen in the top panels vanish to reveal 
nearly diagonal \gmf coefficients in the bottom panels.} 
\label{fig:corr_matrices_subsampled}
\end{figure*}

We next move to the portions of the correlation matrices that correspond to the group multiplicity function, \gmf. Different
multiplicity bins are not as correlated as different scales of \wprp, but there are still significant correlations (coefficients in 
the range 0.4-0.8) in the case of the $M_r<-19$ sample. In the $M_r<-21$ sample \gmf correlations are much weaker 
(coefficients in the range 0.1-0.4). Some of the correlation between bins of \gmf is likely of an indirect nature, due to 
direct correlations between these bins and the overall number density \ngal, which can be quite strong according to 
Fig.~\ref{fig:corr_matrix_Mr19} and Fig.~\ref{fig:corr_matrix_Mr21}. To uncover the intrinsic correlations between the \gmf 
bins, we subsample the 200 mocks (for both samples) to have identical number densities. We then create new 
correlation matrices out of these 200 \ngal matched mocks. In Fig.~\ref{fig:corr_matrices_subsampled}, we show the 
original (top panels) and subsampled (bottom panels) joint correlation matrices for the $M_r<-19$ (left panels) and 
$M_r<-21$ (right panels) samples. The correlation matrices of \wprp do not change because the correlation function is 
not affected by subsampling. However, in both samples we find that, when controlling for \ngal, the correlation matrix of 
\gmf becomes much more diagonal. The matrix becomes completely diagonal in the $M_r<-21$ case, while in the 
$M_r<-19$ case there is some anti-correlation present between low and high multiplicity groups. This anti-correlation is a 
result of the nature of the group-finder since, in the case of constant density, an above average abundance of high 
multiplicity groups in one mock catalogue must come at the expense of low multiplicity groups. From this subsampling test 
we thus learn that the correlations seen in the group multiplicity function are largely due to correlations with the overall 
number density. When the density is higher, the entire \gmf is boosted.

\subsection{Noise in the Correlation Matrix} \label{PCA}

Each correlation matrix $\mathcal{R}$ that we have estimated from 200 mock catalogues contains some degree of noise 
due to the fact that the number of mocks is limited. When we invert the matrix, this noise can amplify and affect the 
calculation of $\chi^2$ values in unpredictable ways. We deal with this problem using a singular value decomposition 
(SVD) approach 
\citep[e.g.,][]{scoccimarro_2000,eisenstein_zaldarriaga_01,gaztanaga_and_scoccimarro_05,norberg_etal_09}. 
Specifically, we decompose the correlation matrix into its principle components by finding the eigenvectors 
$\mathcal{E}_i$ and eigenvalues $\lambda_i$ that satisfy the equations
\begin{equation}
\mathcal{R}\mathcal{E}_i = \lambda_i\mathcal{E}_i .
\end{equation}
This rotates the space of our measurements into a basis where the eigenvectors are uncorrelated (i.e., where the 
correlation matrix is diagonal). We can then sort the eigenvectors by their eigenvalues and trim out those with low 
eigenvalues, which contain much of the noise, but little information. Following \citet{gaztanaga_and_scoccimarro_05}, 
we only keep eigenvectors for which $\lambda_i^2$ is approximately larger than the resolution with which $\mathcal{R}$ 
is measured, which is
\begin{equation}
\lambda_i^2 \gtrsim \sqrt{\frac{2}{N_\mathrm{mocks}}} ,
\end{equation}
where, in our case, $N_\mathrm{mocks}=200$. This procedure effectively reduces the number of data points that we use 
for fitting models. For the $M_r<-19$ correlation matrix, we trim 12 of the 23 eigenvectors and thus only keep 11 data 
points in the new orthogonal measurement space. For the $M_r<-21$ matrix, we only trim 5 of the 21 eigenvectors and 
thus keep 16 data points. The more drastic trimming for the low luminosity sample is a direct result of the higher amount 
of correlation present in its correlation matrix. For all the joint fits to \ngal, \wprp, and \gmf that we perform and describe 
in the next section, we run two versions: one with the full correlation matrix and one after trimming out noisy eigenvectors 
(which we label as PCA in the text). The PCA fits throw away some of the signal present in the data, but the resulting 
$\chi^2$ values are more reliable and so these are the more conservative results.

\section{Model Fitting} \label{Fitting}

Our primary objective is to accurately model galaxy clustering statistics and thus test the standard $\Lambda$CDM + 
halo model. To do this, we explore the \hod parameter space using a MCMC method. 

\subsection{Computing the Likelihood Function}

For each parameter combination we compute the likelihood function $P(\mathrm{data} | \mathrm{model})$. Since the 
error distributions are Gaussian, the likelihood function is proportional to $\mathrm{exp(-\chi^2/2)}$. We compute the 
$\chi^2$ statistic as follows
\begin{equation}
\chi^2 = \sum_{ij} \chi_i \mathcal{R}_{ij}^{-1} \chi_j ,
\label{eqn:chi_2}
\end{equation}
where $\mathcal{R}^{-1}$ is the inverted correlation matrix and 
\begin{equation}
\chi_i = \frac{D_i-M_i}{\sigma_i} ,
\label{eqn:chi_i}
\end{equation}
where $D_i$ is the SDSS measurement for data point $i$, $M_i$ is the model prediction for that measurement, and 
$\sigma_i$ is the error in the SDSS measurement. When we use the PCA analysis to reduce noise introduced by 
inverting the correlation matrix, we calculate $\chi^2$ differently. Since the resulting eigenvectors are uncorrelated, 
$\chi^2$ is simply a sum of the $\chi_i^2$ contributions for the eigenvectors we have kept (where $D_i$ and $M_i$ are 
recomputed for the new orthogonal space).

Ideally, each time we moved to a new point in the \hod parameter space, we would compute $\chi^2$ using updated 
values of $\sigma_i$ and an updated matrix $\mathcal{R}$ that were generated from the new model. In other words, we 
would have to create a new set of 200 mock catalogues. This would increase the computational requirement of the MCMC 
chain by two orders of magnitude and is currently unfeasible. We thus make the approximation that our fiducial \hod, 
which reproduces the \wprp measured in the SDSS sample, is representative of the errors and correlation matrix in the 
parameter-space of interest. Specifically, we adopt the fractional errors given by the fiducial mock catalogues and scale 
them by the SDSS measurements $D_i$ to obtain
\begin{equation}
\sigma_i = \frac{\sigma_{\mathrm{mock},i}}{M_{\mathrm{mock},i}} \times D_i ,
\label{eqn:sigma_i}
\end{equation}
where $M_{\mathrm{mock},i}$ and $\sigma_{\mathrm{mock},i}$ are the mean and standard deviation of data point $i$, 
as measured from our 200 fiducial mock catalogues. $\sigma_i$ are then the scaled SDSS errors that we use when 
computing $\chi^2$ in equation~(\ref{eqn:chi_i}).  

Given that we fix $\mathcal{R}$ and $\sigma_i$ using the fiducial mocks, the only ingredient we must compute for each 
parameter combination within the MCMC is the model prediction $M_i$. As was discussed in \S~\ref{Nbody_mcmc}, we 
use a substantially larger volume to compute $M_i$ than the SDSS volume that was used to compute $D_i$. For \wprp, 
we use a single mock catalogue made from a whole simulation cube, which has approximately 13 times more volume than 
the corresponding SDSS sample. For \gmf we use the mean of eight mock catalogues, each of which has the same volume 
as the corresponding SDSS sample. Furthermore, we reduce cosmic variance errors by an additional factor of $\sim2$ 
by carefully selecting which simulation boxes we use to compute $M_i$. The uncertainties in the model prediction $M_i$ 
are thus sub-dominant compared to the uncertainties in $D_i$ and we can safely ignore them when we compute the 
model likelihood.

\subsection{Running and Analyzing the MCMC Chains}

For each new set of \hod parameters, we must perform the following operations: (1) populate the halo catalogues from two 
simulation boxes with galaxies according to the \hod; (2) compute \wprp on one of these boxes; (3) create eight SDSS-
like mock catalogues from the two boxes; (4) run the friends-of-friends group finder on these eight mocks and measure the
mean \gmf. The codes used for these operations are heavily optimised and take approximately 15 seconds of wallclock 
time on a single TACC Stampede compute core (for details on the software
implementation of \wprp, see \citealt{corrfunc_sinha_lehman}). Though this is remarkably fast given the computations involved, it is slow enough that 
we need to use an efficient and parallel MCMC algorithm to perform the $\sim10^5$ evaluations of model likelihood 
required for each parameter search.

We use the code \emcee~\citep{foreman-mackey_etal_13} to generate the MCMC samples. The algorithm employs a 
number of ``walkers" that probe different points in parameter space and can be run in parallel. The MCMC chain is built 
from a series of iterations where the walkers in each iteration learn from previous iterations and improve their efficiency. 
We run a total of ten MCMC chains for each luminosity sample. Two of these are model fits to \wprp or \gmf alone, while 
the remaining eight are for joint fits to both statistics, using different combinations of assumed cosmology, halo definition, 
and whether we apply PCA to reduce noise in the correlation matrix (2 cosmological models $\times$ 2 halo definitions 
$\times$ 2 PCA cases). We note that the number density \ngal is explicitly included as a statistic in all chains. For each 
chain, we use 500 walkers and $\sim1000$ iterations for a total of $\sim500,000$ \hod evaluations. We check for 
convergence in each parameter by demanding that its probability distribution is stable across iterations and we find that 
in all cases the chains converge within $200-600$ iterations.

When running the MCMC, we adopt uniform priors for all five \hod parameters within the following allowed ranges. For 
the $M_r<-19$ sample, the ranges are $\logMmin: 11-12.2$, $\siglogM: 0.001-1$, $\logM_0: 6-14$, $\logMone: 12-14$, 
$\alpha: 0.001-2$. For the $M_r<-21$ sample, the ranges are $\logMmin: 12-14$, $\siglogM: 0.001-1$, $\logM_0: 6-15$, 
$\logM_1: 13-15$, $\alpha: 0.001-2$. The motivation for the lower limits on \logMmin and the upper limits on \siglogM is 
to avoid a scenario where we place mock galaxies in halos that are not sufficiently resolved. 

When a chain is complete, we throw out the first $200-600$ iterations since these retain memory of the starting locations 
of the 500 walkers (the ``burn-in" phase). We then explore the posterior probability distribution for each of our five \hod 
parameters, as well as the joint distributions for different combinations of parameter pairs. We record the median and 
the percentiles containing 68\% of chain values for each parameter and list those as our main marginalised parameter constraints. 
Moreover, we record the parameter values for the best-fit model, which is simply the point in the whole MCMC chain with 
the lowest value of $\chi^2$. To assess goodness-of-fit, we use the \pvalue that is associated with this best-fit value of 
$\chi^2$. The \pvalue represents the probability that a sample randomly drawn from from the best-fit model could have a 
$\chi^2$ value greater than the one exhibited by the SDSS. In other words, the \pvalue is the probability that the SDSS is 
consistent with the best-fit model. Our ability to estimate reliable goodness-of-fit probabilities stems from the fact that we 
have all the systematic errors of the analysis under control, and represents one of the main advances of this work.

\subsection{Fits to the Correlation Function $\wprp$}\label{sec:wponly}

\begin{figure*}
\centering
\includegraphics[clip=true,width=0.75\linewidth]{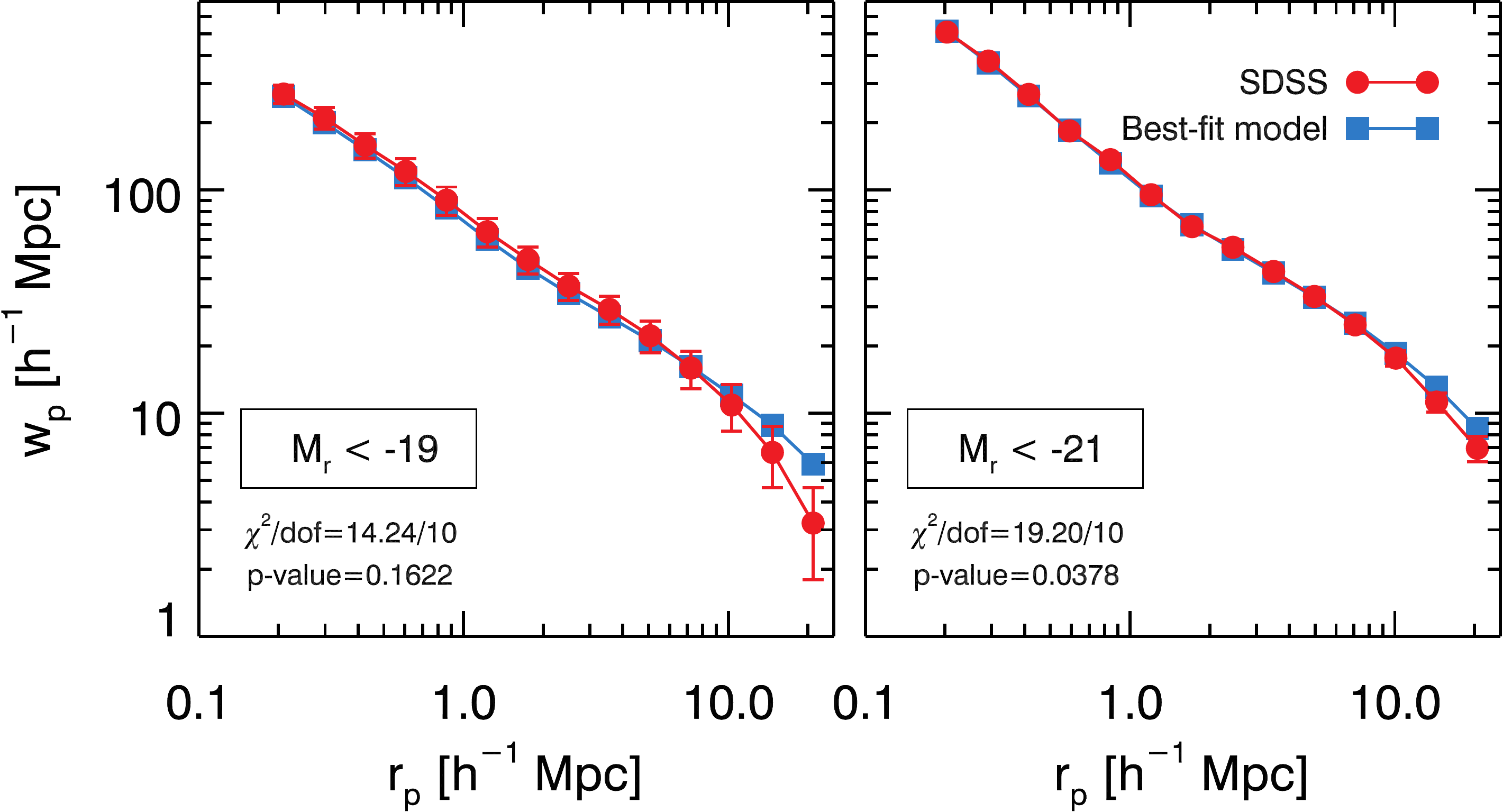}
\caption{\small Projected correlation function \wprp for the best-fit \hod model when the 
model is only fit to \wprp and the galaxy number density. Blue points show the best-fit 
model, while red points show the SDSS measurements in the case of the $M_r<-19$
({\it left panel}) and $M_r<-21$ ({\it right panel}) samples. Error bars on the SDSS
measurements are estimated from the dispersion among 200 mock catalogues (shown
in Fig.~\ref{fig:sdss_wp_fit_for_corr_matrix}). The $\chi^2$, degrees of freedom and 
\pvalues are listed in the panels. The model shown here assumes the LasDamas
cosmology and the virial halo definition (Mvir) and does not include PCA reduction.}
\label{fig:wp_single_chains}
\end{figure*}

Before modelling the correlation function and the group multiplicity function jointly, we model each statistic alone in order 
to compare the constraining power of the statistics. First, we model the correlation function measurements together with 
the number density. This is comparable to what other authors have done in modelling the SDSS DR7 data using both 
analytic \citep{zehavi_etal_11} and mock-based \citep{zentner_etal_16} methods. For this analysis we adopt the 
LasDamas cosmological model and the virial halo definition (Mvir). We also use the original correlation matrices, rather 
than the PCA versions that have trimmed eigenvectors in order to reduce noise. Since we are only fitting a model to \ngal 
and \wprp, we only use the first $15\times15$ portion of the correlation matrices shown in 
Fig.~\ref{fig:corr_matrix_Mr19} and Fig.~\ref{fig:corr_matrix_Mr21}. 

Fig.~\ref{fig:wp_single_chains} shows \wprp for the best-fit model compared to the SDSS data for the two luminosity 
samples we consider in this work. The error bars in the plot show $\sigma_i$ as estimated using 
equation~(\ref{eqn:sigma_i}). The best-fit model appears by-eye to provide a decent match to the SDSS \wprp. However, 
visual inspection can be a highly misleading way to assess goodness of fit when data points are highly correlated, as is 
the case here. We thus rely on values of $\chi^2$ and corresponding \pvalues. The best-fit values of $\chi^2$ are 14.2 
and 19.2 for the $M_r<-19$ and $M_r<-21$ samples, respectively. Both samples have 10 degrees of freedom (14 bins 
for \wprp + 1 for \ngal $-$ 5 free \hod parameters). The resulting \pvalue for the $-19$ sample is 0.162, which is 
acceptable. The \pvalue for the $-21$ sample is 0.038, which reveals a tension slightly larger than $2\sigma$, but 
certainly not enough to warrant ruling out the model. It is notable that the low luminosity sample shows the best 
goodness-of-fit even though it looks like a worse fit by visual inspection of Fig.~\ref{fig:wp_single_chains}. This is due 
to the high degree of correlation between bins in the $-19$ measurement of \wprp.

\subsection{Fits to the Group Multiplicity Function $\gmf$}\label{sec:gmfonly}

\begin{figure*}
\centering
\includegraphics[clip=true,width=0.75\linewidth]{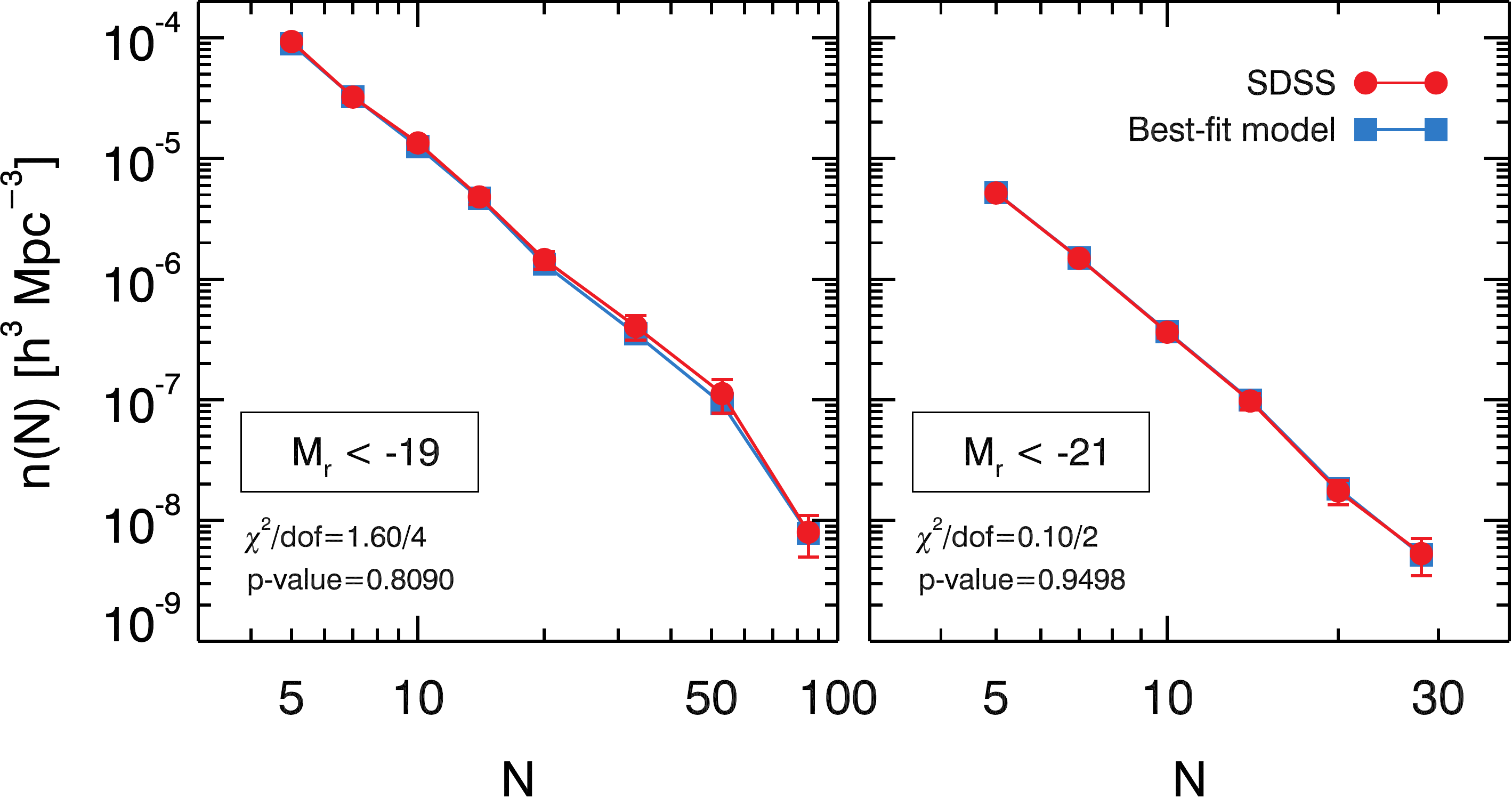}
\caption{\small Group multiplicity function \gmf for the best-fit \hod model when the 
model is only fit to \gmf and the galaxy number density. Blue points show the best-fit 
model, while red points show the SDSS measurements in the case of the $M_r<-19$
({\it left panel}) and $M_r<-21$ ({\it right panel}) samples. Error bars on the SDSS
measurements are estimated from the dispersion among 200 mock catalogues (shown
in Fig.~\ref{fig:sdss_gmf_fit_for_corr_matrix}). The $\chi^2$, degrees of freedom and 
\pvalues are listed in the panels. The model shown here assumes the LasDamas
cosmology and the virial halo definition (Mvir) and does not include PCA reduction.}
\label{fig:gmf_single_chains}
\end{figure*}

We now model the group multiplicity function measurements together with the number density. This is the first time this 
statistic is being used for \hod modelling of SDSS data. As before, for this analysis we adopt the LasDamas cosmological 
model, the virial halo definition (Mvir), and the original non-PCA correlation matrices. Since we are only fitting a model to 
\ngal and \gmf, we only use a $9\times9$ subset of the full correlation matrix shown in Fig.~\ref{fig:corr_matrix_Mr19}  
and a $7\times7$ subset of the matrix shown in Fig.~\ref{fig:corr_matrix_Mr21}. 

Fig.~\ref{fig:gmf_single_chains} shows \gmf for the best-fit model compared to the SDSS data for the two luminosity 
samples we consider in this work. The error bars in the plot show $\sigma_i$ as estimated using 
equation~(\ref{eqn:sigma_i}). The best-fit model appears by-eye to provide a perfect match to the SDSS \gmf and the 
\pvalues estimated from the $\chi^2$ values confirm this. The best-fit values of $\chi^2$ are 1.6 and 0.1 for the 
$M_r<-19$ and $M_r<-21$ samples, respectively. The corresponding number of degrees of freedom for the two samples 
are 4 and 2  (8 or 6 bins for \gmf + 1 for \ngal $-$ 5 free \hod parameters). The resulting \pvalues are 0.81 and 0.95, 
which are large enough to suggest that the model has too much freedom. A simpler \hod form with fewer free parameters 
might be equally successful at reproducing the observed \gmf. 

Since this is the first time that the group multiplicity function is being used to constrain \hod parameters, it is worth
comparing its constraining power to that of the correlation function. In the case of the $M_r<-19$ sample, fitting to \gmf
yields similar marginalised uncertainties in \logMmin and $\siglogM$ compared to fitting to \wprp, but a 20\% smaller 
uncertainty in \logMone  and a 35\% smaller uncertainty in $\alpha$. In the case of the $M_r<-21$ sample however, 
fitting to \gmf leads to substantially weaker constraints, with uncertainties in \logMmin, \siglogM, \logMone, and $\alpha$ 
blowing up by 45-100\% compared to \wprp. This difference is partly due to the more correlated nature of \wprp in the low 
luminosity sample, which reduces its statistical power compared to the high luminosity sample. Since we are not 
proposing replacing \wprp as a statistic to fit to, a more relevant way to assess the value of \gmf is to see how \hod 
constraints improve when fitting to both statistics compared to \wprp alone. We investigate this next.

\subsection{Joint Fits to Both $\wprp$ and $\gmf$}

\begin{figure*}
\centering
\includegraphics[clip=true,width=0.7\linewidth]{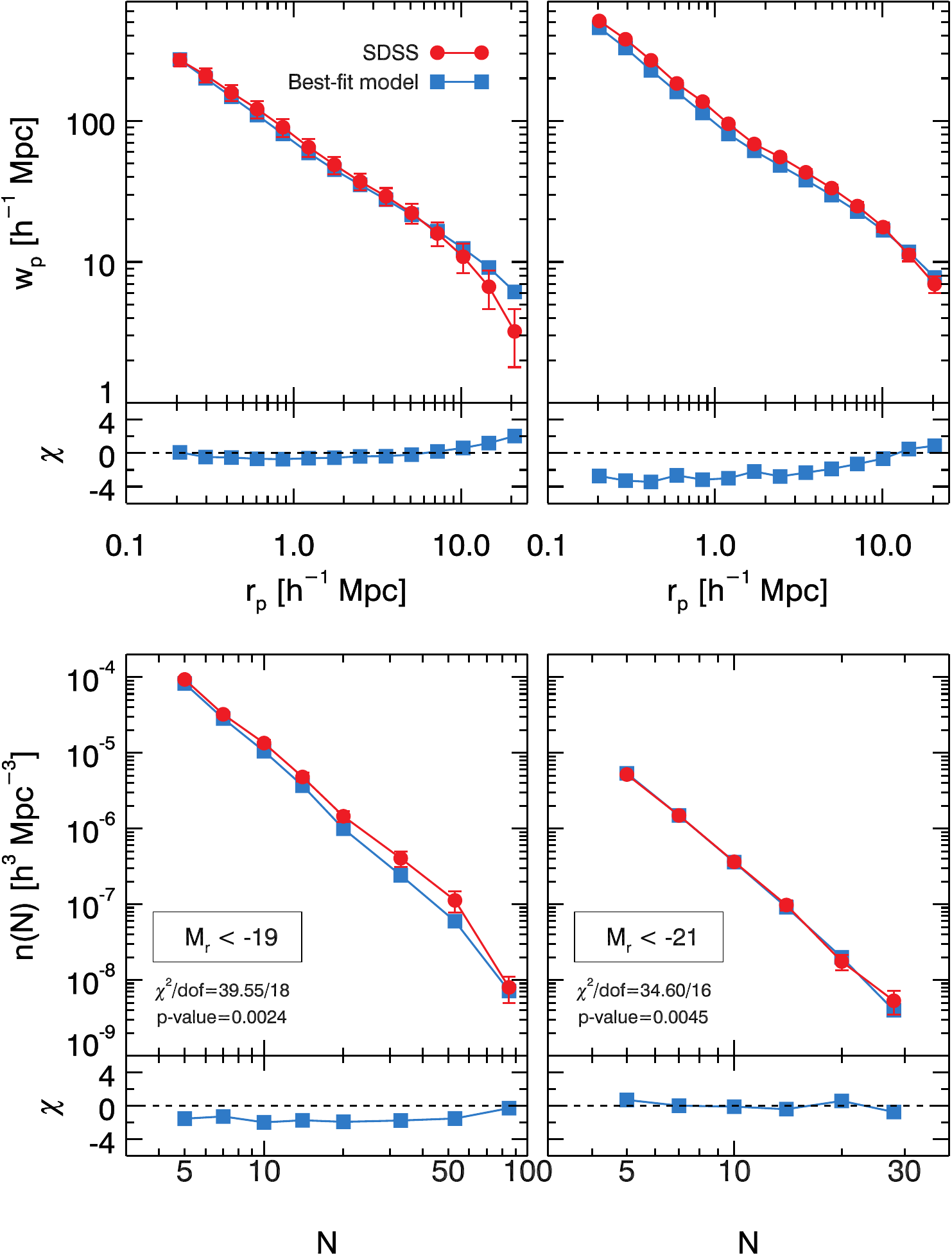}
\caption{\small Projected correlation function \wprp ({\it top panels}) and group multiplicity 
function \gmf ({\it bottom panels}) for the best-fit \hod model when the model is jointly fit to 
\wprp, \gmf, and the galaxy number density. Blue points show the best-fit model, while 
red points show the SDSS measurements in the case of the $M_r<-19$ ({\it left panels}) 
and $M_r<-21$ ({\it right panels}) samples. Error bars on the SDSS measurements 
are estimated from the dispersion among 200 mock catalogues (shown in 
Figs.~\ref{fig:sdss_wp_fit_for_corr_matrix} and~\ref{fig:sdss_gmf_fit_for_corr_matrix}). 
The bottom section of each panel shows $\chi$, which is the difference between the best-fit 
and SDSS measurements, divided by the standard deviation of the mocks. The $\chi^2$, 
degrees of freedom, and \pvalues are listed in the panels. The model shown here assumes 
the LasDamas cosmology and the virial halo definition (Mvir) and does not include PCA reduction.}
\label{fig:joint_best_fit}
\end{figure*}  

In this section, we present results from the MCMC chain that simultaneously fits \wprp, \gmf, and the galaxy number 
density. Once again, for this analysis we adopt the LasDamas cosmological model, the virial halo definition (Mvir), and 
the original non-PCA correlation matrices (the full correlation matrices shown in Fig.~\ref{fig:corr_matrix_Mr19}  
and Fig.~\ref{fig:corr_matrix_Mr21}). In the next sections we will explore a different cosmology and halo definition, 
as well as the effect of reducing noise in the correlation matrices via PCA. 

Fig.~\ref{fig:joint_best_fit} shows both statistics for the best-fit model compared to the SDSS data, for the two 
luminosity samples we consider in this work. It is clear from the figure that jointly fitting \wprp and \gmf is a significantly 
more challenging task than fitting each statistic individually. The best-fit model for the $M_r<-19$ sample (top panels) 
reproduces \wprp reasonably well, but fails at matching \gmf, while the reverse is true for the $M_r<-21$ sample
(bottom panels). The corresponding values of $\chi^2$ are 39.6 and 34.6 and the number of degrees of freedom are 18 
and 16, respectively. The resulting \pvalues are 0.0024 and 0.0045, which suggest that the model is ruled out at the 
$\sim3\sigma$ level when tested against either luminosity sample. The best-fit model results for this and all subsequent
MCMC chains are listed in Table~\ref{table:bestfit}.

\begin{figure*}
\centering
\includegraphics[clip=true,width=0.75\linewidth]{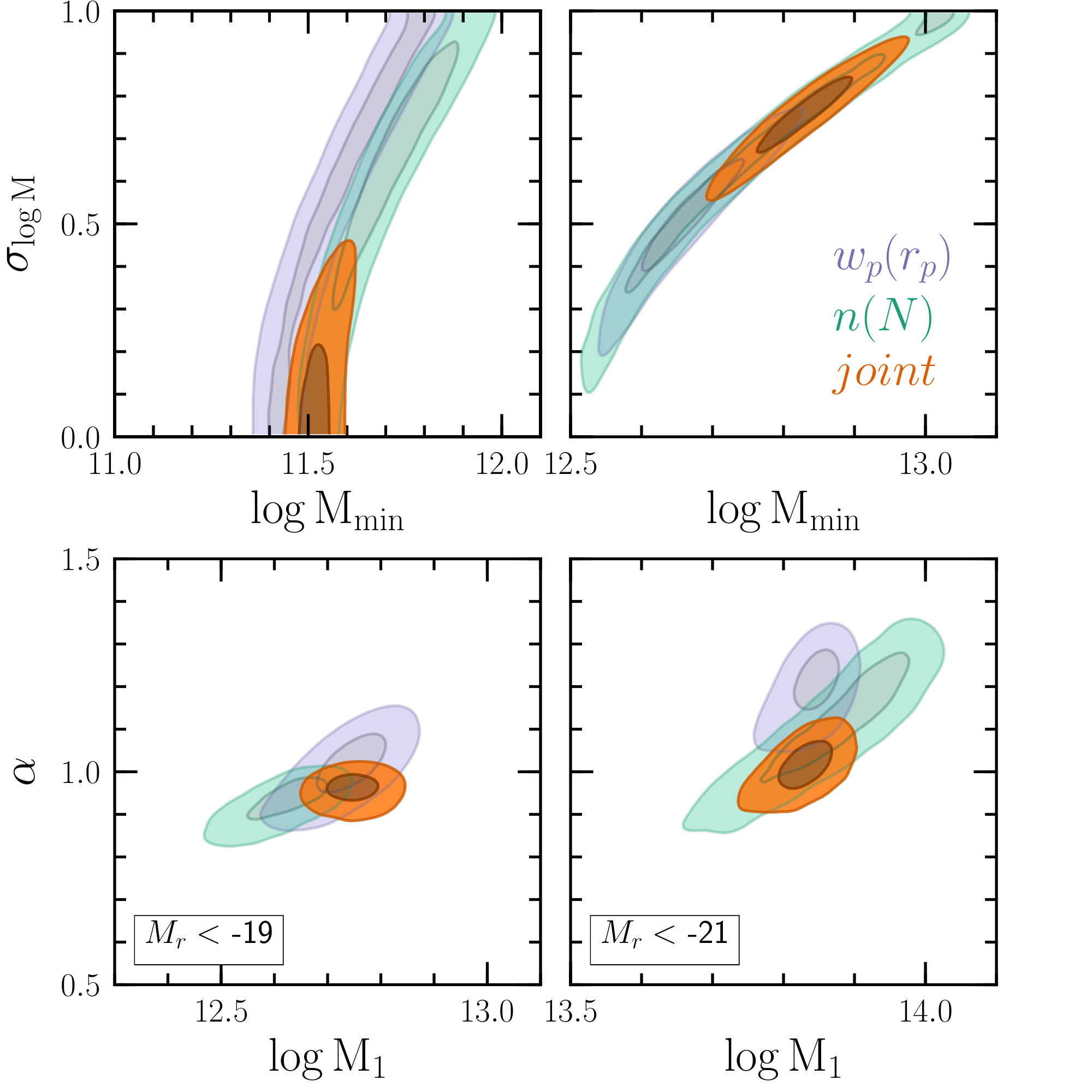}
\caption{\small Posterior probability distributions for \hod parameters 
when the model is fit to SDSS measurements in the case of the 
$M_r<-19$ ({\it left panels}) and $M_r<-21$ ({\it right panels}) 
samples. Top panels show the joint distribution of central galaxy 
parameters \logMmin and \siglogM, while bottom panels show the 
joint distribution of satellite galaxy parameters \logMone and $\alpha$. 
In each case, the probability distributions are marginalised over the 
remaining three \hod parameters. In each panel, blue, green and orange 
contours show results when the model is fit to measurements of \wprp 
only, \gmf only, and both statistics jointly (together with the galaxy number 
density in all cases). The contours show the regions of parameter space 
that contain 68\% and 95\% of the MCMC probability. The model shown 
here assumes the LasDamas cosmology and the virial halo definition (Mvir) and does not include PCA reduction.}
\label{fig:joint_parameter_conf}
\end{figure*}  

To better understand the tension between fitting \wprp and \gmf, it is helpful to study the posterior probability distributions
for the \hod parameters. Fig.~\ref{fig:joint_parameter_conf} shows joint distributions for central galaxy \hod 
parameters \logMmin vs. \siglogM, and satellite parameters \logMone vs. $\alpha$. We do not show results for $M_0$ 
because it is always very poorly constrained. In each case we compare results when fitting our model to \wprp only (blue 
contours), \gmf only (green contours), and both statistics jointly (orange contours). The galaxy number density is also 
used as a constraint in all cases.

First, let us examine the parameter constraints for the \wprp and \gmf only chains. In the case of the $M_r<-19$ sample, 
the \gmf chain prefers lower values of \logMone and $\alpha$, and a higher value of \logMmin, compared to the \wprp 
chain. To understand this, we can look back at Fig.~\ref{fig:sdss_gmf_fit_for_corr_matrix}, which shows \gmf 
for the mock catalogues that we used to make the joint correlation matrices (and that matched the SDSS \wprp by design). 
The top panel of that figure shows that the model predictions for \gmf are lower than the SDSS \gmf at all $N$ except for 
the largest-$N$ bin. In other words, a \hod model that matches the SDSS \wprp under-predicts \gmf by almost 
2-$\sigma$. To fit the SDSS \gmf, we would thus need to boost \gmf at all $N$. This can be achieved primarily by 
reducing \logMone (with an adjustment to $\alpha$). However, reducing \logMone increases \ngal because, in the 
$M_r<-19$ sample, the satellite fraction is $\sim 0.3$ so changes in the satellite occupation can contribute a significant 
change in \ngal. This increase in \ngal has to be compensated by increasing \logMmin as well.\footnote{Changing 
\logMmin only mildly affects \gmf since the linking length changes adaptively as \ngal$^{-1/3}$. The effect is stochastic 
and affects mostly the small-$N$ groups.} As a result, the parameter constraints for the \gmf only chain have 
systematically lower \logMone and higher \logMmin than for the \wprp only chain.

In the case of the $M_r<-21$ sample, the \gmf chain prefers a somewhat lower value of $\alpha$, but similar values of 
the other three parameters, compared to the \wprp chain. The bottom panel of 
Fig.~\ref{fig:sdss_gmf_fit_for_corr_matrix} shows that the \hod model that matches the SDSS \wprp predicts a \gmf 
with a slope that is too shallow (it underpredicts the abundance of small-$N$ groups but overpredicts the abundance of 
large $N$-groups). To fit the SDSS \gmf, we would thus primarily need to lower $\alpha$ (with a small adjustment to 
\logMone). This change to the satellite occupation does not significantly affect \ngal because the satellite fraction in the 
$M_r<-21$ sample is only $\sim 0.1$. Consequently, the central galaxy parameters do not need to change.

Fig.~\ref{fig:joint_parameter_conf} shows that the constraints on \hod parameters when fitting jointly to \wprp and \gmf
are significantly tighter than when fitting \wprp alone. In the case of the $M_r<-19$ sample, adding group statistics 
reduces the marginalised uncertainties in \logMmin, \siglogM, \logMone, and $\alpha$ by 40-60\%. In the case of the
$M_r<-21$ sample, while the uncertainty in \logMmin does not change and the uncertainty in \logMone only improves by 
10\%, the uncertainties in \siglogM and $\alpha$ improve by 30\%. The means and standard deviations of the 
marginalised posterior distributions for all \hod parameters in this and all subsequent MCMC chains are listed in 
Table~\ref{table:hodvalues}. These improved parameter constraints, along with the lower best-fit \pvalues, demonstrate 
the power of combining new clustering statistics with \wprp into \hod modelling of the galaxy distribution.

\subsection{Accounting for Noise in the Correlation Matrix}

Our results jointly fitting \wprp, \gmf, and the galaxy number density of $M_r<-19$ and $-20$ galaxies with a LasDamas 
cosmological model, Mvir halos, and the standard 5-parameter \hod, show that the model is ruled out at the 
$\sim 3\sigma$ level for both low and high luminosity samples. Before we conclude that our model assumptions are 
incorrect, we need to make sure that the best-fit \pvalues are robust. As discussed in \S~\ref{PCA}, noise in our 
estimated correlation matrices can bias our fit results in difficult to predict ways. We thus follow the PCA procedure 
outlined in that section to trim noisy eigenvectors from the correlation matrices and we re-run all of our MCMC chains.

Table~\ref{table:bestfit} shows the best-fit results for the PCA chains. In the case of the $M_r<-19$ sample, the best-fit
\pvalue increases by a factor of $\sim$30, from 0.0024 to 0.0751. In other words, the tension between model and 
data reduces from 3$\sigma$ to 1.8$\sigma$ and we no longer rule it out. For the $M_r<-21$ sample, however, the 
results do not change significantly and the tension between model and data remains strong. The difference between the 
two samples is due to the much more correlated nature of the low luminosity measurements, as seen in their correlation
matrices (Figs~\ref{fig:corr_matrix_Mr19} and~\ref{fig:corr_matrix_Mr21}) . Table~\ref{table:hodvalues} shows the effect 
that our PCA analysis has on the \hod parameter constraints. For the $M_r<-19$ sample, the constraints degrade 
considerably, with 1$\sigma$ uncertainties in \logMmin, \siglogM, \logMone, and $\alpha$ increasing by factors of 4.3, 
2.7, 1.6, and 1.7, respectively. Parameter uncertainties for the $M_r<-21$ sample are unaffected. In summary, the PCA
approach leads to no change for the high luminosity sample, and weaker but more reliable constraints for the low luminosity sample. We adopt these as our main results 
moving forward.

\subsection{Varying the Halo Definition and Cosmology}

\begin{figure}
\centering
\includegraphics[clip=true,width=\linewidth]{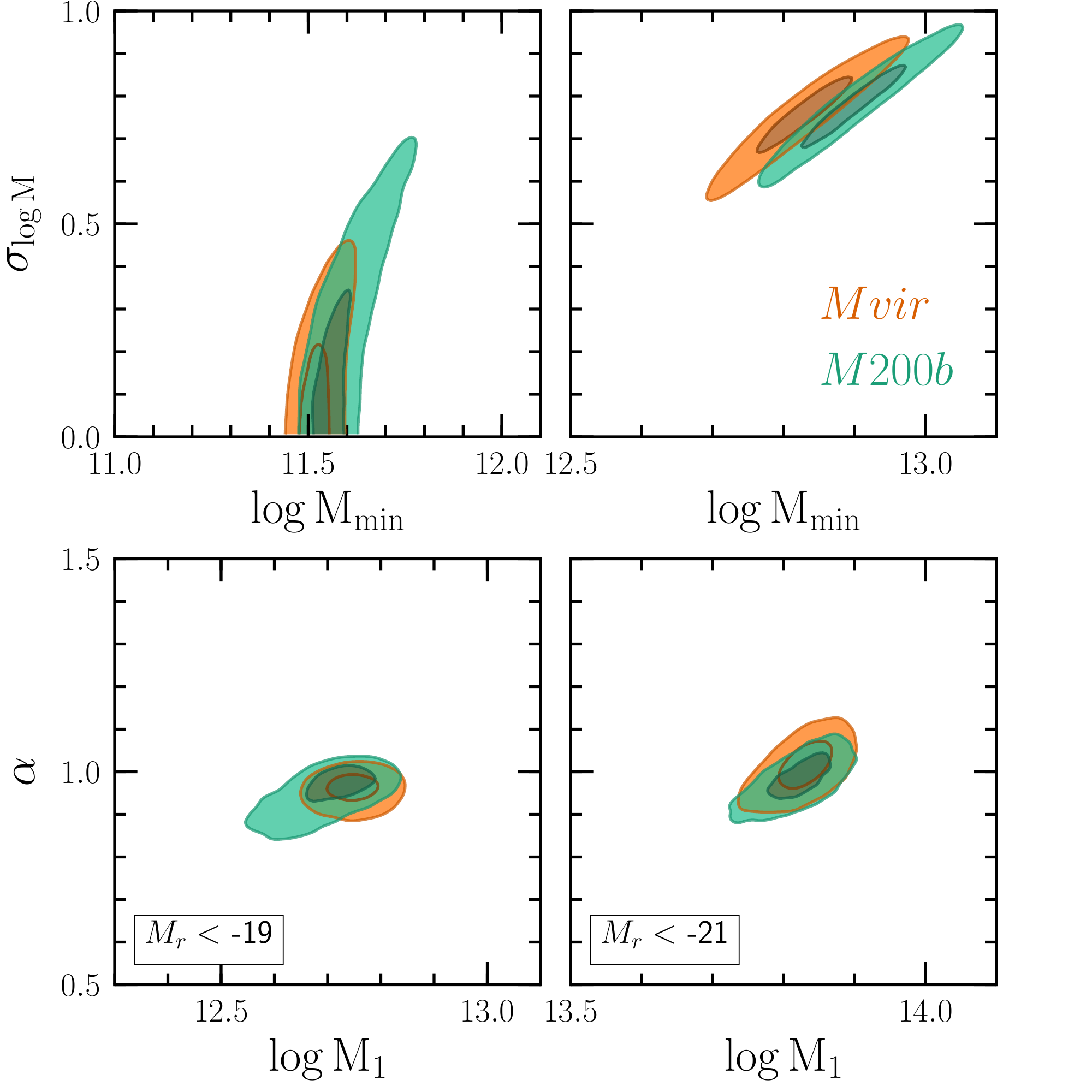}
\caption{\small Posterior probability distributions for \hod parameters 
comparing the cases of two different halo definitions. Panel layout
is identical to that in Fig.~\ref{fig:joint_parameter_conf}. In each panel, 
orange and green contours show results for the Mvir and M200b halo 
definitions, respectively. In both cases, the model shown assumes 
the LasDamas cosmology and does not include PCA reduction.}
\label{fig:different_delta}
\end{figure}  

In the halo model framework, the clustering of galaxies is determined by two main ingredients: the halo distribution and 
the way in which galaxies populate halos, i.e., the \hod. The halo distribution depends on the assumed cosmological 
model and halo definition. The \hod is comprised of various assumptions, such as the functional form of the mean galaxy
occupation, the scatter about the mean, and the spatial and velocity distribution of galaxies within halos. In this section,
we probe the sensitivity of our results to the assumed halo definition and cosmological model. Specifically, we introduce 
a different halo definition (M200b) and different cosmological model (Planck), and we run MCMC chains for all the 
combinations of these choices. We can then study how sensitive the results are to these assumptions.

First, we consider a change in the halo definition. In principle, any reasonable halo definition should be able to 
successfully model the galaxy distribution. However, some halo definitions might require more complicated \hod 
parameterisations than others. For a given parameterisation (like the 5-parameter model we adopt in this work), some
halo definitions likely work better than others. This dependence of \hod modelling on halo definition is a research area 
that has not been previously explored so the test we perform here represents the first step in that direction.  

Our fiducial halo definition, Mvir, corresponds to values of the mean halo over-density (with respect to the mean density) 
$\Delta=351$ and 321 for the LasDamas cosmological model at the two median redshifts of the $M_r<-19$ and -21 
samples \citep{bryan_norman_98}. For the Planck cosmological model that we consider below, the corresponding values 
are $\Delta=315$ and 292. Changing the halo definition to M200b, which corresponds to $\Delta=200$, makes all halos 
approximately 20\% larger in radius.\footnote{This calculation assumes the \citet[][NFW]{navarro_etal_97} density 
profile.} As a result, satellite galaxies placed inside halos will be more spatially extended. Moreover, some halos that 
were previously classified as host halos will now be classified as subhalos, thus changing the probability that they 
receive a galaxy. These changes will certainly alter \wprp and perhaps \gmf for a fixed \hod model, but the question is 
whether a different set of \hod parameters can compensate for this.
 
Fig.~\ref{fig:different_delta} shows posterior probability distributions for \hod parameters for the two halo definitions,
when the model is fit jointly to measurements of \wprp, \gmf, and \ngal. The main effect of changing the halo definition 
is to shift \logMmin to larger values. This is required to preserve the number density since all halos grow in mass under
the M200b definition. The satellite occupation does not change, however. The best-fit results in Table~\ref{table:bestfit}
show that adopting this lower $\Delta$ definition leads to similar quality fits for the $M_r<-19$ sample (\pvalue decreases
from 0.0751 to 0.0653), but substantially worse fits for the $M_r<-21$ sample (\pvalue decreases from 0.0056 to 
0.0017). Therefore, lowering $\Delta$ from the Mvir definition does not alleviate the tension between model and data that
we find here.

\begin{figure}
\centering
\includegraphics[clip=true,width=\linewidth]{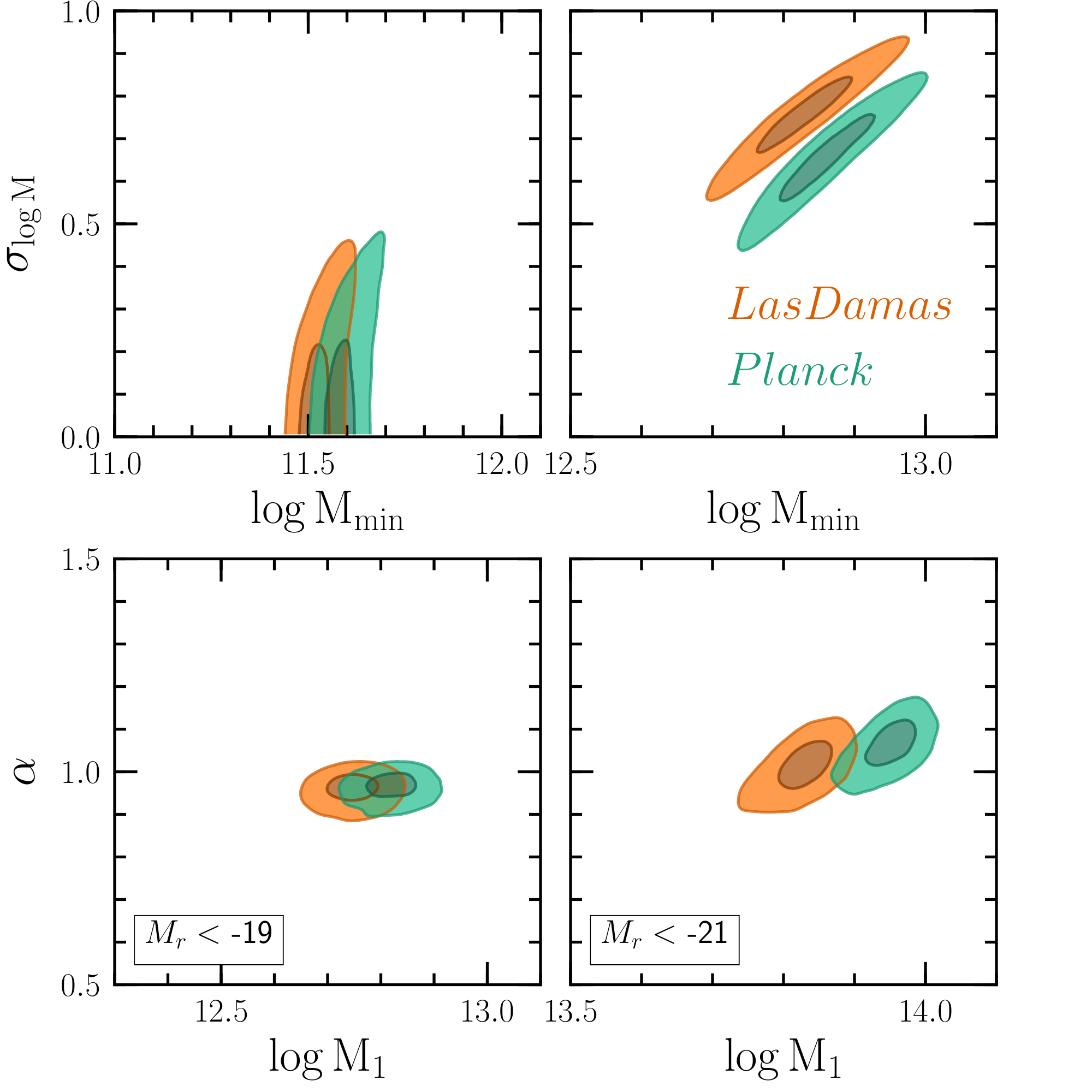}%
\caption{\small Posterior probability distributions for \hod parameters 
comparing the cases of two different cosmological models. Panel layout
is identical to that in Fig.~\ref{fig:joint_parameter_conf}. In each panel, 
orange and green contours show results for the LasDamas and Planck
cosmological models, respectively. In both cases, the model shown assumes 
the Mvir halo definition and does not include PCA reduction.}
\label{fig:different_cosmology}
\end{figure}  

Next, we consider a change in the cosmological model. Since cosmology and the \hod are not degenerate 
\citep{zheng_weinberg_07}, it is possible that some models work better than others. We compare our LasDamas model
to the Planck model. The primary difference between the two is that Planck has a higher value of $\Om$ (0.302 vs. 0.25), 
but there are also differences in $\Ob$ and $n_s$. Fig.~\ref{fig:different_cosmology} shows posterior probability 
distributions of \hod parameters for the two cosmologies, when the model is fit jointly to measurements of \wprp, \gmf, 
and \ngal (assuming the Mvir halo definition). The main effect of adopting the Planck cosmology is to shift the whole 
mean galaxy occupation to higher mass (i.e., increase both \logMmin and \logMone). This shift compensates for the 
higher halo masses due to the larger value of $\Om$. Table~\ref{table:bestfit} reveals that all the fits improve 
considerably with the Planck cosmology. In the case of the $M_r<-19$ sample, the \pvalue shows a modest increase 
from 0.0751 to 0.1087. However, in the case of the $M_r<-21$ sample, the \pvalue grows from 0.0056 to 0.0229.

While changing the halo definition did not relieve the tension between model and data, changing the cosmological 
model does just that. The tension in the $M_r<-21$ sample when adopting the Planck model and the Mvir definition
reduces to the $2.3\sigma$ level, which is not high enough to warrant ruling the model out. This sensitivity of $\chi^2$
to the assumed cosmology illustrates the power of small-scale clustering to constrain cosmological models. 
Unfortunately, to properly probe the cosmological parameter space would require a very large number of simulations and 
is not a trivial exercise.

\begin{table*}
\renewcommand{\arraystretch}{1.20}
\caption{\small Best-fit \hod Model Results. The table lists the best-fit values of the five \hod parameters 
(columns 5-9) and the corresponding values of $\chi^2$, number of degrees of freedom and \pvalues (columns 10-12)
for all the combinations of galaxy sample, assumed cosmology, assumed halo definition, and choice of reducing 
noise in the correlation matrix via PCA (columns 1-4).}
\begin{center}
\begin{tabular}{ccccccrccccc}
\toprule
\multicolumn{1}{c}{\textbf{Sample}}                   &
\multicolumn{1}{c}{\textbf{Cosmology}}                & 
\multicolumn{1}{c}{\textbf{Halo def}}                 & 
\multicolumn{1}{c}{\textbf{PCA}}                      & 
\multicolumn{1}{c}{$\boldsymbol{\rm \log M_{min}}$}    & 
\multicolumn{1}{c}{$\boldsymbol{\rm \sigma_{\log M}}$} & 
\multicolumn{1}{c}{$\boldsymbol{\rm \log M_0}$}       & 
\multicolumn{1}{c}{$\boldsymbol{\rm \log M_1}$}       & 
\multicolumn{1}{c}{$\boldsymbol{\rm \alpha}$}         & 
\multicolumn{1}{c}{$\boldsymbol{\rm \chi^2}$}         & 
\multicolumn{1}{c}{\textbf{d.o.f.}}                   & 
\multicolumn{1}{c}{\textbf{p-value}}                  \\
\midrule
\multirow{8}{*}{$\boldsymbol{M_r<-19}$} &  LasDamas &  Mvir  & No  & 11.53  &  0.13  &  9.61  & 12.75  &  0.97  & 39.55 &   18 &       0.0024 \\
                                        &  LasDamas &  Mvir  & Yes & 11.36  &  0.14  & 12.25  & 12.56  &  0.90  & 11.46 &    6 &       0.0751 \\
                                        &  LasDamas &  M200b & No  & 11.53  &  0.20  & 11.88  & 12.61  &  0.92  & 36.63 &   18 &       0.0059 \\
                                        &  LasDamas &  M200b & Yes & 11.42  &  0.12  & 12.25  & 12.58  &  0.92  & 11.85 &    6 &       0.0653 \\
\cmidrule(lr){2-12}
                                        &  Planck   &  Mvir  & No  & 11.58  &  0.05  & 11.14  & 12.80  &  0.95  & 38.72 &   18 &       0.0031 \\
                                        &  Planck   &  Mvir  & Yes & 11.40  &  0.12  & 12.38  & 12.64  &  0.92  & 10.40 &    6 &       0.1087 \\
                                        &  Planck   &  M200b & No  & 11.62  &  0.28  & 11.66  & 12.77  &  0.96  & 35.52 &   18 &       0.0081 \\
                                        &  Planck   &  M200b & Yes & 11.50  &  0.22  & 12.17  & 12.69  &  0.94  & 10.74 &    6 &       0.0968 \\
                                   
\midrule
\multirow{8}{*}{$\boldsymbol{M_r<-21}$} &  LasDamas &  Mvir  & No  & 12.82  &  0.75  &  8.97  & 13.83  &  1.06  & 34.60 &   16 &       0.0045 \\
                                        &  LasDamas &  Mvir  & Yes & 12.81  &  0.72  & 11.52  & 13.83  &  1.03  & 26.41 &   11 &       0.0056 \\
                                        &  LasDamas &  M200b & No  & 12.87  &  0.75  &  7.51  & 13.83  &  1.03  & 36.64 &   16 &       0.0024 \\
                                        &  LasDamas &  M200b & Yes & 12.89  &  0.76  &  8.65  & 13.83  &  1.02  & 29.74 &   11 &       0.0017 \\
\cmidrule(lr){2-12}
                                        &  Planck   &  Mvir  & No  & 12.91  &  0.73  & 10.21  & 13.93  &  1.07  & 28.96 &   16 &       0.0242 \\
                                        &  Planck   &  Mvir  & Yes & 12.83  &  0.60  &  9.00  & 13.97  &  1.08  & 22.20 &   11 &       0.0229 \\
                                        &  Planck   &  M200b & No  & 12.87  &  0.61  & 11.35  & 13.96  &  1.06  & 30.76 &   16 &       0.0144 \\
                                        &  Planck   &  M200b & Yes & 12.89  &  0.64  &  9.19  & 13.95  &  1.07  & 23.95 &   11 &       0.0130 \\
\bottomrule
\end{tabular}
\end{center}
\label{table:bestfit}
\end{table*}

\begin{table*}
\renewcommand{\arraystretch}{1.20}
\caption{\small Marginalised \hod Parameter Constraints. Similar to Table~\ref{table:bestfit}, except that listed \hod
values show the median, and the upper and lower limits corresponding to the 84 and 16 percentiles of the parameter values from the MCMC chain.}
\begin{center}
\begin{tabular}{ccccccrcc}
\toprule
\multicolumn{1}{c}{\textbf{Sample}}                    &
\multicolumn{1}{c}{\textbf{Cosmology}}                 & 
\multicolumn{1}{c}{\textbf{Halo def}}                  & 
\multicolumn{1}{c}{\textbf{PCA}}                       & 
\multicolumn{1}{c}{$\boldsymbol{\rm \log M_{min}}$}    & 
\multicolumn{1}{c}{$\boldsymbol{\rm \sigma_{\log M}}$} & 
\multicolumn{1}{c}{$\boldsymbol{\rm \log M_0}$}        & 
\multicolumn{1}{c}{$\boldsymbol{\rm \log M_1}$}        & 
\multicolumn{1}{c}{$\boldsymbol{\rm \alpha}$}          \\
\midrule
\multirow{8}{*}{$\boldsymbol{M_r<-19}$} & LasDamas &     Mvir  &   No  & 11.53$^{+0.05}_{-0.04}$   &  0.15$^{+0.17}_{-0.11}$   &  8.95$^{+2.05}_{-2.02}$   & 12.75$^{+0.04}_{-0.05}$   &  0.96$^{+0.03}_{-0.04}$  \\
                                        & LasDamas &     Mvir  &  Yes  & 11.59$^{+0.15}_{-0.15}$   &  0.60$^{+0.30}_{-0.41}$   & 10.78$^{+1.42}_{-3.24}$   & 12.63$^{+0.08}_{-0.08}$   &  0.93$^{+0.05}_{-0.06}$  \\
                                        & LasDamas &    M200b  &   No  & 11.59$^{+0.09}_{-0.05}$   &  0.24$^{+0.26}_{-0.17}$   & 10.72$^{+1.22}_{-3.20}$   & 12.71$^{+0.06}_{-0.08}$   &  0.96$^{+0.04}_{-0.06}$  \\
                                        & LasDamas &    M200b  &  Yes  & 11.65$^{+0.16}_{-0.16}$   &  0.59$^{+0.30}_{-0.41}$   & 10.74$^{+1.51}_{-3.22}$   & 12.64$^{+0.07}_{-0.09}$   &  0.94$^{+0.04}_{-0.06}$  \\
\cmidrule(lr){2-9}
                                        & Planck   &     Mvir  &   No  & 11.60$^{+0.05}_{-0.04}$   &  0.15$^{+0.17}_{-0.11}$   &  9.10$^{+2.01}_{-2.10}$   & 12.82$^{+0.04}_{-0.05}$   &  0.96$^{+0.03}_{-0.03}$  \\
                                        & Planck   &     Mvir  &  Yes  & 11.64$^{+0.17}_{-0.18}$   &  0.61$^{+0.29}_{-0.40}$   & 11.80$^{+0.75}_{-3.86}$   & 12.68$^{+0.09}_{-0.12}$   &  0.92$^{+0.05}_{-0.07}$  \\
                                        & Planck   &    M200b  &  No   & 11.63$^{+0.06}_{-0.04}$   &  0.21$^{+0.20}_{-0.14}$   &  9.90$^{+1.86}_{-2.65}$   & 12.79$^{+0.05}_{-0.05}$   &  0.97$^{+0.02}_{-0.04}$  \\
                                        & Planck   &    M200b  &  Yes  & 11.70$^{+0.16}_{-0.17}$   &  0.58$^{+0.32}_{-0.41}$   & 11.09$^{+1.30}_{-3.51}$   & 12.71$^{+0.08}_{-0.09}$   &  0.94$^{+0.04}_{-0.06}$  \\
\midrule
\multirow{8}{*}{$\boldsymbol{M_r<-21}$} & LasDamas &     Mvir  &   No  & 12.83$^{+0.07}_{-0.07}$   &  0.76$^{+0.09}_{-0.09}$   &  9.62$^{+2.10}_{-2.38}$   & 13.83$^{+0.04}_{-0.04}$   &  1.01$^{+0.05}_{-0.06}$  \\
                                        & LasDamas &     Mvir  &  Yes  & 12.80$^{+0.07}_{-0.06}$   &  0.72$^{+0.09}_{-0.09}$   &  9.46$^{+2.19}_{-2.34}$   & 13.84$^{+0.03}_{-0.04}$   &  1.03$^{+0.05}_{-0.06}$  \\
                                        & LasDamas &    M200b  &   No  & 12.90$^{+0.07}_{-0.07}$   &  0.78$^{+0.09}_{-0.10}$   &  9.42$^{+2.09}_{-2.28}$   & 13.82$^{+0.04}_{-0.05}$   &  0.98$^{+0.05}_{-0.05}$  \\
                                        & LasDamas &    M200b  &  Yes  & 12.86$^{+0.07}_{-0.07}$   &  0.72$^{+0.09}_{-0.10}$   &  9.47$^{+2.09}_{-2.30}$   & 13.85$^{+0.04}_{-0.04}$   &  1.01$^{+0.05}_{-0.05}$  \\
\cmidrule(lr){2-9}
                                        & Planck   &     Mvir  &   No  & 12.86$^{+0.07}_{-0.06}$   &  0.66$^{+0.10}_{-0.10}$   &  9.54$^{+2.29}_{-2.39}$   & 13.95$^{+0.03}_{-0.04}$   &  1.06$^{+0.05}_{-0.06}$  \\
                                        & Planck   &     Mvir  &  Yes  & 12.84$^{+0.07}_{-0.07}$   &  0.61$^{+0.11}_{-0.12}$   &  9.73$^{+2.17}_{-2.52}$   & 13.96$^{+0.03}_{-0.04}$   &  1.08$^{+0.05}_{-0.06}$  \\
                                        & Planck   &    M200b  &   No  & 12.90$^{+0.07}_{-0.07}$   &  0.66$^{+0.10}_{-0.11}$   &  9.87$^{+1.99}_{-2.49}$   & 13.95$^{+0.03}_{-0.04}$   &  1.05$^{+0.05}_{-0.05}$  \\
                                        & Planck   &    M200b  &  Yes  & 12.89$^{+0.07}_{-0.07}$   &  0.63$^{+0.11}_{-0.11}$   &  9.63$^{+2.15}_{-2.45}$   & 13.96$^{+0.03}_{-0.04}$   &  1.07$^{+0.05}_{-0.05}$  \\
\bottomrule
\end{tabular}
\end{center}
\label{table:hodvalues}
\end{table*}

Looking at all the (PCA) chain results in Table~\ref{table:bestfit}, we see that the model that works best for both 
luminosity samples is the Planck Mvir model. This model combined with the ``vanilla'' \hod model described in 
\S~\ref{HOD} predicts \wprp, \gmf, and \ngal that are consistent with SDSS measurements for low luminosity galaxies, 
but show $2.3\sigma$ tension for high luminosity galaxies. To further alleviate this tension we would likely have to relax 
some of the assumptions built into the \hod model. However, the most important result of this paper is that we have 
achieved a sufficiently robust modelling methodology that we can now test halo models in a statistical sense.

\section{Future Modelling Improvements} \label{Improvements}

In this section we discuss improvements to our modelling pipeline that we leave for future work. Improvements come in 
two types: changes that reduce systematic errors and thus make the modelling more accurate, and changes that add 
freedom to the model being tested. Regarding improvements to the model accuracy, we believe that our methodology 
represents the most accurate \hod modelling to-date given that we compute statistics from realistic mock catalogues, we 
compute covariance matrices from a large ensemble of mocks, we account for noise in the matrices, etc. Nevertheless, 
there are three potential sources of systematic error that need further testing. 

First, we do not include fibre collision incompleteness in our modelling pipeline. We apply the nearest neighbour correction 
to the SDSS data, but we do not attempt to model how this correction might fail. We have argued that, for the scales in 
\wprp and multiplicities in \gmf that we consider in this paper, this is sufficient. However, a proper modelling of fibre 
collisions may become necessary if we extend the analysis to different scales or statistics. Including fibre collisions
in the mocks is non-trivial because doing it correctly requires mocks of the full flux-limited SDSS sample. However, it may
be possible to approximate the effect with our existing simulations via semi-analytic methods.

Second, the resolution of the simulations that we use to construct the covariance matrix of the $M_r<-21$ sample is fairly
low. We expect that this will not have a large impact on the scatter among many simulation realisations, but this needs to
be tested more thoroughly. Ideally, we need a proper convergence test, but that would require a very large number of 
additional simulations (50 boxes per resolution tested). More realistically, we could use fewer mock catalogues of smaller 
volume each to get a feel for how the covariance matrix depends on resolution.

Third, we construct the correlation matrix from a fiducial \hod model and assume that it is fixed as we explore the \hod 
parameter space. Ideally, to estimate $P(\mathrm{data} | \mathrm{model})$, we should be recomputing the correlation 
matrix at each new point in parameter space. This is prohibitively expensive because it would require construction 
and analysis of 200 mocks at each unique link of the MCMC chain, instead of the $\sim 10$ that we use now. If testing 
shows that varying the correlation matrix is necessary, we can explore ways to interpolate between a sparse set of 
matrices within the parameter space. 

There are a number of extensions to the ``vanilla''  \hod model that we plan to implement for the purpose of giving the 
model more freedom and to probe interesting aspects of the galaxy-halo connection. For example, we can drop the 
assumption that satellite occupation follows Poisson statistics, which is not necessarily the case 
\citep{mbk_etal_10,mao_etal_2015}. Another extension is to drop the assumption that the central galaxy is at rest at
the halo centre and that satellite galaxies trace the spatial and velocity distributions of dark matter within halos. Several
studies, both theoretical \citep[e.g.,][]{berlind_etal_03} and observational \citep[e.g.,][]{vandenbosch_etal_05}, have 
shown that these are not good assumptions. An \hod model with built-in velocity bias for centrals and satellites was 
recently used by \citet{guo_etal_15a}. Finally, we can drop the assumption that galaxy occupation statistics only depend 
on halo mass and account for galaxy assembly bias~\citep{croton_etal_06}. Though there are theoretical reasons to 
expect some level of assembly bias for luminosity threshold samples \citep[e.g.,][]{zentner_etal_05,zehavi_assembly_bias_17}, 
there is not yet strong observational evidence for this. To incorporate assembly bias, we could adopt the ``decorated \hod'' model of 
\citet{hearin_etal_16}, which was recently used to model SDSS data by \citet{zentner_etal_16}. Naturally, as we add 
freedom to the \hod model, we can also add additional clustering statistics. A great advantage of our mock-based 
modelling methodology is that it can easily incorporate new statistics.

\section{Summary and Discussion} \label{Summary}

In this paper, we have developed an accurate mock-based \hod modelling framework and have applied it to 
measurements of the projected correlation function \wprp, the group multiplicity function \gmf, and the galaxy number 
density \ngal of two luminosity threshold samples in the SDSS DR7. Features of the modelling framework include (1) 
construction of realistic mock galaxy catalogues by populating dark matter halos in cosmological N-body simulations with 
galaxies, and applying redshift distortions and survey selection functions; (2) calculation of model predictions by running 
the same analysis codes on mock catalogues as on the SDSS data, and averaging over enough mocks so that statistical 
errors in the model are negligible; (3) estimation of errors and covariances via 200 independent mock catalogues; (4) 
reduction of noise in the covariance matrices via an eigenmode analysis; (5) parameter search using the \texttt{emcee} 
MCMC code that includes $\sim5\times10^5$ evaluations of likelihood; (6) new blazing fast analysis codes that make this 
data-intensive approach feasible.

The specific model we have tested in this paper is the $\Lambda$CDM + ``vanilla'' \hod model, whereby the dark matter 
halo population is given by cosmological N-body simulations that contain only dark matter, and galaxies are placed within 
halos according to a simple 5-parameter \hod model \citep{zheng_etal_07} that contains no spatial or velocity bias of 
centrals or satellites and no galaxy assembly bias. We have tested two cosmological models (LasDamas and Planck) 
and two halo definitions (Mvir and M200b). Our main results are the following.
\begin{itemize}
\item
The model is successful at fitting either \wprp or \gmf (plus \ngal in both cases) for both luminosity samples. However, 
the regions of \hod parameter space selected are different depending on which clustering statistic is used. In terms of 
constraining power, \gmf yields tighter \hod constraints for low luminosity galaxies, while \wprp is better at constraining 
parameters for high luminosity galaxies. When all three statistics are used jointly, \hod constraints tighten significantly
compared to the common case of only using \wprp and \ngal.
\item
The model struggles to jointly fit \wprp, \gmf, and \ngal, demonstrating the power of combining multiple clustering statistics 
for ruling out models. Adopting a different halo definition does not make a big difference (though Mvir is slightly preferred 
over M200b), but changing the cosmological model does. When adopting the LasDamas cosmology, the model is ruled 
out at the $3\sigma$ level when tested against either luminosity sample. However, when adopting the Planck cosmology,
the model is consistent with the clustering of low luminosity galaxies and exhibits $2.3\sigma$ tension with the clustering 
of high luminosity galaxies.
\end{itemize}

Most importantly, we have demonstrated that it is possible to use galaxy clustering on small scales to perform sensitive
statistical tests of cosmology + halo models. This is made possible by our fully numerical mock-based methodology 
combined with fast analysis codes and a careful treatment of systematic and statistical errors. Though most halo model 
analyses in the literature use analytic models, there are a few studies that have adopted mock-based methods like ours.
The first such studies modelled the projected correlation function of red galaxies in the the high redshift 
\citep{white_etal_11} and low redshift \citep{parejko_etal_13} samples of the BOSS survey. \citet{zheng_guo_16} 
developed a numerical method in which halo pairs and particle pairs within halos are measured in a N-body simulation 
and tabulated as a function of halo mass and separation. The correlation function of galaxies can then be calculated 
accurately by appropriately weighting these functions by the \hod. This method was applied to the projected and redshift
space correlation functions of BOSS and SDSS galaxies in order to constrain the velocity bias of central and satellite 
galaxies~\citep{guo_etal_15a,guo_etal_15b,guo_etal_15c}. The downside of this methodology is that it cannot be 
extended to other statistics beyond pair (or triple) counts, like the group multiplicity function. The most similar modelling 
methodology to the one we present in this paper was developed as part of the powerful \texttt{halotools} software 
package \citep{hearin_etal_16b}. This was recently used by \citet{zentner_etal_16} to model the clustering of SDSS 
galaxies with the goal of constraining galaxy assembly bias. Our analysis improves on \citet{zentner_etal_16} in the 
following ways. (1) we include the group multiplicity function as a constraint; (2) we directly adopt the spatial and velocity 
distribution of particles within halos to place satellite galaxies, while they assume a NFW profile and a Gaussian velocity distribution; 
(3) we use independent mock catalogues to estimate the covariance matrix, while they use Jackknife resampling; (4) we 
eliminate noise in the covariance matrix via PCA; (5) they adopt a Poisson error on the galaxy number density, which 
ignores cosmic variance and is only $\sim0.33\%$, while the correct errors that we obtain from our 200 mock catalogues 
are 8\% and 2.5\% for the $M_r<-19$ and $-18$ samples, respectively. This work therefore represents the most accurate 
modelling of SDSS galaxies to-date.

Our hope is that with these and future improvements to the accuracy of modelling together with an optimal set of 
statistics, galaxy clustering on small scales will definitively measure the galaxy-halo connection, including second-order 
features like assembly bias. Moreover, small scale clustering has the potential to become a standard test of cosmological 
models. Whether constraints on cosmology can compete with other probes remains to be seen.

\section{Acknowledgements}
We would like to thank the anonymous referee for constructive comments that
helped improve the paper. We would also like to thank Adam Szewciw for running
some importance sampling tests to make sure our results are not qualitatively
affected by the choice of the correlation matrix. We sincerely thank Jeremy Tinker, Frank van den Bosch, Risa Wechsler, David
Weinberg, Andrew Zentner, Andrew Hearin, Doug Watson and Zheng
Zheng for valuable discussions over the course of this project. We thank Daniel Foreman-Mackey for the \texttt{emcee} 
MCMC software. We thank Qingqing Mao for software to compute FoF halo centres
and Matt Becker for sharing \texttt{uber-LGadget2} that was used to run the
ConsueloHD and CarmenHD simulations. The mock catalogues used in this 
paper were produced by the LasDamas project (http://lss.phy.vanderbilt.edu/lasdamas/); we thank NSF XSEDE for 
providing the computational resources for LasDamas. MS especially acknowledges the fantastic \texttt{largemem} nodes 
with 1 TB of RAM at Stampede (TACC) without which the MCMC chains could not have been done. Some of the computational 
facilities used in this project were provided by the Vanderbilt Advanced Computing Center for Research and Education 
(ACCRE). This project has been supported by the National Science Foundation (NSF) through a Career Award 
(AST-1151650). Parts of this research were conducted by the Australian Research Council Centre of Excellence for All
Sky Astrophysics in 3 Dimensions (ASTRO 3D), through project number
CE170100013. This research has made use of NASA's Astrophysics Data
System and has used \texttt{python}~(\url{https://www.python.org/}),
\texttt{numpy}~\citep{van2011numpy}, \texttt{matplotlib}~\citep{Hunter:2007}, \texttt{GSL}~(\url{http://www.gnu.org/software/gsl/}),
\texttt{ChainConsumer}~\citep{Hinton2016}, and \texttt{The (Astronomy) Acknowledgment Generator}~(\url{http://astrofrog.github.io/acknowledgment-generator/}).

\bibliographystyle{mnras}
\bibliography{master}

\begin{thebibliography}{}
\makeatletter
\relax
\def\mn@urlcharsother{\let\do\@makeother \do\$\do\&\do\#\do\^\do\_\do\%\do\~}
\def\mn@doi{\begingroup\mn@urlcharsother \@ifnextchar [ {\mn@doi@}
  {\mn@doi@[]}}
\def\mn@doi@[#1]#2{\def\@tempa{#1}\ifx\@tempa\@empty \href
  {http://dx.doi.org/#2} {doi:#2}\else \href {http://dx.doi.org/#2} {#1}\fi
  \endgroup}
\def\mn@eprint#1#2{\mn@eprint@#1:#2::\@nil}
\def\mn@eprint@arXiv#1{\href {http://arxiv.org/abs/#1} {{\tt arXiv:#1}}}
\def\mn@eprint@dblp#1{\href {http://dblp.uni-trier.de/rec/bibtex/#1.xml}
  {dblp:#1}}
\def\mn@eprint@#1:#2:#3:#4\@nil{\def\@tempa {#1}\def\@tempb {#2}\def\@tempc
  {#3}\ifx \@tempc \@empty \let \@tempc \@tempb \let \@tempb \@tempa \fi \ifx
  \@tempb \@empty \def\@tempb {arXiv}\fi \@ifundefined
  {mn@eprint@\@tempb}{\@tempb:\@tempc}{\expandafter \expandafter \csname
  mn@eprint@\@tempb\endcsname \expandafter{\@tempc}}}

\bibitem[\protect\citeauthoryear{{Abazajian} et~al.,}{{Abazajian}
  et~al.}{2005}]{abazajian_etal_05}
{Abazajian} K.,  et~al., 2005, \aj, \href
  {http://adsabs.harvard.edu/cgi-bin/nph-bib_query?bibcode=2005AJ....129.1755A&db_key=AST}
  {129, 1755}

\bibitem[\protect\citeauthoryear{{Abazajian} et~al.,}{{Abazajian}
  et~al.}{2009}]{abazajian_etal_09}
{Abazajian} K.~N.,  et~al., 2009, \mn@doi [\apjs]
  {10.1088/0067-0049/182/2/543}, \href
  {http://adsabs.harvard.edu/abs/2009ApJS..182..543A} {182, 543}

\bibitem[\protect\citeauthoryear{{Abbas} et~al.,}{{Abbas}
  et~al.}{2010}]{abbas_etal_10}
{Abbas} U.,  et~al., 2010, \mn@doi [\mnras] {10.1111/j.1365-2966.2010.16764.x},
  \href {http://adsabs.harvard.edu/abs/2010MNRAS.406.1306A} {406, 1306}

\bibitem[\protect\citeauthoryear{{Baugh}, {Benson}, {Cole}, {Frenk}  \&
  {Lacey}}{{Baugh} et~al.}{1999}]{baugh_etal_99}
{Baugh} C.~M.,  {Benson} A.~J.,  {Cole} S.,  {Frenk} C.~S.,   {Lacey} C.~G.,
  1999, \mn@doi [\mnras] {10.1046/j.1365-8711.1999.02590.x}, \href
  {http://adsabs.harvard.edu/abs/1999MNRAS.305L..21B} {305, L21}

\bibitem[\protect\citeauthoryear{{Behroozi}, {Wechsler}  \& {Wu}}{{Behroozi}
  et~al.}{2013}]{behroozi_etal_13}
{Behroozi} P.~S.,  {Wechsler} R.~H.,   {Wu} H.-Y.,  2013, \mn@doi [\apj]
  {10.1088/0004-637X/762/2/109}, \href
  {http://adsabs.harvard.edu/abs/2013ApJ...762..109B} {762, 109}

\bibitem[\protect\citeauthoryear{{Benson}, {Cole}, {Frenk}, {Baugh}  \&
  {Lacey}}{{Benson} et~al.}{2000}]{benson_etal_00}
{Benson} A.~J.,  {Cole} S.,  {Frenk} C.~S.,  {Baugh} C.~M.,   {Lacey} C.~G.,
  2000, \mn@doi [\mnras] {10.1046/j.1365-8711.2000.03101.x}, \href
  {http://adsabs.harvard.edu/abs/2000MNRAS.311..793B} {311, 793}

\bibitem[\protect\citeauthoryear{{Berlind} \& {Weinberg}}{{Berlind} \&
  {Weinberg}}{2002}]{berlind_weinberg_02}
{Berlind} A.~A.,  {Weinberg} D.~H.,  2002, \apj, \href
  {http://adsabs.harvard.edu/cgi-bin/nph-bib_query?bibcode=2002ApJ...575..587B&db_key=AST}
  {575, 587}

\bibitem[\protect\citeauthoryear{{Berlind} et~al.,}{{Berlind}
  et~al.}{2003}]{berlind_etal_03}
{Berlind} A.~A.,  et~al., 2003, \apj, \href
  {http://adsabs.harvard.edu/cgi-bin/nph-bib_query?bibcode=2003ApJ...593....1B&db_key=AST}
  {593, 1}

\bibitem[\protect\citeauthoryear{{Berlind} et~al.,}{{Berlind}
  et~al.}{2006}]{berlind_etal_06}
{Berlind} A.~A.,  et~al., 2006, \mn@doi [\apjs] {10.1086/508170}, \href
  {http://adsabs.harvard.edu/abs/2006ApJS..167....1B} {167, 1}

\bibitem[\protect\citeauthoryear{{Beutler} et~al.,}{{Beutler}
  et~al.}{2013}]{beutler_etal_13}
{Beutler} F.,  et~al., 2013, \mn@doi [\mnras] {10.1093/mnras/sts637}, \href
  {http://adsabs.harvard.edu/abs/2013MNRAS.429.3604B} {429, 3604}

\bibitem[\protect\citeauthoryear{{Blake}, {Collister}  \& {Lahav}}{{Blake}
  et~al.}{2008}]{blake_etal_08}
{Blake} C.,  {Collister} A.,   {Lahav} O.,  2008, \mn@doi [\mnras]
  {10.1111/j.1365-2966.2007.11925.x}, \href
  {http://adsabs.harvard.edu/abs/2008MNRAS.385.1257B} {385, 1257}

\bibitem[\protect\citeauthoryear{{Blanton}}{{Blanton}}{2006}]{blanton_etal_06}
{Blanton} M.~R.,  2006, \mn@doi [\apj] {10.1086/505628}, \href
  {http://adsabs.harvard.edu/abs/2006ApJ...648..268B} {648, 268}

\bibitem[\protect\citeauthoryear{{Blanton}, {Lin}, {Lupton}, {Maley}, {Young},
  {Zehavi}  \& {Loveday}}{{Blanton} et~al.}{2003a}]{blanton_etal_03a}
{Blanton} M.~R.,  {Lin} H.,  {Lupton} R.~H.,  {Maley} F.~M.,  {Young} N.,
  {Zehavi} I.,   {Loveday} J.,  2003a, \aj, \href
  {http://adsabs.harvard.edu/cgi-bin/nph-bib_query?bibcode=2003AJ....125.2276B&db_key=AST}
  {125, 2276}

\bibitem[\protect\citeauthoryear{{Blanton} et~al.,}{{Blanton}
  et~al.}{2003b}]{blanton_etal_03b}
{Blanton} M.~R.,  et~al., 2003b, \aj, \href
  {http://adsabs.harvard.edu/cgi-bin/nph-bib_query?bibcode=2003AJ....125.2348B&db_key=AST}
  {125, 2348}

\bibitem[\protect\citeauthoryear{{Blanton} et~al.,}{{Blanton}
  et~al.}{2005}]{blanton_etal_05}
{Blanton} M.~R.,  et~al., 2005, \mn@doi [\aj] {10.1086/429803}, \href
  {http://adsabs.harvard.edu/abs/2005AJ....129.2562B} {129, 2562}

\bibitem[\protect\citeauthoryear{{Boylan-Kolchin}, {Springel}, {White}  \&
  {Jenkins}}{{Boylan-Kolchin} et~al.}{2010}]{mbk_etal_10}
{Boylan-Kolchin} M.,  {Springel} V.,  {White} S.~D.~M.,   {Jenkins} A.,  2010,
  \mn@doi [\mnras] {10.1111/j.1365-2966.2010.16774.x}, \href
  {http://adsabs.harvard.edu/abs/2010MNRAS.406..896B} {406, 896}

\bibitem[\protect\citeauthoryear{{Brown} et~al.,}{{Brown}
  et~al.}{2008}]{brown_etal_08}
{Brown} M.~J.~I.,  et~al., 2008, \mn@doi [\apj] {10.1086/589538}, \href
  {http://adsabs.harvard.edu/abs/2008ApJ...682..937B} {682, 937}

\bibitem[\protect\citeauthoryear{{Bryan} \& {Norman}}{{Bryan} \&
  {Norman}}{1998}]{bryan_norman_98}
{Bryan} G.~L.,  {Norman} M.~L.,  1998, \mn@doi [\apj] {10.1086/305262}, \href
  {http://adsabs.harvard.edu/abs/1998ApJ...495...80B} {495, 80}

\bibitem[\protect\citeauthoryear{{Bullock}, {Wechsler}  \&
  {Somerville}}{{Bullock} et~al.}{2002}]{bullock_etal_02}
{Bullock} J.~S.,  {Wechsler} R.~H.,   {Somerville} R.~S.,  2002, \mn@doi
  [\mnras] {10.1046/j.1365-8711.2002.04959.x}, \href
  {http://adsabs.harvard.edu/abs/2002MNRAS.329..246B} {329, 246}

\bibitem[\protect\citeauthoryear{{Cacciato}, {van den Bosch}, {More}, {Mo}  \&
  {Yang}}{{Cacciato} et~al.}{2013}]{cacciato_etal_13}
{Cacciato} M.,  {van den Bosch} F.~C.,  {More} S.,  {Mo} H.,   {Yang} X.,
  2013, \mn@doi [\mnras] {10.1093/mnras/sts525}, \href
  {http://adsabs.harvard.edu/abs/2013MNRAS.430..767C} {430, 767}

\bibitem[\protect\citeauthoryear{{Campbell}, {van den Bosch}, {Hearin},
  {Padmanabhan}, {Berlind}, {Mo}, {Tinker}  \& {Yang}}{{Campbell}
  et~al.}{2015}]{campbell_etal_15}
{Campbell} D.,  {van den Bosch} F.~C.,  {Hearin} A.,  {Padmanabhan} N.,
  {Berlind} A.,  {Mo} H.~J.,  {Tinker} J.,   {Yang} X.,  2015, \mn@doi [\mnras]
  {10.1093/mnras/stv1091}, \href
  {http://adsabs.harvard.edu/abs/2015MNRAS.452..444C} {452, 444}

\bibitem[\protect\citeauthoryear{{Colless} et~al.,}{{Colless}
  et~al.}{2001}]{colless_etal_01}
{Colless} M.,  et~al., 2001, \mn@doi [\mnras]
  {10.1046/j.1365-8711.2001.04902.x}, \href
  {http://adsabs.harvard.edu/cgi-bin/nph-bib_query?bibcode=2001MNRAS.328.1039C&db_key=AST}
  {328, 1039}

\bibitem[\protect\citeauthoryear{{Collister} \& {Lahav}}{{Collister} \&
  {Lahav}}{2005}]{collister_lahav_05}
{Collister} A.~A.,  {Lahav} O.,  2005, \mn@doi [\mnras]
  {10.1111/j.1365-2966.2005.09172.x}, \href
  {http://adsabs.harvard.edu/abs/2005MNRAS.361..415C} {361, 415}

\bibitem[\protect\citeauthoryear{{Cooray}}{{Cooray}}{2006}]{cooray_06}
{Cooray} A.,  2006, \mn@doi [\mnras] {10.1111/j.1365-2966.2005.09747.x}, \href
  {http://adsabs.harvard.edu/abs/2006MNRAS.365..842C} {365, 842}

\bibitem[\protect\citeauthoryear{{Cooray} \& {Sheth}}{{Cooray} \&
  {Sheth}}{2002}]{cooray_sheth_02}
{Cooray} A.,  {Sheth} R.,  2002, \physrep, \href
  {http://adsabs.harvard.edu/cgi-bin/nph-bib_query?bibcode=2002PhR...372....1C&db_key=AST}
  {372, 1}

\bibitem[\protect\citeauthoryear{{Crocce}, {Pueblas}  \&
  {Scoccimarro}}{{Crocce} et~al.}{2006}]{crocce_etal_06}
{Crocce} M.,  {Pueblas} S.,   {Scoccimarro} R.,  2006, \mn@doi [\mnras]
  {10.1111/j.1365-2966.2006.11040.x}, \href
  {http://adsabs.harvard.edu/abs/2006MNRAS.373..369C} {373, 369}

\bibitem[\protect\citeauthoryear{{Croton}, {Gao}  \& {White}}{{Croton}
  et~al.}{2007}]{croton_etal_06}
{Croton} D.~J.,  {Gao} L.,   {White} S.~D.~M.,  2007, \mn@doi [\mnras]
  {10.1111/j.1365-2966.2006.11230.x}, \href
  {http://adsabs.harvard.edu/abs/2007MNRAS.374.1303C} {374, 1303}

\bibitem[\protect\citeauthoryear{{Cui}, {Borgani}, {Dolag}, {Murante}  \&
  {Tornatore}}{{Cui} et~al.}{2012}]{cui_etal_12}
{Cui} W.,  {Borgani} S.,  {Dolag} K.,  {Murante} G.,   {Tornatore} L.,  2012,
  \mn@doi [\mnras] {10.1111/j.1365-2966.2012.21037.x}, \href
  {http://adsabs.harvard.edu/abs/2012MNRAS.423.2279C} {423, 2279}

\bibitem[\protect\citeauthoryear{{Davis}, {Efstathiou}, {Frenk}  \&
  {White}}{{Davis} et~al.}{1985}]{davis_etal_85}
{Davis} M.,  {Efstathiou} G.,  {Frenk} C.~S.,   {White} S.~D.~M.,  1985, \apj,
  \href
  {http://adsabs.harvard.edu/cgi-bin/nph-bib_query?bibcode=1985ApJ...292..371D&db_key=AST}
  {292, 371}

\bibitem[\protect\citeauthoryear{{Dawson} et~al.,}{{Dawson}
  et~al.}{2013}]{dawson_etal_13}
{Dawson} K.~S.,  et~al., 2013, \mn@doi [\aj] {10.1088/0004-6256/145/1/10},
  \href {http://adsabs.harvard.edu/abs/2013AJ....145...10D} {145, 10}

\bibitem[\protect\citeauthoryear{{Eisenstein} \& {Zaldarriaga}}{{Eisenstein} \&
  {Zaldarriaga}}{2001}]{eisenstein_zaldarriaga_01}
{Eisenstein} D.~J.,  {Zaldarriaga} M.,  2001, \mn@doi [\apj] {10.1086/318226},
  \href {http://adsabs.harvard.edu/abs/2001ApJ...546....2E} {546, 2}

\bibitem[\protect\citeauthoryear{{Eisenstein} et~al.,}{{Eisenstein}
  et~al.}{2005}]{eisenstein_etal_05}
{Eisenstein} D.~J.,  et~al., 2005, \mn@doi [\apj] {10.1086/466512}, \href
  {http://adsabs.harvard.edu/abs/2005ApJ...633..560E} {633, 560}

\bibitem[\protect\citeauthoryear{{Foreman-Mackey}, {Hogg}, {Lang}  \&
  {Goodman}}{{Foreman-Mackey} et~al.}{2013}]{foreman-mackey_etal_13}
{Foreman-Mackey} D.,  {Hogg} D.~W.,  {Lang} D.,   {Goodman} J.,  2013, \mn@doi
  [\pasp] {10.1086/670067}, \href
  {http://adsabs.harvard.edu/abs/2013PASP..125..306F} {125, 306}

\bibitem[\protect\citeauthoryear{{Gardner}, {Connolly}  \& {McBride}}{{Gardner}
  et~al.}{2007}]{gardner_etal_07}
{Gardner} J.~P.,  {Connolly} A.,   {McBride} C.,  2007, in {Shaw} R.~A.,
  {Hill} F.,   {Bell} D.~J.,  eds,  Astronomical Society of the Pacific
  Conference Series Vol. 376, Astronomical Data Analysis Software and Systems
  XVI. p.~69

\bibitem[\protect\citeauthoryear{{Gazta{\~n}aga} \&
  {Scoccimarro}}{{Gazta{\~n}aga} \&
  {Scoccimarro}}{2005}]{gaztanaga_and_scoccimarro_05}
{Gazta{\~n}aga} E.,  {Scoccimarro} R.,  2005, \mn@doi [\mnras]
  {10.1111/j.1365-2966.2005.09234.x}, \href
  {http://adsabs.harvard.edu/abs/2005MNRAS.361..824G} {361, 824}

\bibitem[\protect\citeauthoryear{{Guo} et~al.,}{{Guo}
  et~al.}{2014}]{guo_etal_14}
{Guo} H.,  et~al., 2014, \mn@doi [\mnras] {10.1093/mnras/stu763}, \href
  {http://adsabs.harvard.edu/abs/2014MNRAS.441.2398G} {441, 2398}

\bibitem[\protect\citeauthoryear{{Guo} et~al.,}{{Guo}
  et~al.}{2015a}]{guo_etal_15a}
{Guo} H.,  et~al., 2015a, \mn@doi [\mnras] {10.1093/mnras/stu2120}, \href
  {http://adsabs.harvard.edu/abs/2015MNRAS.446..578G} {446, 578}

\bibitem[\protect\citeauthoryear{{Guo} et~al.,}{{Guo}
  et~al.}{2015b}]{guo_etal_15b}
{Guo} H.,  et~al., 2015b, \mn@doi [\mnras] {10.1093/mnrasl/slv020}, \href
  {http://adsabs.harvard.edu/abs/2015MNRAS.449L..95G} {449, L95}

\bibitem[\protect\citeauthoryear{{Guo} et~al.,}{{Guo}
  et~al.}{2015c}]{guo_etal_15c}
{Guo} H.,  et~al., 2015c, \mn@doi [\mnras] {10.1093/mnras/stv1966}, \href
  {http://adsabs.harvard.edu/abs/2015MNRAS.453.4368G} {453, 4368}

\bibitem[\protect\citeauthoryear{{Guzik} \& {Seljak}}{{Guzik} \&
  {Seljak}}{2002}]{guzik_seljak_02}
{Guzik} J.,  {Seljak} U.,  2002, \mn@doi [\mnras]
  {10.1046/j.1365-8711.2002.05591.x}, \href
  {http://adsabs.harvard.edu/abs/2002MNRAS.335..311G} {335, 311}

\bibitem[\protect\citeauthoryear{{Hamana}, {Ouchi}, {Shimasaku}, {Kayo}  \&
  {Suto}}{{Hamana} et~al.}{2004}]{hamana_etal_04}
{Hamana} T.,  {Ouchi} M.,  {Shimasaku} K.,  {Kayo} I.,   {Suto} Y.,  2004,
  \mn@doi [\mnras] {10.1111/j.1365-2966.2004.07253.x}, \href
  {http://adsabs.harvard.edu/abs/2004MNRAS.347..813H} {347, 813}

\bibitem[\protect\citeauthoryear{{Hearin}, {Zentner}, {Berlind}  \&
  {Newman}}{{Hearin} et~al.}{2013}]{hearin_etal_13}
{Hearin} A.~P.,  {Zentner} A.~R.,  {Berlind} A.~A.,   {Newman} J.~A.,  2013,
  \mn@doi [\mnras] {10.1093/mnras/stt755}, \href
  {http://adsabs.harvard.edu/abs/2013MNRAS.433..659H} {433, 659}

\bibitem[\protect\citeauthoryear{{Hearin} et~al.,}{{Hearin}
  et~al.}{2016a}]{hearin_etal_16b}
{Hearin} A.,  et~al., 2016a, preprint, \href
  {http://adsabs.harvard.edu/abs/2016arXiv160604106H} {} (\mn@eprint {arXiv}
  {1606.04106})

\bibitem[\protect\citeauthoryear{{Hearin}, {Zentner}, {van den Bosch},
  {Campbell}  \& {Tollerud}}{{Hearin} et~al.}{2016b}]{hearin_etal_16}
{Hearin} A.~P.,  {Zentner} A.~R.,  {van den Bosch} F.~C.,  {Campbell} D.,
  {Tollerud} E.,  2016b, \mn@doi [\mnras] {10.1093/mnras/stw840}, \href
  {http://adsabs.harvard.edu/abs/2016MNRAS.460.2552H} {460, 2552}

\bibitem[\protect\citeauthoryear{{Hinton}}{{Hinton}}{2016}]{Hinton2016}
{Hinton} S.~R.,  2016, \mn@doi [The Journal of Open Source Software]
  {10.21105/joss.00045}, \href
  {http://adsabs.harvard.edu/abs/2016JOSS....1...45H} {1, 00045}

\bibitem[\protect\citeauthoryear{Hunter}{Hunter}{2007}]{Hunter:2007}
Hunter J.~D.,  2007, Computing In Science \& Engineering, 9, 90

\bibitem[\protect\citeauthoryear{{Jing}, {Mo}  \& {B{\"o}rner}}{{Jing}
  et~al.}{1998}]{jing_etal_98}
{Jing} Y.~P.,  {Mo} H.~J.,   {B{\"o}rner} G.,  1998, \mn@doi [\apj]
  {10.1086/305209}, \href {http://adsabs.harvard.edu/abs/1998ApJ...494....1J}
  {494, 1}

\bibitem[\protect\citeauthoryear{{Jones} et~al.,}{{Jones}
  et~al.}{2004}]{jones_etal_04}
{Jones} D.~H.,  et~al., 2004, \mn@doi [\mnras]
  {10.1111/j.1365-2966.2004.08353.x}, \href
  {http://adsabs.harvard.edu/abs/2004MNRAS.355..747J} {355, 747}

\bibitem[\protect\citeauthoryear{{Jose}, {Subramanian}, {Srianand}  \&
  {Samui}}{{Jose} et~al.}{2013}]{jose_etal_13}
{Jose} C.,  {Subramanian} K.,  {Srianand} R.,   {Samui} S.,  2013, \mn@doi
  [\mnras] {10.1093/mnras/sts503}, \href
  {http://adsabs.harvard.edu/abs/2013MNRAS.429.2333J} {429, 2333}

\bibitem[\protect\citeauthoryear{{Kauffmann}, {Nusser}  \&
  {Steinmetz}}{{Kauffmann} et~al.}{1997}]{kauffmann_etal_97}
{Kauffmann} G.,  {Nusser} A.,   {Steinmetz} M.,  1997, \mnras, \href
  {http://adsabs.harvard.edu/abs/1997MNRAS.286..795K} {286, 795}

\bibitem[\protect\citeauthoryear{{Kauffmann}, {Colberg}, {Diaferio}  \&
  {White}}{{Kauffmann} et~al.}{1999}]{kauffmann_etal_99}
{Kauffmann} G.,  {Colberg} J.~M.,  {Diaferio} A.,   {White} S.~D.~M.,  1999,
  \mn@doi [\mnras] {10.1046/j.1365-8711.1999.02202.x}, \href
  {http://adsabs.harvard.edu/abs/1999MNRAS.303..188K} {303, 188}

\bibitem[\protect\citeauthoryear{{Kim} et~al.,}{{Kim}
  et~al.}{2014}]{kim_etal_14}
{Kim} J.-W.,  et~al., 2014, \mn@doi [\mnras] {10.1093/mnras/stt2245}, \href
  {http://adsabs.harvard.edu/abs/2014MNRAS.438..825K} {438, 825}

\bibitem[\protect\citeauthoryear{{Kravtsov}, {Berlind}, {Wechsler}, {Klypin},
  {Gottl{\" o}ber}, {Allgood}  \& {Primack}}{{Kravtsov}
  et~al.}{2004}]{kravtsov_etal_04}
{Kravtsov} A.~V.,  {Berlind} A.~A.,  {Wechsler} R.~H.,  {Klypin} A.~A.,
  {Gottl{\" o}ber} S.,  {Allgood} B.,   {Primack} J.~R.,  2004, \apj, \href
  {http://adsabs.harvard.edu/cgi-bin/nph-bib_query?bibcode=2004ApJ...609...35K&db_key=AST}
  {609, 35}

\bibitem[\protect\citeauthoryear{{Lacey} \& {Cole}}{{Lacey} \&
  {Cole}}{1994}]{lacey_cole_94}
{Lacey} C.,  {Cole} S.,  1994, \mn@doi [\mnras] {10.1093/mnras/271.3.676},
  \href {http://adsabs.harvard.edu/abs/1994MNRAS.271..676L} {271, 676}

\bibitem[\protect\citeauthoryear{{Landy} \& {Szalay}}{{Landy} \&
  {Szalay}}{1993}]{landy_szalay_93}
{Landy} S.~D.,  {Szalay} A.~S.,  1993, \mn@doi [\apj] {10.1086/172900}, \href
  {http://adsabs.harvard.edu/cgi-bin/nph-bib_query?bibcode=1993ApJ...412...64L&db_key=AST}
  {412, 64}

\bibitem[\protect\citeauthoryear{{Leauthaud}, {Tinker}, {Behroozi}, {Busha}  \&
  {Wechsler}}{{Leauthaud} et~al.}{2011}]{leauthaud_etal_11}
{Leauthaud} A.,  {Tinker} J.,  {Behroozi} P.~S.,  {Busha} M.~T.,   {Wechsler}
  R.~H.,  2011, \mn@doi [\apj] {10.1088/0004-637X/738/1/45}, \href
  {http://adsabs.harvard.edu/abs/2011ApJ...738...45L} {738, 45}

\bibitem[\protect\citeauthoryear{{Leauthaud} et~al.,}{{Leauthaud}
  et~al.}{2012}]{leauthaud_etal_12}
{Leauthaud} A.,  et~al., 2012, \mn@doi [\apj] {10.1088/0004-637X/744/2/159},
  \href {http://adsabs.harvard.edu/abs/2012ApJ...744..159L} {744, 159}

\bibitem[\protect\citeauthoryear{{Lee}, {Giavalisco}, {Gnedin}, {Somerville},
  {Ferguson}, {Dickinson}  \& {Ouchi}}{{Lee} et~al.}{2006}]{lee_etal_06}
{Lee} K.-S.,  {Giavalisco} M.,  {Gnedin} O.~Y.,  {Somerville} R.~S.,
  {Ferguson} H.~C.,  {Dickinson} M.,   {Ouchi} M.,  2006, \mn@doi [\apj]
  {10.1086/500387}, \href {http://adsabs.harvard.edu/abs/2006ApJ...642...63L}
  {642, 63}

\bibitem[\protect\citeauthoryear{{Magliocchetti} \& {Porciani}}{{Magliocchetti}
  \& {Porciani}}{2003}]{magliocchetti_porciani_03}
{Magliocchetti} M.,  {Porciani} C.,  2003, \mn@doi [\mnras]
  {10.1046/j.1365-2966.2003.07094.x}, \href
  {http://adsabs.harvard.edu/abs/2003MNRAS.346..186M} {346, 186}

\bibitem[\protect\citeauthoryear{{Mandelbaum}, {Li}, {Kauffmann}  \&
  {White}}{{Mandelbaum} et~al.}{2009}]{mandelbaum_etal_09}
{Mandelbaum} R.,  {Li} C.,  {Kauffmann} G.,   {White} S.~D.~M.,  2009, \mn@doi
  [\mnras] {10.1111/j.1365-2966.2008.14235.x}, \href
  {http://adsabs.harvard.edu/abs/2009MNRAS.393..377M} {393, 377}

\bibitem[\protect\citeauthoryear{{Mao}, {Williamson}  \& {Wechsler}}{{Mao}
  et~al.}{2015}]{mao_etal_2015}
{Mao} Y.-Y.,  {Williamson} M.,   {Wechsler} R.~H.,  2015, \mn@doi [\apj]
  {10.1088/0004-637X/810/1/21}, \href
  {http://adsabs.harvard.edu/abs/2015ApJ...810...21M} {810, 21}

\bibitem[\protect\citeauthoryear{{Mar{\'{\i}}n}}{{Mar{\'{\i}}n}}{2011}]{marin_etal_11}
{Mar{\'{\i}}n} F.,  2011, \mn@doi [\apj] {10.1088/0004-637X/737/2/97}, \href
  {http://adsabs.harvard.edu/abs/2011ApJ...737...97M} {737, 97}

\bibitem[\protect\citeauthoryear{{McBride}, {Berlind}, {Scoccimarro},
  {Wechsler}, {Busha}, {Gardner}  \& {van den Bosch}}{{McBride}
  et~al.}{2009}]{McBride_etal_09}
{McBride} C.,  {Berlind} A.,  {Scoccimarro} R.,  {Wechsler} R.,  {Busha} M.,
  {Gardner} J.,   {van den Bosch} F.,  2009, in American Astronomical Society
  Meeting Abstracts \#213. p. 425.06

\bibitem[\protect\citeauthoryear{{McBride}, {Connolly}, {Gardner}, {Scranton},
  {Newman}, {Scoccimarro}, {Zehavi}  \& {Schneider}}{{McBride}
  et~al.}{2011}]{mcbride_etal_11}
{McBride} C.~K.,  {Connolly} A.~J.,  {Gardner} J.~P.,  {Scranton} R.,  {Newman}
  J.~A.,  {Scoccimarro} R.,  {Zehavi} I.,   {Schneider} D.~P.,  2011, \mn@doi
  [\apj] {10.1088/0004-637X/726/1/13}, \href
  {http://adsabs.harvard.edu/abs/2011ApJ...726...13M} {726, 13}

\bibitem[\protect\citeauthoryear{{More}, {van den Bosch}, {Cacciato}, {Mo},
  {Yang}  \& {Li}}{{More} et~al.}{2009}]{more_etal_09}
{More} S.,  {van den Bosch} F.~C.,  {Cacciato} M.,  {Mo} H.~J.,  {Yang} X.,
  {Li} R.,  2009, \mn@doi [\mnras] {10.1111/j.1365-2966.2008.14095.x}, \href
  {http://adsabs.harvard.edu/abs/2009MNRAS.392..801M} {392, 801}

\bibitem[\protect\citeauthoryear{{Moustakas} \& {Somerville}}{{Moustakas} \&
  {Somerville}}{2002}]{moustakas_somerville_02}
{Moustakas} L.~A.,  {Somerville} R.~S.,  2002, \mn@doi [\apj] {10.1086/342133},
  \href {http://adsabs.harvard.edu/abs/2002ApJ...577....1M} {577, 1}

\bibitem[\protect\citeauthoryear{{Narayanan}, {Berlind}  \&
  {Weinberg}}{{Narayanan} et~al.}{2000}]{narayanan_etal_00}
{Narayanan} V.~K.,  {Berlind} A.~A.,   {Weinberg} D.~H.,  2000, \mn@doi [\apj]
  {10.1086/308140}, \href {http://adsabs.harvard.edu/abs/2000ApJ...528....1N}
  {528, 1}

\bibitem[\protect\citeauthoryear{{Navarro}, {Frenk}  \& {White}}{{Navarro}
  et~al.}{1997}]{navarro_etal_97}
{Navarro} J.~F.,  {Frenk} C.~S.,   {White} S.~D.~M.,  1997, \mn@doi [\apj]
  {10.1086/304888}, \href {http://adsabs.harvard.edu/abs/1997ApJ...490..493N}
  {490, 493}

\bibitem[\protect\citeauthoryear{Nelder \& Mead}{Nelder \&
  Mead}{1965}]{nelder_mead_65}
Nelder J.~A.,  Mead R.,  1965, \mn@doi [The Computer Journal]
  {10.1093/comjnl/7.4.308}, 7, 308

\bibitem[\protect\citeauthoryear{{Neyman} \& {Scott}}{{Neyman} \&
  {Scott}}{1952}]{neyman_scott_52}
{Neyman} J.,  {Scott} E.~L.,  1952, \mn@doi [\apj] {10.1086/145599}, \href
  {http://adsabs.harvard.edu/abs/1952ApJ...116..144N} {116, 144}

\bibitem[\protect\citeauthoryear{{Nikoloudakis}, {Shanks}  \&
  {Sawangwit}}{{Nikoloudakis} et~al.}{2013}]{nikoloudakis_etal_13}
{Nikoloudakis} N.,  {Shanks} T.,   {Sawangwit} U.,  2013, \mn@doi [\mnras]
  {10.1093/mnras/sts475}, \href
  {http://adsabs.harvard.edu/abs/2013MNRAS.429.2032N} {429, 2032}

\bibitem[\protect\citeauthoryear{{Norberg}, {Baugh}, {Gazta{\~n}aga}  \&
  {Croton}}{{Norberg} et~al.}{2009}]{norberg_etal_09}
{Norberg} P.,  {Baugh} C.~M.,  {Gazta{\~n}aga} E.,   {Croton} D.~J.,  2009,
  \mn@doi [\mnras] {10.1111/j.1365-2966.2009.14389.x}, \href
  {http://adsabs.harvard.edu/abs/2009MNRAS.396...19N} {396, 19}

\bibitem[\protect\citeauthoryear{{Parejko} et~al.,}{{Parejko}
  et~al.}{2013}]{parejko_etal_13}
{Parejko} J.~K.,  et~al., 2013, \mn@doi [\mnras] {10.1093/mnras/sts314}, \href
  {http://adsabs.harvard.edu/abs/2013MNRAS.429...98P} {429, 98}

\bibitem[\protect\citeauthoryear{{Peacock} \& {Smith}}{{Peacock} \&
  {Smith}}{2000}]{peacock_smith_00}
{Peacock} J.~A.,  {Smith} R.~E.,  2000, \mnras, \href
  {http://adsabs.harvard.edu/cgi-bin/nph-bib_query?bibcode=2000MNRAS.318.1144P&db_key=AST}
  {318, 1144}

\bibitem[\protect\citeauthoryear{{Piscionere}, {Berlind}, {McBride}  \&
  {Scoccimarro}}{{Piscionere} et~al.}{2015}]{piscionere_etal_14}
{Piscionere} J.~A.,  {Berlind} A.~A.,  {McBride} C.~K.,   {Scoccimarro} R.,
  2015, \mn@doi [\apj] {10.1088/0004-637X/806/1/125}, \href
  {http://adsabs.harvard.edu/abs/2015ApJ...806..125P} {806, 125}

\bibitem[\protect\citeauthoryear{{Planck Collaboration} et~al.,}{{Planck
  Collaboration} et~al.}{2014}]{planck_etal_14}
{Planck Collaboration} et~al., 2014, \mn@doi [\aap]
  {10.1051/0004-6361/201321591}, \href
  {http://adsabs.harvard.edu/abs/2014A%26A...571A..16P} {571, A16}

\bibitem[\protect\citeauthoryear{{Reid}, {Seo}, {Leauthaud}, {Tinker}  \&
  {White}}{{Reid} et~al.}{2015}]{reid_etal_14}
{Reid} B.~A.,  {Seo} H.-J.,  {Leauthaud} A.,  {Tinker} J.~L.,   {White} M.,
  2015, \mn@doi [\mnras] {10.1093/mnras/stu1391}, \href
  {http://adsabs.harvard.edu/abs/2014MNRAS.444..476R} {444, 476}

\bibitem[\protect\citeauthoryear{{Richardson}, {Chatterjee}, {Zheng}, {Myers}
  \& {Hickox}}{{Richardson} et~al.}{2013}]{richardson_etal_13}
{Richardson} J.,  {Chatterjee} S.,  {Zheng} Z.,  {Myers} A.~D.,   {Hickox} R.,
  2013, \mn@doi [\apj] {10.1088/0004-637X/774/2/143}, \href
  {http://adsabs.harvard.edu/abs/2013ApJ...774..143R} {774, 143}

\bibitem[\protect\citeauthoryear{{Scherrer} \& {Bertschinger}}{{Scherrer} \&
  {Bertschinger}}{1991}]{scherrer_bertchinger_91}
{Scherrer} R.~J.,  {Bertschinger} E.,  1991, \mn@doi [\apj] {10.1086/170658},
  \href {http://adsabs.harvard.edu/abs/1991ApJ...381..349S} {381, 349}

\bibitem[\protect\citeauthoryear{{Scherrer} \& {Weinberg}}{{Scherrer} \&
  {Weinberg}}{1998}]{scherrer_weinberg_98}
{Scherrer} R.~J.,  {Weinberg} D.~H.,  1998, \mn@doi [\apj] {10.1086/306113},
  \href {http://adsabs.harvard.edu/abs/1998ApJ...504..607S} {504, 607}

\bibitem[\protect\citeauthoryear{{Scoccimarro}}{{Scoccimarro}}{1998}]{scoccimarro_98}
{Scoccimarro} R.,  1998, \mn@doi [\mnras] {10.1046/j.1365-8711.1998.01845.x},
  \href {http://adsabs.harvard.edu/abs/1998MNRAS.299.1097S} {299, 1097}

\bibitem[\protect\citeauthoryear{{Scoccimarro}}{{Scoccimarro}}{2000}]{scoccimarro_2000}
{Scoccimarro} R.,  2000, \mn@doi [\apj] {10.1086/317248}, \href
  {http://adsabs.harvard.edu/abs/2000ApJ...544..597S} {544, 597}

\bibitem[\protect\citeauthoryear{{Scoccimarro}, {Sheth}, {Hui}  \&
  {Jain}}{{Scoccimarro} et~al.}{2001}]{scoccimarro_etal_01}
{Scoccimarro} R.,  {Sheth} R.~K.,  {Hui} L.,   {Jain} B.,  2001, \mn@doi [\apj]
  {10.1086/318261}, \href
  {http://adsabs.harvard.edu/cgi-bin/nph-bib_query?bibcode=2001ApJ...546...20S&db_key=AST}
  {546, 20}

\bibitem[\protect\citeauthoryear{{Seljak}}{{Seljak}}{2000}]{seljak_00}
{Seljak} U.,  2000, \mnras, \href
  {http://adsabs.harvard.edu/cgi-bin/nph-bib_query?bibcode=2000MNRAS.318..203S&db_key=AST}
  {318, 203}

\bibitem[\protect\citeauthoryear{{Sinha} \& {Garrison}}{{Sinha} \&
  {Garrison}}{2017}]{corrfunc_sinha_lehman}
{Sinha} M.,  {Garrison} L.,  2017, {Corrfunc: Blazing fast correlation
  functions on the CPU}, Astrophysics Source Code Library (\mn@eprint {ascl}
  {1703.003})

\bibitem[\protect\citeauthoryear{{Spergel} et~al.,}{{Spergel}
  et~al.}{2007}]{spergel_etal_07}
{Spergel} D.~N.,  et~al., 2007, \mn@doi [\apjs] {10.1086/513700}, \href
  {http://adsabs.harvard.edu/abs/2007ApJS..170..377S} {170, 377}

\bibitem[\protect\citeauthoryear{{Springel}}{{Springel}}{2005}]{springel_05}
{Springel} V.,  2005, \mn@doi [\mnras] {10.1111/j.1365-2966.2005.09655.x},
  \href
  {http://adsabs.harvard.edu/cgi-bin/nph-bib_query?bibcode=2005MNRAS.364.1105S&db_key=AST}
  {364, 1105}

\bibitem[\protect\citeauthoryear{{Strauss} et~al.,}{{Strauss}
  et~al.}{2002}]{strauss_etal_02}
{Strauss} M.~A.,  et~al., 2002, \aj, \href
  {http://adsabs.harvard.edu/cgi-bin/nph-bib_query?bibcode=2002AJ....124.1810S&db_key=AST}
  {124, 1810}

\bibitem[\protect\citeauthoryear{{Tegmark} et~al.,}{{Tegmark}
  et~al.}{2004}]{tegmark_etal_04a}
{Tegmark} M.,  et~al., 2004, \apj, \href
  {http://adsabs.harvard.edu/cgi-bin/nph-bib_query?bibcode=2004ApJ...606..702T&db_key=AST}
  {606, 702}

\bibitem[\protect\citeauthoryear{{Tinker} \& {Wetzel}}{{Tinker} \&
  {Wetzel}}{2010}]{tinker_wetzel_10}
{Tinker} J.~L.,  {Wetzel} A.~R.,  2010, \mn@doi [\apj]
  {10.1088/0004-637X/719/1/88}, \href
  {http://adsabs.harvard.edu/abs/2010ApJ...719...88T} {719, 88}

\bibitem[\protect\citeauthoryear{{Tinker}, {Weinberg}, {Zheng}  \&
  {Zehavi}}{{Tinker} et~al.}{2005}]{tinker_etal_05}
{Tinker} J.~L.,  {Weinberg} D.~H.,  {Zheng} Z.,   {Zehavi} I.,  2005, \mn@doi
  [\apj] {10.1086/432084}, \href
  {http://adsabs.harvard.edu/cgi-bin/nph-bib_query?bibcode=2005ApJ...631...41T&db_key=AST}
  {631, 41}

\bibitem[\protect\citeauthoryear{{Tinker}, {Wechsler}  \& {Zheng}}{{Tinker}
  et~al.}{2010}]{tinker_etal_10}
{Tinker} J.~L.,  {Wechsler} R.~H.,   {Zheng} Z.,  2010, \mn@doi [\apj]
  {10.1088/0004-637X/709/1/67}, \href
  {http://adsabs.harvard.edu/abs/2010ApJ...709...67T} {709, 67}

\bibitem[\protect\citeauthoryear{{Tinker}, {Leauthaud}, {Bundy}, {George},
  {Behroozi}, {Massey}, {Rhodes}  \& {Wechsler}}{{Tinker}
  et~al.}{2013}]{tinker_etal_13}
{Tinker} J.~L.,  {Leauthaud} A.,  {Bundy} K.,  {George} M.~R.,  {Behroozi} P.,
  {Massey} R.,  {Rhodes} J.,   {Wechsler} R.~H.,  2013, \mn@doi [\apj]
  {10.1088/0004-637X/778/2/93}, \href
  {http://adsabs.harvard.edu/abs/2013ApJ...778...93T} {778, 93}

\bibitem[\protect\citeauthoryear{{Vale} \& {Ostriker}}{{Vale} \&
  {Ostriker}}{2004}]{vale_ostriker_04}
{Vale} A.,  {Ostriker} J.~P.,  2004, \mn@doi [\mnras]
  {10.1111/j.1365-2966.2004.08059.x}, \href
  {http://adsabs.harvard.edu/abs/2004MNRAS.353..189V} {353, 189}

\bibitem[\protect\citeauthoryear{Van Der~Walt, Colbert  \& Varoquaux}{Van
  Der~Walt et~al.}{2011}]{van2011numpy}
Van Der~Walt S.,  Colbert S.~C.,   Varoquaux G.,  2011, Computing in Science \&
  Engineering, 13, 22

\bibitem[\protect\citeauthoryear{{Wake}, {Croom}, {Sadler}  \&
  {Johnston}}{{Wake} et~al.}{2008}]{wake_etal_08}
{Wake} D.~A.,  {Croom} S.~M.,  {Sadler} E.~M.,   {Johnston} H.~M.,  2008,
  \mn@doi [\mnras] {10.1111/j.1365-2966.2008.14039.x}, \href
  {http://adsabs.harvard.edu/abs/2008MNRAS.391.1674W} {391, 1674}

\bibitem[\protect\citeauthoryear{{Wake} et~al.,}{{Wake}
  et~al.}{2011}]{wake_etal_11}
{Wake} D.~A.,  et~al., 2011, \mn@doi [\apj] {10.1088/0004-637X/728/1/46}, \href
  {http://adsabs.harvard.edu/abs/2011ApJ...728...46W} {728, 46}

\bibitem[\protect\citeauthoryear{{Warren}, {Abazajian}, {Holz}  \&
  {Teodoro}}{{Warren} et~al.}{2006}]{warren_etal_06}
{Warren} M.~S.,  {Abazajian} K.,  {Holz} D.~E.,   {Teodoro} L.,  2006, \mn@doi
  [\apj] {10.1086/504962}, \href
  {http://adsabs.harvard.edu/cgi-bin/nph-bib_query?bibcode=2006ApJ...646..881W&db_key=AST}
  {646, 881}

\bibitem[\protect\citeauthoryear{{Watson}, {Berlind}, {McBride}  \&
  {Masjedi}}{{Watson} et~al.}{2010}]{watson_etal_10}
{Watson} D.~F.,  {Berlind} A.~A.,  {McBride} C.~K.,   {Masjedi} M.,  2010,
  \mn@doi [\apj] {10.1088/0004-637X/709/1/115}, \href
  {http://adsabs.harvard.edu/abs/2010ApJ...709..115W} {709, 115}

\bibitem[\protect\citeauthoryear{{Watson}, {Berlind}  \& {Zentner}}{{Watson}
  et~al.}{2011}]{watson_etal_11}
{Watson} D.~F.,  {Berlind} A.~A.,   {Zentner} A.~R.,  2011, \mn@doi [\apj]
  {10.1088/0004-637X/738/1/22}, \href
  {http://adsabs.harvard.edu/abs/2011ApJ...738...22W} {738, 22}

\bibitem[\protect\citeauthoryear{{Watson}, {Berlind}, {McBride}, {Hogg}  \&
  {Jiang}}{{Watson} et~al.}{2012}]{watson_etal_12}
{Watson} D.~F.,  {Berlind} A.~A.,  {McBride} C.~K.,  {Hogg} D.~W.,   {Jiang}
  T.,  2012, \mn@doi [\apj] {10.1088/0004-637X/749/1/83}, \href
  {http://adsabs.harvard.edu/abs/2012ApJ...749...83W} {749, 83}

\bibitem[\protect\citeauthoryear{{White} et~al.,}{{White}
  et~al.}{2011}]{white_etal_11}
{White} M.,  et~al., 2011, \mn@doi [\apj] {10.1088/0004-637X/728/2/126}, \href
  {http://adsabs.harvard.edu/abs/2011ApJ...728..126W} {728, 126}

\bibitem[\protect\citeauthoryear{{Yan}, {Madgwick}  \& {White}}{{Yan}
  et~al.}{2003}]{yan_etal_03}
{Yan} R.,  {Madgwick} D.~S.,   {White} M.,  2003, \mn@doi [\apj]
  {10.1086/379067}, \href {http://adsabs.harvard.edu/abs/2003ApJ...598..848Y}
  {598, 848}

\bibitem[\protect\citeauthoryear{{Yang}, {Mo}  \& {van den Bosch}}{{Yang}
  et~al.}{2003}]{yang_etal_03}
{Yang} X.,  {Mo} H.~J.,   {van den Bosch} F.~C.,  2003, \mn@doi [\mnras]
  {10.1046/j.1365-8711.2003.06254.x}, \href
  {http://adsabs.harvard.edu/abs/2003MNRAS.339.1057Y} {339, 1057}

\bibitem[\protect\citeauthoryear{{York} et~al.,}{{York}
  et~al.}{2000}]{york_etal_00}
{York} D.~G.,  et~al., 2000, \aj, \href
  {http://adsabs.harvard.edu/cgi-bin/nph-bib_query?bibcode=2000AJ....120.1579Y&db_key=AST}
  {120, 1579}

\bibitem[\protect\citeauthoryear{{Zehavi} et~al.,}{{Zehavi}
  et~al.}{2002}]{zehavi_etal_02}
{Zehavi} I.,  et~al., 2002, \mn@doi [\apj] {10.1086/339893}, \href
  {http://adsabs.harvard.edu/cgi-bin/nph-bib_query?bibcode=2002ApJ...571..172Z&db_key=AST}
  {571, 172}

\bibitem[\protect\citeauthoryear{{Zehavi} et~al.,}{{Zehavi}
  et~al.}{2004}]{zehavi_etal_04}
{Zehavi} I.,  et~al., 2004, \apj, \href
  {http://adsabs.harvard.edu/cgi-bin/nph-bib_query?bibcode=2004ApJ...608...16Z&db_key=AST}
  {608, 16}

\bibitem[\protect\citeauthoryear{{Zehavi} et~al.,}{{Zehavi}
  et~al.}{2005}]{zehavi_etal_05}
{Zehavi} I.,  et~al., 2005, \apj, \href
  {http://adsabs.harvard.edu/cgi-bin/nph-bib_query?bibcode=2004ApJ...608...16Z&db_key=AST}
  {630, 1}

\bibitem[\protect\citeauthoryear{{Zehavi} et~al.,}{{Zehavi}
  et~al.}{2011}]{zehavi_etal_11}
{Zehavi} I.,  et~al., 2011, \mn@doi [\apj] {10.1088/0004-637X/736/1/59}, \href
  {http://adsabs.harvard.edu/abs/2011ApJ...736...59Z} {736, 59}

\bibitem[\protect\citeauthoryear{{Zehavi}, {Contreras}, {Padilla}, {Smith},
  {Baugh}  \& {Norberg}}{{Zehavi} et~al.}{2017}]{zehavi_assembly_bias_17}
{Zehavi} I.,  {Contreras} S.,  {Padilla} N.,  {Smith} N.~J.,  {Baugh} C.~M.,
  {Norberg} P.,  2017, preprint, \href
  {http://adsabs.harvard.edu/abs/2017arXiv170607871Z} {} (\mn@eprint {arXiv}
  {1706.07871})

\bibitem[\protect\citeauthoryear{{Zentner}, {Berlind}, {Bullock}, {Kravtsov}
  \& {Wechsler}}{{Zentner} et~al.}{2005}]{zentner_etal_05}
{Zentner} A.~R.,  {Berlind} A.~A.,  {Bullock} J.~S.,  {Kravtsov} A.~V.,
  {Wechsler} R.~H.,  2005, \apj, \href
  {http://adsabs.harvard.edu/cgi-bin/nph-bib_query?bibcode=2005ApJ...624..505Z&db_key=AST}
  {624, 505}

\bibitem[\protect\citeauthoryear{{Zentner}, {Hearin}  \& {van den
  Bosch}}{{Zentner} et~al.}{2014}]{zentner_etal_14}
{Zentner} A.~R.,  {Hearin} A.~P.,   {van den Bosch} F.~C.,  2014, \mn@doi
  [\mnras] {10.1093/mnras/stu1383}, \href
  {http://adsabs.harvard.edu/abs/2014MNRAS.443.3044Z} {443, 3044}

\bibitem[\protect\citeauthoryear{{Zentner}, {Hearin}, {van den Bosch}, {Lange}
  \& {Villarreal}}{{Zentner} et~al.}{2016}]{zentner_etal_16}
{Zentner} A.~R.,  {Hearin} A.,  {van den Bosch} F.~C.,  {Lange} J.~U.,
  {Villarreal} A.,  2016, preprint, \href
  {http://adsabs.harvard.edu/abs/2016arXiv160607817Z} {} (\mn@eprint {arXiv}
  {1606.07817})

\bibitem[\protect\citeauthoryear{{Zheng}}{{Zheng}}{2004}]{zheng_04}
{Zheng} Z.,  2004, \mn@doi [\apj] {10.1086/421542}, \href
  {http://adsabs.harvard.edu/abs/2004ApJ...610...61Z} {610, 61}

\bibitem[\protect\citeauthoryear{{Zheng} \& {Guo}}{{Zheng} \&
  {Guo}}{2016}]{zheng_guo_16}
{Zheng} Z.,  {Guo} H.,  2016, \mn@doi [\mnras] {10.1093/mnras/stw523}, \href
  {http://adsabs.harvard.edu/abs/2016MNRAS.458.4015Z} {458, 4015}

\bibitem[\protect\citeauthoryear{{Zheng} \& {Weinberg}}{{Zheng} \&
  {Weinberg}}{2007}]{zheng_weinberg_07}
{Zheng} Z.,  {Weinberg} D.~H.,  2007, \mn@doi [\apj] {10.1086/512151}, \href
  {http://adsabs.harvard.edu/abs/2007ApJ...659....1Z} {659, 1}

\bibitem[\protect\citeauthoryear{{Zheng} et~al.,}{{Zheng}
  et~al.}{2005}]{zheng_etal_05}
{Zheng} Z.,  et~al., 2005, \mn@doi [\apj] {10.1086/466510}, \href
  {http://adsabs.harvard.edu/abs/2005ApJ...633..791Z} {633, 791}

\bibitem[\protect\citeauthoryear{{Zheng}, {Coil}  \& {Zehavi}}{{Zheng}
  et~al.}{2007}]{zheng_etal_07}
{Zheng} Z.,  {Coil} A.~L.,   {Zehavi} I.,  2007, \mn@doi [\apj]
  {10.1086/521074}, \href {http://adsabs.harvard.edu/abs/2007ApJ...667..760Z}
  {667, 760}

\bibitem[\protect\citeauthoryear{{Zheng}, {Zehavi}, {Eisenstein}, {Weinberg}
  \& {Jing}}{{Zheng} et~al.}{2009}]{zheng_etal_09}
{Zheng} Z.,  {Zehavi} I.,  {Eisenstein} D.~J.,  {Weinberg} D.~H.,   {Jing}
  Y.~P.,  2009, \mn@doi [\apj] {10.1088/0004-637X/707/1/554}, \href
  {http://adsabs.harvard.edu/abs/2009ApJ...707..554Z} {707, 554}

\bibitem[\protect\citeauthoryear{{van den Bosch}, {Yang}  \& {Mo}}{{van den
  Bosch} et~al.}{2003}]{vandenbosch_etal_03}
{van den Bosch} F.~C.,  {Yang} X.,   {Mo} H.~J.,  2003, \mn@doi [\mnras]
  {10.1046/j.1365-8711.2003.06335.x}, \href
  {http://adsabs.harvard.edu/abs/2003MNRAS.340..771V} {340, 771}

\bibitem[\protect\citeauthoryear{{van den Bosch}, {Weinmann}, {Yang}, {Mo},
  {Li}  \& {Jing}}{{van den Bosch} et~al.}{2005}]{vandenbosch_etal_05}
{van den Bosch} F.~C.,  {Weinmann} S.~M.,  {Yang} X.,  {Mo} H.~J.,  {Li} C.,
  {Jing} Y.~P.,  2005, \mn@doi [\mnras] {10.1111/j.1365-2966.2005.09260.x},
  \href {http://adsabs.harvard.edu/abs/2005MNRAS.361.1203V} {361, 1203}

\bibitem[\protect\citeauthoryear{{van den Bosch} et~al.,}{{van den Bosch}
  et~al.}{2007}]{vandenbosch_etal_07}
{van den Bosch} F.~C.,  et~al., 2007, \mn@doi [\mnras]
  {10.1111/j.1365-2966.2007.11493.x}, \href
  {http://adsabs.harvard.edu/abs/2007MNRAS.376..841V} {376, 841}

\makeatother
\end{thebibliography}

\bsp	
\label{lastpage}

\end{document}